\documentclass[prl,aps,epsf,twocolumn,longbibliography]{revtex4-1}
\usepackage{times}
\usepackage{graphicx}
\usepackage{float}
\usepackage{latexsym,amsmath,amssymb,bm,euscript}
\usepackage{color}
\usepackage{epstopdf}
\usepackage[colorlinks=true,linkcolor=blue,citecolor=blue,urlcolor=blue]{hyperref}
\usepackage{inputenc}

\begin{document}

\title{Robust $d$-Wave Superconductivity in the Square-Lattice $t$-$J$ Model} 

\author{Shoushu Gong$^1$}
\email{shoushu.gong$@$buaa.edu.cn}

\author{W. Zhu$^2$}
\email{zhuwei$@$westlake.edu.cn}

\author{D. N. Sheng$^3$}
\email{donna.sheng1$@$csun.edu}
\affiliation{$^1$Department of Physics, Beihang University, Beijing 100191, China\\
$^2$ School of Science, Westlake University, Hangzhou 310024, China, and \\
Institute of Natural Sciences, Westlake Institute of Advanced Study, Hangzhou 310024, China, and \\
Key Laboratory for Quantum Materials of Zhejiang Province, Westlake University, Hangzhou 310024, China\\
$^3$Department of Physics and Astronomy, California State University Northridge, California 91330, USA}

\begin{abstract}
Unravelling competing orders emergent in doped Mott insulators and their interplay with unconventional superconductivity is one of the major challenges in condensed matter physics. To explore possible superconducting state in doped Mott insulator, we study the square-lattice $t$-$J$ model with both the nearest-neighbor and next-nearest-neighbor electron hoppings and spin interactions. By using the state-of-the-art density matrix renormalization group calculation with imposing charge $U(1)$ and spin $SU(2)$ symmetries on the six-leg cylinders, we establish a quantum phase diagram including three phases: a stripe charge density wave phase, a superconducting phase without static charge order, and a superconducting phase coexistent with a weak charge stripe order. Crucially, we demonstrate that the superconducting phase has a power-law pairing correlation that decays much slower than the charge density and spin correlations, which is a quasi-1D descendant of the uniform d-wave superconductor in two dimensions. These findings reveal that enhanced charge and spin fluctuations with optimal doping is able to produce robust d-wave superconductivity in doped Mott insulators, providing a foundation for connecting theories of superconductivity to models of strongly correlated systems.
\end{abstract}

\maketitle

{\it Introduction.---}
To understand the emergence of unconventional superconductivity (SC) is one of the major challenges of modern physics~\cite{Zaanen2015,Taillefer2019}. 
Despite intensive studies in the past 30 years, it remains elusive if a robust SC state can emerge in the electron systems with strong repulsive interaction.
Since the SC phase is usually realized by doping the parent antiferromagnetic compounds such as cuprate-based materials, the Hubbard model and the closely related $t$-$J$ model are taken as canonical models for studying SC
in strongly correlated systems~\cite{Zaanen2015,Taillefer2019,PALee2006,Fukuyama2008,Weng1997}. 
Lacking of well controlled analytical solutions in two dimensions (2D), unbiased computational studies play an important role in establishing the quantum phases in such models.
So far, the common consensus is that charge and spin intertwined orders are dominant in lightly doped Hubbard and $t$-$J$ models on the square lattice, while SC  correlations are relatively weak on wider 
systems \cite{SWhite1998,SWhite1999,SWhite2003,Sorella2002,Fehske2005,Corboz2014,LeBlanc2015,BXZheng2017,Noack2017,EWHuang2017,Imada2018,Corboz2019,HCJiang2018,MPQin2020}. 
The inconsistency of these results with the insight from experimental observations, i.e. a SC ``dome" throughout a range of doping parent antiferromagnetic compounds, poses a fundamental challenge to our understanding of strongly correlated electron systems~\cite{Zaanen2015,Taillefer2019}. 

Intuitively, introducing the next-nearest-neighbor hopping $t_2$ to the basic Hubbard or $t$-$J$ models should be more realistic for describing materials \cite{Pavarini2001,Tanaka2004,ZXShen1998}, which may help to weaken charge order and enhance SC \cite{SWhite1999,Corboz2019,SWhite2009,YCChen2004,Dagotto2001,Bejas2012,Walter2014}.
Specifically, recent studies of the $t_1$-$t_2$ Hubbard model on the width-4 cylinder observed a quasi-long-range SC correlation \cite{HCJiang2017,HCJiang2019,YFJiang2020}, which coexists with the power-law charge density correlation in the form of the Luther-Emery liquid \cite{Emery1974,Balents1996,Kivelson2004,Kivelson2020,Jiang2020c}. 
However, a more recent numerical study suggested that there can be different d-wave symmetries in such a system, and a plaquette d-wave correlation may be favored on the width-4 cylinder, which does not represent a true d-wave SC order in the 2D limit \cite{Chung2020}. 
This work also highlights the importance of going to wider systems, which is an essential step towards understanding the competing orders in the 2D limit. 

To make a significant progress towards understanding SC in 2D strongly correlated systems, we study the quantum phases in lightly doped square-lattice $t$-$J$ model using the state-of-the-art density matrix renormalization group (DMRG)~\cite{SWhite1992,SWhite1993}, and demonstrate a global phase diagram on the width-6 cylinder by tuning doping level $\delta$ and hopping ratio $t_2 / t_1$.
We identify three distinct phases: a stripe charge density wave (CDW) phase, a uniform d-wave SC phase, and a SC phase coexistent with a weak CDW order.
The intermediate uniform SC phase occupies a large portion of the phase diagram upon increasing doping level.
The SC correlation has a power-law quasi-long-range order with the Luttinger exponent reaching a small value $K_{sc} \approx 0.36$ and the ordinary d-wave symmetry, which dominates over other correlations.
Crucially, through a rigorous bond-dimension scaling, we provide compelling evidence that the SC phase is the quasi-1D descendant of a robust 2D superconductor.
These results offer strong evidence that SC order can overtake the tendency of other orderings in a doped Mott insulator, based on which we discuss some insight for doping-induced quantum phase transitions and compare with experimental observations in the cuprate systems.

\begin{figure}[t]
	\includegraphics[width=0.32\linewidth]{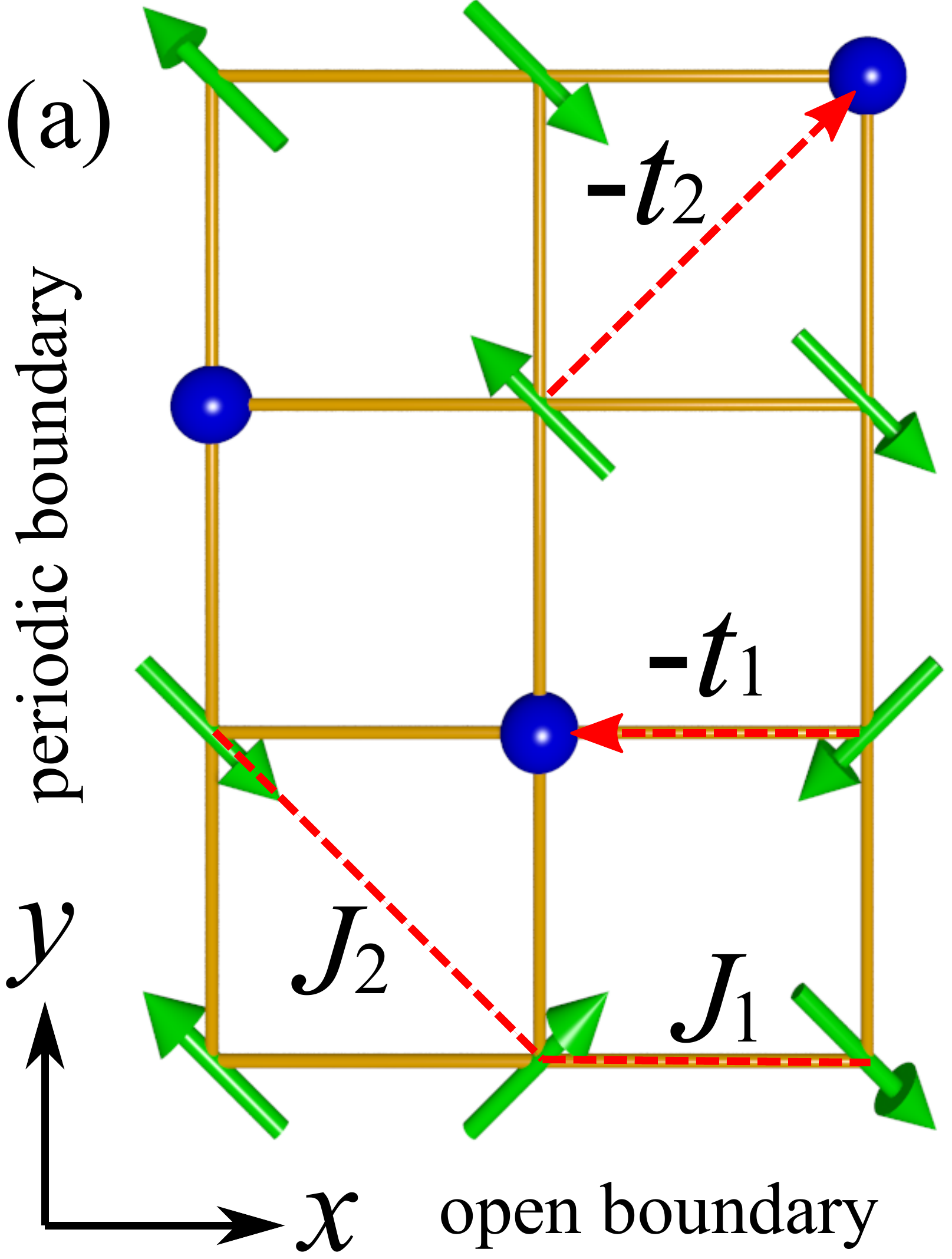}
	\includegraphics[width=0.59\linewidth]{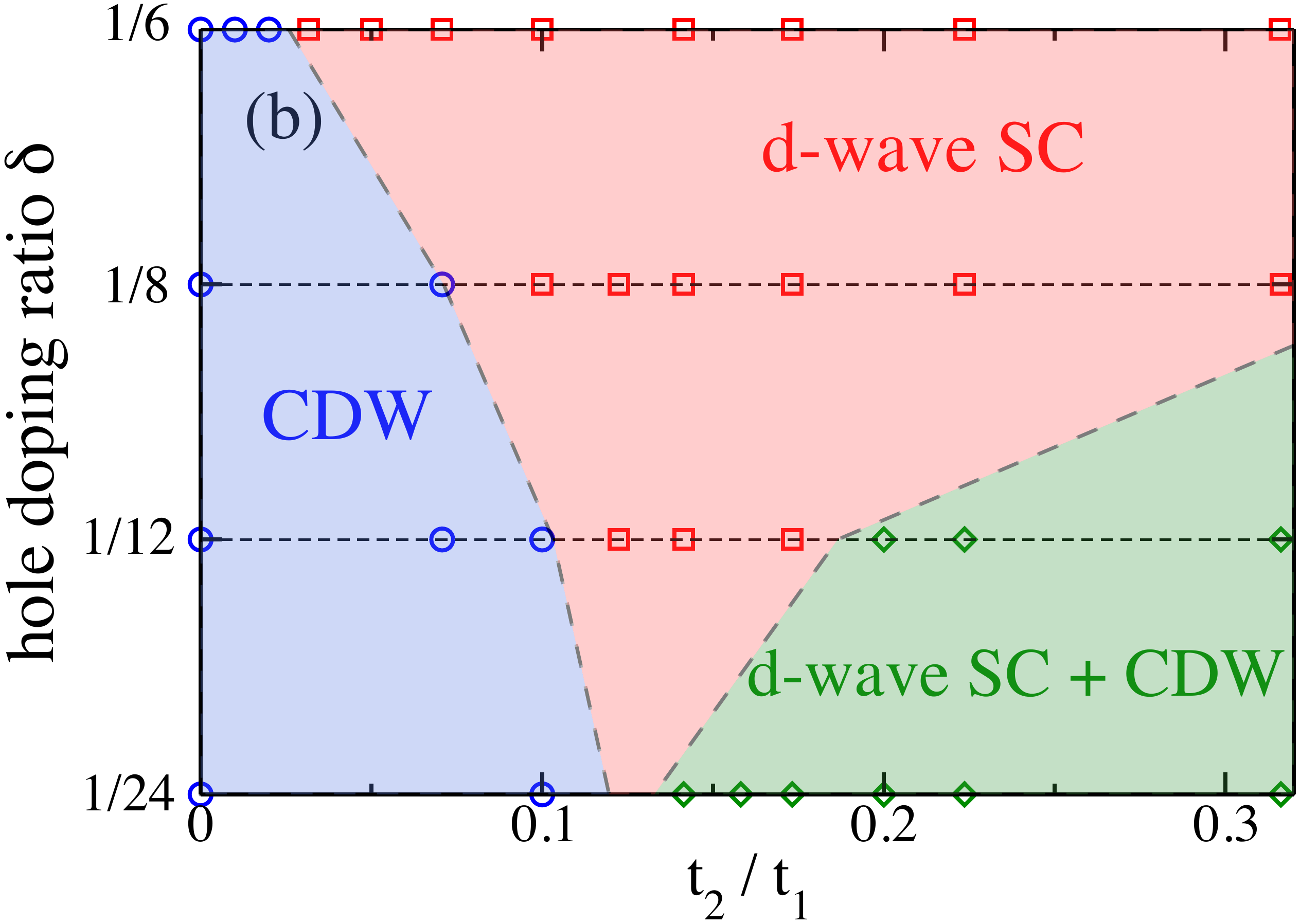}
	\includegraphics[width=1\linewidth]{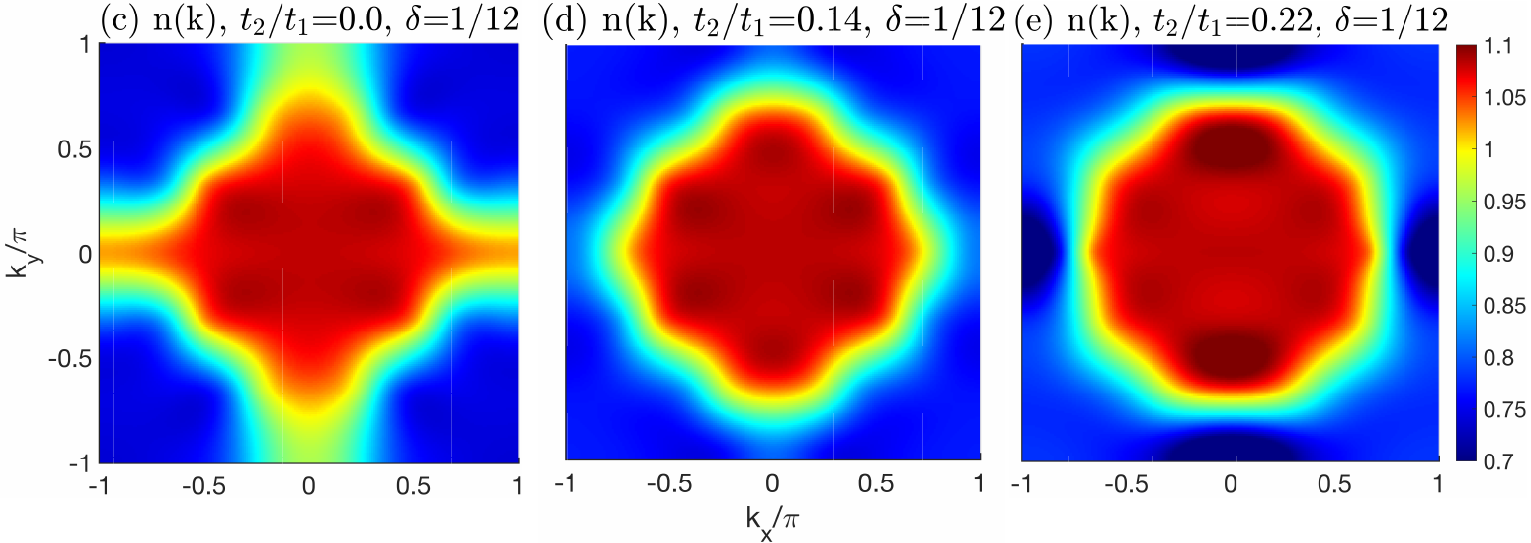}
	\caption{Global quantum phase diagram.
		(a) Schematic plot of the $t$-$J$ model on the square lattice, where arrows and circles respectively denote electrons and doped holes. The model has both the nearest-neighbor and the next-nearest-neighbor hoppings ($t_1$ and $t_2$) and spin exchange ($J_1$ and $J_2$) interactions. (b) Quantum phase diagram of the model obtained on the $L_y = 6$ cylinder based on the static charge density pattern shown in Fig.~\ref{fig:cdw}. For $0 \leq t_2 / t_1 \leq 0.32$ and doping level $1/24 \leq \delta \leq 1/6$, we identify a CDW phase, a uniform d-wave SC phase, and a coexistent d-wave SC and CDW (SC + CDW) phase. The Luttinger exponents of SC pairing and density correlations cross over between different phases. Momentum distribution functions $n(\bf k)$ for (c) CDW phase, (d) uniform SC phase, and (e) SC + CDW coexistent phase.
	}
	\label{fig:phase}
\end{figure}

{\it Solving the $t$-$J$ model with DMRG.---} 
The extended $t$-$J$ model is defined as
\begin{equation*}\label{Hamiltonian}
H = -\sum_{\{ij\},\sigma}t_{ij}(\hat{c}^{\dagger}_{i,\sigma} \hat{c}_{j,\sigma} + h.c.)  + \sum_{\{ij\}} J_{ij} (\hat{\bf S}_i \cdot \hat{\bf S}_j - \frac{1}{4} \hat{n}_i \hat{n}_j),
\end{equation*}
where $\hat{c}^{\dagger}_{i,\sigma}$ and $\hat{c}_{i,\sigma}$ are the creation and annihilation operators for the electron with spin $\sigma$ ($\sigma = \pm 1/2$) at the site $i$, $\hat{\bf S}_{i}$ is the spin-$1/2$ operator, and $\hat{n}_i \equiv \sum_{\sigma} \hat{c}^{\dagger}_{i,\sigma} \hat{c}_{i,\sigma}$ is the electron number operator.
We consider the nearest-neighbor (NN) and next-nearest-neighbor (NNN) hoppings ($t_1$ and $t_2$) and interactions ($J_1$ and $J_2$), as shown in Fig.~\ref{fig:phase}(a).
We choose $t_1 / J_1 = 3.0$, $J_2 / J_1 = (t_2 / t_1)^2$ \cite{YFJiang2020} and  focus on the region with $0 \leq t_2 / t_1 \leq 0.32$ and hole doping level $1/24 \leq \delta \leq 1/6$ which is the optimal region for the SC in the cuprates \cite{Pavarini2001,Tanaka2004,ZXShen1998}.

By advancing the DMRG simulations with $U(1) \times SU(2)$ symmetries \cite{McCulloch2002} (also see Supple. Mat. \cite{sm}), we study the system on a cylinder with the periodic boundary conditions along the circumference direction ($y$) and the open boundary along the axis direction ($x$), where $L_y$ and $L_x$ denote the lattice sites along these two directions. 
We keep the bond dimensions up to $D = 20000$ $SU(2)$ multiplets, which is equivalent to about $60000$ $U(1)$ states (it is about double of the previous standard in the literatures for the $t$-$J$ model \cite{HCJiang2018,YFJiang2020}) and thus allows us to obtain accurate results on the $L_y = 6$ cylinder with the truncation error near $1\times 10^{-6}$ \cite{sm}.

\begin{figure}[b]
	\includegraphics[width=0.85\linewidth]{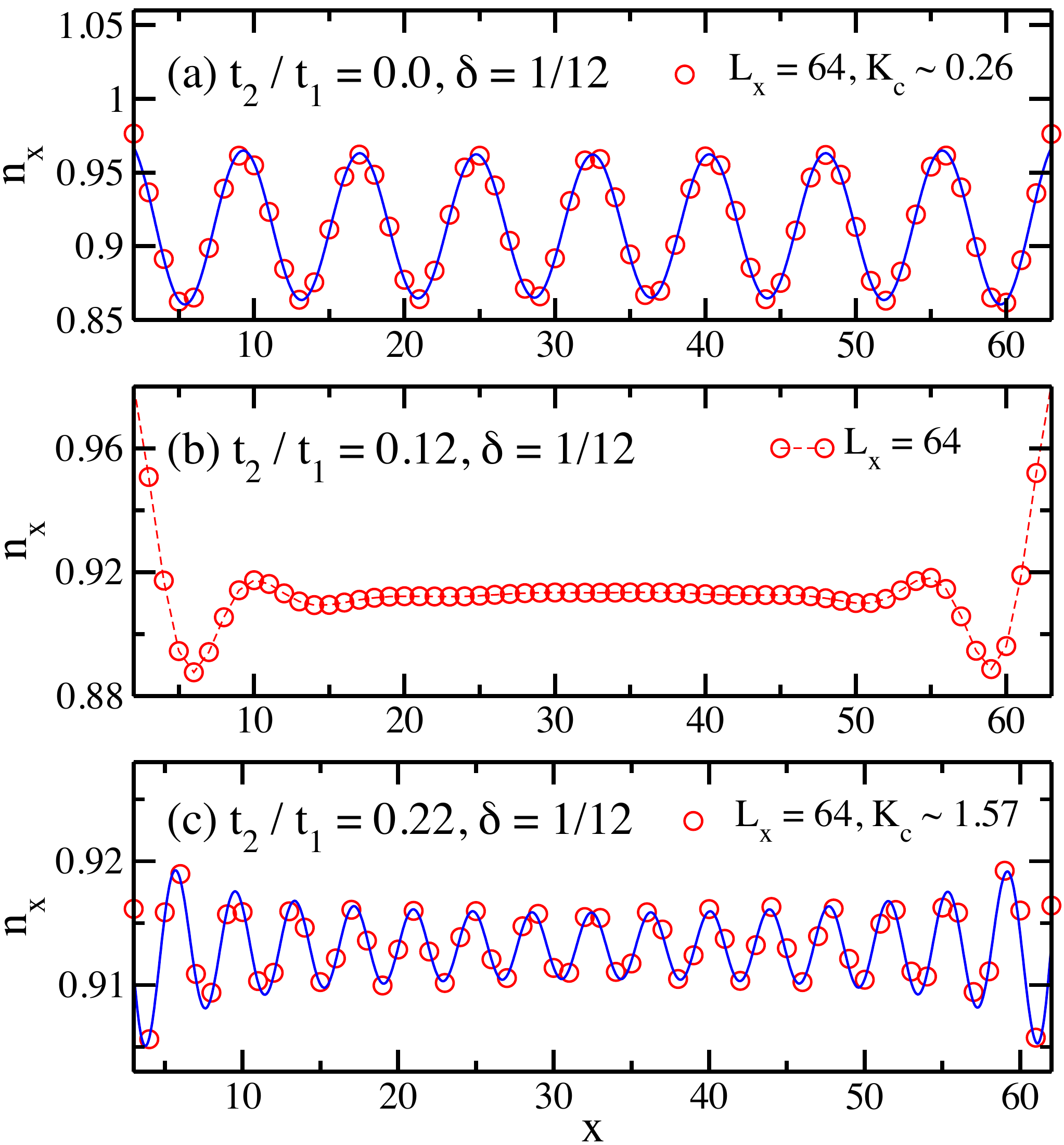}
	\caption{Charge density profiles.
		The charge density distributions $n_x = \sum^{L_y}_{y=1}\langle \hat{n}_{x,y} \rangle / L_y$ on the $L_x=64, L_y=6$ cylinder for 
		(a) CDW phase, (b) SC phase, and (c) SC + CDW phase. The blue lines are fitting curves to the function $n_x=n_0+A_{cdw}\cos(Qx+\phi)$, where $A_{cdw}=A_0(x^{-K_c/2}+(L_x+1-x)^{-K_c/2})$ and $Q$ are the CDW amplitude and wave vector, respectively. $\phi$ is a phase shift. 
	}
	\label{fig:cdw}
\end{figure}

\begin{figure}[b]
	\includegraphics[width=1\linewidth]{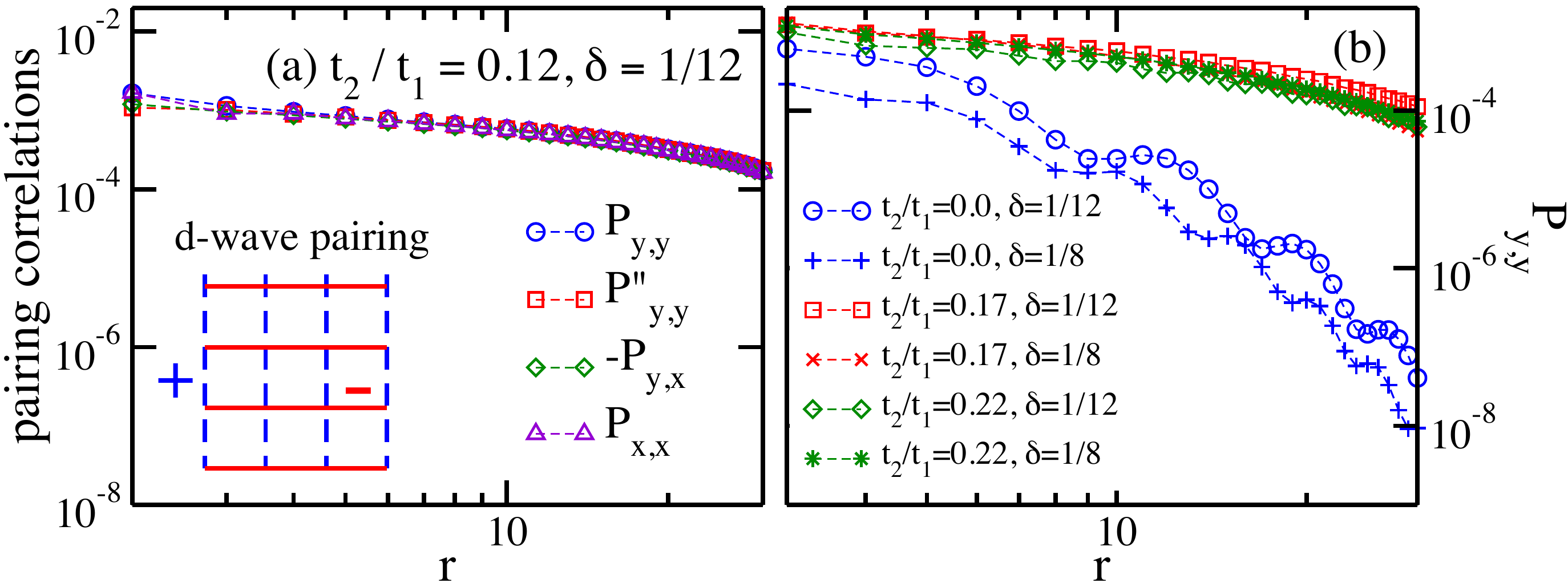}
	\caption{SC pairing correlations.
		(a) Various kinds of pairing correlations along different bond directions:  vertical-vertical correlation $P_{y,y}$ (blue) and $P_{y,y}^{''}$ (red),  horizontal-horizontal correlation $P_{x,x}$ (purple), and vertical-horizontal correlation $P_{y,x}$ (green). 
		The inset shows the pattern of the d-wave symmetry.
		(b) Double-logarithmic plot of the pairing correlations $P_{y,y}$ for different $t_2 / t_1$ at $\delta = 1/12, 1/8$. 
	}
	\label{fig:sc}
\end{figure}

{\it Quantum phase diagram.---}
Figure~\ref{fig:phase} presents the phase diagram as a function of $t_2/t_1$ and doping level $\delta$ based on comprehensive simulations of cylinder systems with $L_x = 48, 64$ and $L_y = 6$. We identify three phases with different charge density distributions: a CDW phase (light purple), a d-wave SC phase without static charge order (red), and a SC + CDW coexistent phase (green).
In the CDW phase, we identify stripe orders with wavelength $\lambda \simeq  4/(L_y \delta)$ depending on doping level (Fig.~\ref{fig:cdw}(a)), consistent with previous results~\cite{BXZheng2017,Imada2018,Corboz2019}. 
Meanwhile, SC pairing correlations are weak and become very small at long distance near $t_2 = 0$ (Fig.~\ref{fig:sc}(b)). 
In the SC phase, we find uniform charge density without static charge order (Fig.~\ref{fig:cdw}(b)), but with a strong quasi-long-range SC order of the ordinary d-wave symmetry (Fig.~\ref{fig:sc}(a)).
For the coexistent phase, we also find a dominant quasi-long-range SC order (Fig.~\ref{fig:sc}(b)), which cooperates with a weak stripe order with wavelength $\lambda \simeq 2/(L_y \delta)$ (Fig.~\ref{fig:cdw}(c)).  

The intermediate uniform SC phase is the key finding in this paper.  
Interestingly, the window of the d-wave SC phase gradually spans with increasing doping level, inducing the doping-tuned CDW (or SC + CDW coexistent phase) to a uniform SC phase transition.
As we will discuss below, this picture could be relevant to experimental observations in cuprates. 
In the following, we turn to the identification of these phases.

{\it Charge density wave.---}
Since the charge density of the ground state is uniform along the $y$ direction due to translational symmetry and shows distinct  behaviors along the $x$ direction for different phases, we define the averaged charge density for each column as $n_x = \sum^{L_y}_{y=1}\langle \hat{n}_{x,y} \rangle / L_y$ and show the density profiles in Fig.~\ref{fig:cdw}.
In the CDW phase, we identify an approximate periodic density modulation with the wavelength $\lambda \simeq 4/(L_y\delta)$ doping dependent. For example, at $t_2=0, \delta=1/12$, the density profile has $\lambda \simeq 8$, i.e. each stripe is filled with four holes (or $n^h_{str} = 4$ in average,  see Fig.~\ref{fig:cdw}(a)). 
In contrast, in the coexistent phase we find a charge modulation with $\lambda \simeq 4$ (Fig.~\ref{fig:cdw}(c)), which contains two holes $n^h_{str} = 2$ on average per stripe, regardless of the doping level.
Thus, a quasi-long-range SC occurs likely in the coexistent phase as the charge modulation with two holes ($n^h_{str}=2$) may be plausible for pairing~\cite{YFJiang2020,Zaanen1989,Vojta2009,Machida1989}.
Importantly, in addition to the aforementioned charge ordered phases, we find a uniform charge density phase with vanishing-small density modulation in the bulk of system (see Fig.~\ref{fig:cdw}(b)) and \cite{sm}).
The absence of static charge order indicates that CDW is very weak and thus may give way to a robust SC.

\begin{figure}[t]
	\includegraphics[width=0.9\linewidth]{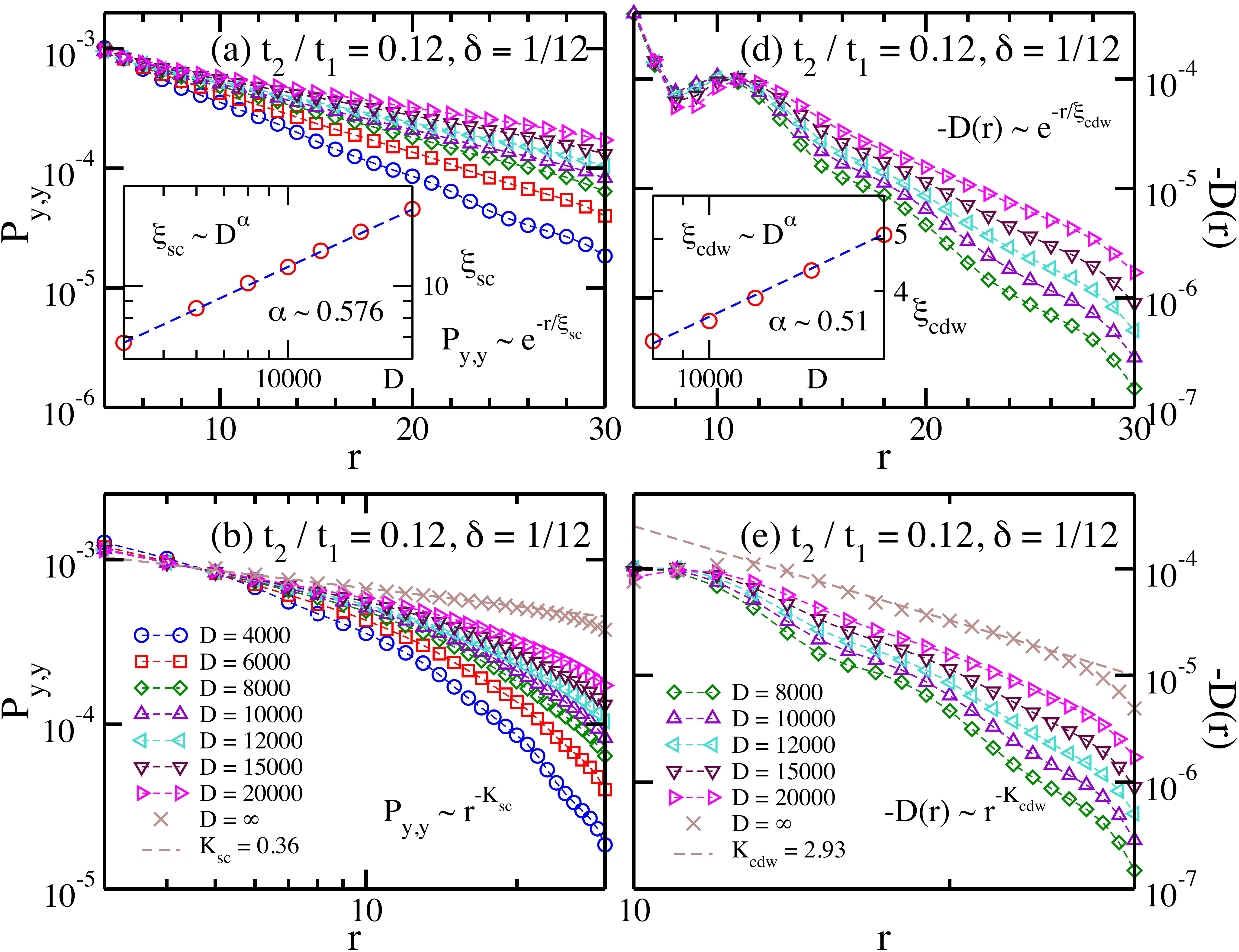}
	\includegraphics[width=0.9\linewidth]{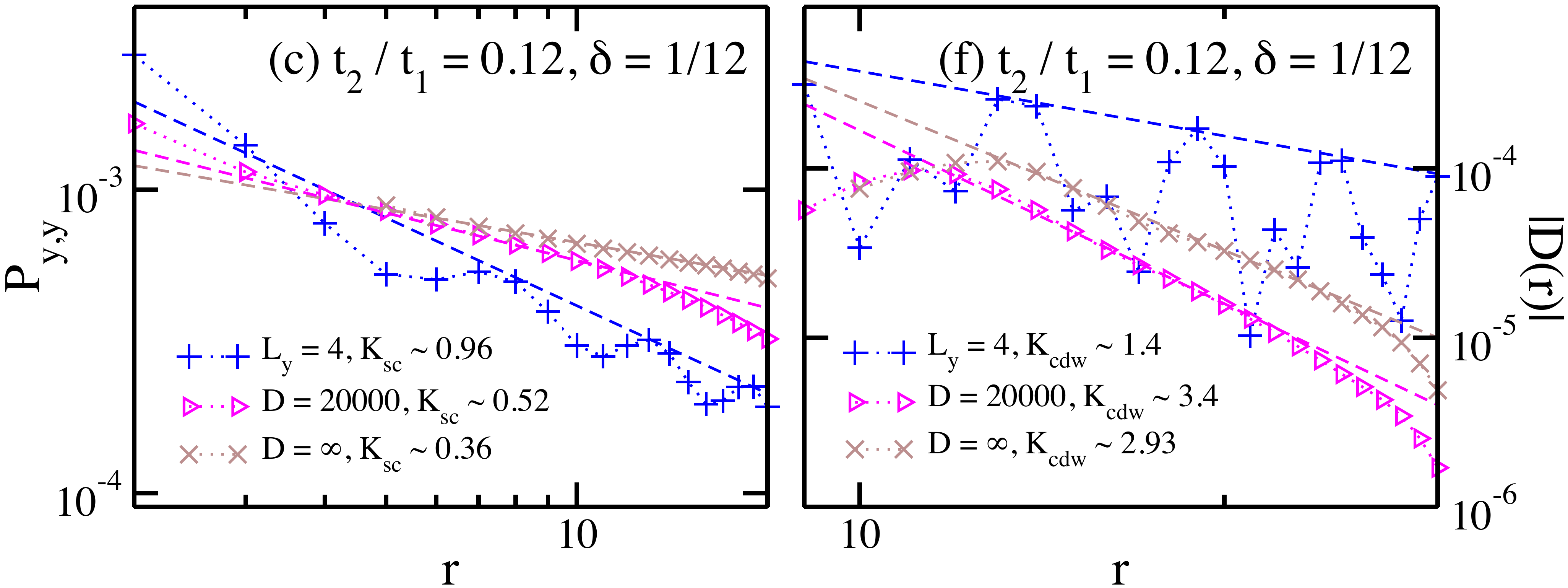}
	\caption{Scaling of correlations in the SC phase.
		(a) Semi-logarithmic plot of $P_{y,y}$ obtained by different bond dimensions $D$. The inset shows the dependence of the correlation length $\xi_{sc}$ on $D$, where $\xi_{sc}$ is obtained by fitting $P_{y,y} \sim \exp(-r/\xi_{sc})$. In the range of $D = 4000 - 20000$ (equivalent to $U(1)$ $D = 12000 - 60000$), $\xi_{sc}$ fits to $\xi_{sc} \sim D^\alpha$ with $\alpha = 0.576$.
		(b) Double-logarithmic plot of $P_{y,y}$ with the same data in the subfigure (a). The dashed crossed line denotes the power-law fitting of the extrapolated $D \rightarrow \infty$ results. (c) Comparing the pairing correlations on the $L_y = 4, 6$ cylinders.  
		(d-f) Similar plots for the density-density correlation function $D(r)$. 
	}
	\label{fig:scaling}
\end{figure}

{\it SC pairing correlation and d-wave symmetry.---} 
We examine the SC by measuring the dominant spin-singlet pairing correlations 
$P_{\alpha,\beta}(\mathbf r) = \langle \hat \Delta^\dagger_{\alpha}(\mathbf r_0) \hat \Delta_{\beta}(\mathbf r_0 + \mathbf r)\rangle$, 
where the pairing operator is defined on two NN sites $\mathbf r_1$ and $\mathbf r_2=\mathbf r_1 +\mathbf e_{\alpha}$ and $\hat \Delta_{\alpha}(\mathbf r_1)=(c_{\mathbf r_1 \uparrow}c_{\mathbf r_2 \downarrow}-c_{\mathbf r_1 \downarrow}c_{\mathbf r_2 \uparrow})/\sqrt{2}$ ($\mathbf{e}_{\alpha=x,y}$ denote the unit lengths along $x$- and $y$-direction, respectively).
We consider correlation decay along the $x$ direction with distance $r$.

\begin{figure}[t]
	\includegraphics[width=1\linewidth]{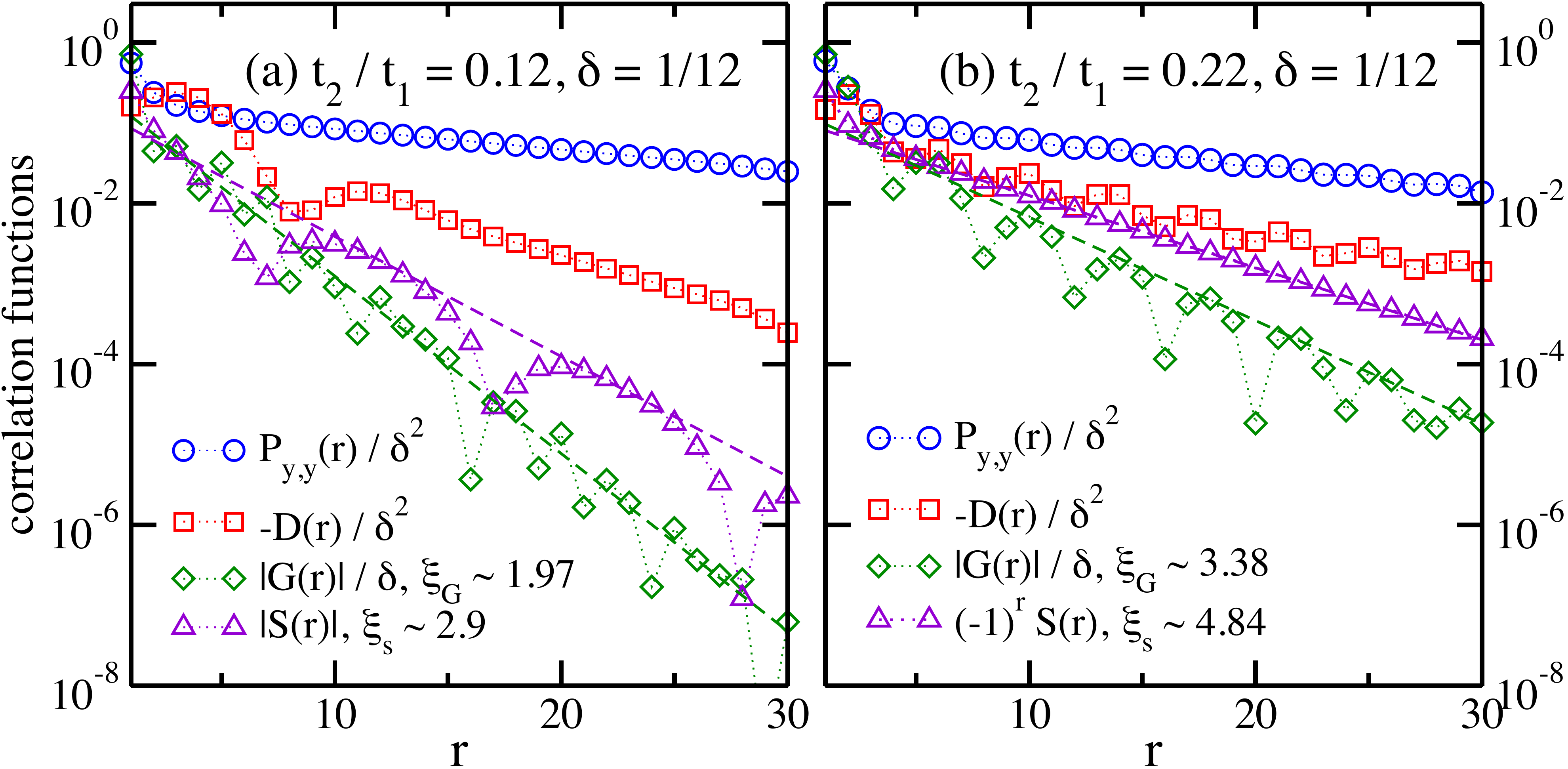}
	\caption{Correlations in the SC and SC + CDW phases.
		Comparison among the pairing correlation $P_{y,y}(r)$, density correlation $D(r)$, spin correlation $S(r)=\langle \mathbf S_x \cdot \mathbf S_{x+r}\rangle$, and single particle correlation $G(r) = \langle  \sum_{\sigma} c^\dagger_{x,\sigma}  c_{x+r,\sigma}\rangle$ for (a) SC phase and (b) SC + CDW phase. The correlations are rescaled to make a direct comparison.
	}
	\label{fig:compare}
\end{figure}

First, we discuss the SC pairing symmetry by inspecting the different pairing correlations shown in Fig.~\ref{fig:sc}(a).
While two kinds of the vertical-vertical correlations $P_{y,y}$ (blue, for two $y$-bonds in the same chains), $P_{y,y}^{''}$ (red, for two $y$-bonds with one relative lattice shift in the $y$-direction) and the horizontal-horizontal correlation $P_{x,x}$ (purple) are always positive, the vertical-horizontal correlations $P_{y,x}$ (green) are negative.
Thus, the pairing order parameters should have the opposite signs for the $x$-bond and $y$-bond, respectively. 
Furthermore, the pairing term has no phase shift along both directions, showing a conventional d-wave pairing symmetry as depicted by the inset of Fig.~\ref{fig:sc}(a).
In addition, the magnitudes of the pairing correlations are insensitive to bond orientations, showing a spatially uniform feature of the SC order. 
Second, by tuning $t_2 / t_1$, the pairing correlations are enhanced and become strong in the SC phase, signaling the developed quasi-long-range order. 
Such pairing correlations remain stable for the larger $t_2 / t_1$ entering the SC + CDW phase as shown in Fig.~\ref{fig:sc}(b) for $\delta=1/12$. 
Third, in the SC and SC + CDW phases, we identify that the pairing correlation dominates over all other competing charge and spin correlations, as evidenced by Fig.~\ref{fig:compare} for $\delta = 1/12, t_2 / t_1 = 0.12$ (SC phase)  and $0.22$ (SC + CDW phase). 
All above features strongly support a robust d-wave pairing nature in the SC and the SC + CDW phases.

To clarify the presence of quasi-long-range SC order, we further investigate the decay behavior of pairing correlations using two different ways. 
As DMRG method represents the ground state as a Matrix product state with a finite bond dimension, the correlations at long distance usually decay exponentially on wider systems \cite{SCHOLLWOCK2011}, which would recover the true nature of correlations in the infinite bond dimension limit.
Therefore, we first fit the raw data of pairing correlations for various bond dimensions using the exponential function $P_{y,y}(r) \sim \exp(-r/\xi_{sc})$, as shown in Fig.~\ref{fig:scaling}(a). 
One can see that the correlation length $\xi_{sc}$ monotonically grows  as the bond dimension increases. 
We find a power-law dependence $\xi_{sc} \sim D^\alpha$ (see the inset of Fig.~\ref{fig:scaling}(a)) for the bond dimension up to $D = 20000$, indicating that $\xi_{sc}$ tends to diverge in the $D \rightarrow \infty$ limit and a true quasi-long-range order is expected. 
In the second method, the obtained SC correlations are extrapolated to the $D \rightarrow \infty$ limit first~\cite{MPQin2020,HCJiang2019}, using a second-order polynomial function of $1/D$ for the data points of $D = 8000 - 20000$ (Fig.~\ref{fig:scaling}(b)). 
We find that the extrapolated pairing correlations over a wide range of distance collapse to a power-law decay function $P_{y,y}(r)\sim r^{-K_{sc}}$, with a Luttinger exponent $K_{sc} \approx 0.36$.
In Fig.~\ref{fig:scaling}(c), we compare the power-law SC correlations on the $L_y=4$ and $6$ systems, which give the exponent $K_{sc} \approx 0.96$ for $L_y=4$ and $0.36$ for $L_y=6$.
It is clear that the pairing correlations are significantly enhanced for $L_y = 6$ and we find that $K_{sc} < 1$ is a common feature in the SC phase~\cite{sm}.
It signals that the SC order, which becomes stronger and tends to be stabilized on larger system sizes, should survive in the 2D limit.
Thus, this uniform SC state can be regarded as the quasi-1D descendant of a 2D superconductor \cite{Balents1996}.

Last but not least, we compare density correlations with SC pairing correlations. 
We identify a power-law behavior of density correlations with a much higher exponent $K_{cdw} \approx 2.93$ (see Fig.~\ref{fig:scaling}(d,e)). 
In Fig.~\ref{fig:scaling}(c,f) we show that SC pairing and density correlations behave differently going from width-4 to width-6 cylinder: while SC correlations are greatly enhanced ($K_{sc}$ reduces from $0.96$ to $0.36$), density correlations are strongly suppressed with $K_{cdw}$ increasing from $1.4$ to $2.93$.
Notice that $K_{sc} < 0.5$ and $K_{cdw} > 2$ imply that the SC susceptibility diverges whereas the CDW susceptibility remains finite on the ladder systems~\cite{Kivelson2004}.   
This trend indicates that the SC order may grow stronger with increasing system width, thus we anticipate a robust uniform SC phase without a CDW instability in the 2D limit. 
Furthermore, we have carefully confirmed that the single-particle and spin correlations all decay exponentially in the uniform SC phase (see Fig.~\ref{fig:compare}(a) and~\cite{sm}).

In comparison, in the SC + CDW phase the SC order is found to cooperate with a weak stripe order, qualitatively consistent with the results of the width-4 Hubbard model or $t$-$J$ model~\cite{HCJiang2019,YFJiang2020}. 
Quantitatively, SC correlations still dominate all other correlations (see Fig.~\ref{fig:compare}(b)) with the Luttinger exponents $K_{sc} < K_{cdw} < 2$ (see Supple. Mat.~\cite{sm}).

{\it Fermi surface evolution.---}
Lastly we measure the electron distribution function in the momentum space $n(\mathbf k) = \sum_{i,j,\sigma}\langle c^\dagger_{i,\sigma}c_{j,\sigma} \rangle e^{i\mathbf k \cdot (\mathbf r_i - \mathbf r_j)} / (L_x L_y)$ to study the evolution of electronic structure. 
We identify that the normal and SC phases have distinct topologies of $n(\mathbf k)$: In the normal CDW phase (Fig.~\ref{fig:phase}(c)), 
the size of the electron pocket near the $\Gamma = (0,0)$ point expands eventually covering a large portion of the Brillouin zone with a clear nematic distortion of Fermi surface from the unidirectional stripe order. In the SC and SC + CDW phases (Fig.~\ref{fig:phase}(d-e)), electronic states form a closed Fermi surface with approximate $C_4$ symmetry and  an isolated electron pocket centers around the $\Gamma$ point. Such a change of the Fermi surface topology is robust for all doping levels~\cite{sm}. We conjecture that the Fermi surface topology may be related to the emergence of quantum criticality between the CDW and SC phase, which we leave for future study.

{\it Summary and Discussion.---}
We have presented a comprehensive study of a doped Mott insulator by further advancing the state-of-the-art DMRG computations, which allows us to identify a robust superconductivity on wider cylindrical systems. 
We map out a global phase diagram in terms of doping level and the NNN electron hopping strength.
We identify two  SC phases, either with or without a static CDW order. 
The remarkable result found on the wider system is that, by suppressing charge and spin orders, a uniform SC phase with the ordinary d-wave pairing symmetry emerges. We carefully established that the SC pairing correlation is the strongest correlation with robust quasi-long-range order and a small power exponent. The density correlations also decay with a power-law behavior, but have a large exponent, indicating a special limit of Luther-Emery liquid where the CDW correlations cannot compete with the SC correlations.
Such a uniform d-wave SC state has been sought for decades, and the current numerical identification provides convincing evidence for the emergent of such a  state in strongly correlated electron systems with only repulsive interactions.

As the width-6 system has reduced ring and plaquette correlations around the cylinder~\cite{Chung2020}, it may be a better representation of 2D system. Intuitively, our phase diagram on the 6-leg system turns out to resemble the essential features of the cuprate compounds~\cite{Taillefer2019}. 
For instance, upon increasing the hole doping level, two different possibilities could occur: the system could be driven from the normal state to a uniform SC phase directly, or it could first go into a SC + CDW coexistent phase and then it takes another transition into a uniform SC state.
This picture provides an intuitive understanding that CDW order often but not always appears in the underdoped regime with the onset of superconductivity, which may depend on the ratio $t_2 / t_1$ and other properties of materials.

{\it Acknowledgments.---}
We acknowledge stimulating discussions with L. Balents, H. C. Jiang, S.A. Kivelson and R. H. He. This work was supported by the NSFC grants 11834014, 11874078, and the Fundamental Research Funds for the Central Universities (S.S.G.).  W.Z. was supported by the foundation of Westlake University. This work was also supported by the U.S. Department of Energy, Office of Science, Advanced Scientific Computing Research and Basic Energy Sciences, Materials Sciences and Engineering Division, Scientific Discovery through Advanced Computing (SciDAC) program under the grant number DE-AC02-76SF00515 (D.N.S.).

{\it Note added.---}
At the final stage of preparing this work, we notice an arXiv preprint focusing on larger positive $t_2$ regime \cite{HCJiang2021} and another preprint studying the phase diagram with both negative and positive $t_2$ \cite{STJiang2021}. The superconducting state found in Ref.~\cite{HCJiang2021} has the similar pairing correlation and density correlation power exponents as those in our SC + CDW state. The enhanced spin correlations with growing system circumference in the SC + CDW phase also agree with the observation in Ref.~\cite{STJiang2021} in the same parameter region.

\bibliography{tjmodel}

\clearpage
\appendix
\widetext
\begin{center}
	\textbf{\large Supplementary Materials for: ``Robust d-wave superconductivity in the square-lattice $t$-$J$ model''}
\end{center}

\vspace{1mm}

\renewcommand\thefigure{\thesection S\arabic{figure}}
\renewcommand\theequation{\thesection S\arabic{equation}}

\setcounter{figure}{0} 
\setcounter{equation}{0}

In the supplemental materials, we provide more numerical results to support the conclusions we have discussed in the main text. In Sec. 1, we show the charge density distributions for various parameter points in the three phases. In Sec. 2, we present more data on pairing correlations and density correlations. In Sec. 3, the single-particle correlations and electron densities in momentum space are discussed. In Sec. 4, we show spin correlations and static spin structure factor. In Sec. 5, we also demonstrate the correlation functions  in the charge density wave (CDW) phase. In Sec. 6, we further show the details of the extrapolation of physical quantities with increasing bond dimension. In Sec. 7, we introduce technical details on implementing spin rotational $SU(2)$ symmetry in the density matrix renormalization group (DMRG) simulation of the $t-J$ model.

\section{1. Charge density profile}

In the main text, we have shown the charge density profiles in the different quantum phases at the doping ratio $\delta = 1/12$.
Here, we show more examples to support our findings.
Following the main text, we define the averaged charge density on each column as $n_x = \sum^{L_y}_{y=1}\langle \hat{n}_{x,y} \rangle / L_y$, where $\hat{n}_{x,y} \equiv \sum_{\sigma} \hat{c}^{\dagger}_{(x,y),\sigma} \hat{c}_{(x,y),\sigma}$ is the electron density operator defined on the site $(x,y)$.
With the translational symmetry along the $y$ direction on cylinder geometry, $\langle \hat{n}_{x,y} \rangle$ is independent of $y$ for any given column $x$.

In the d-wave superconducting (SC) phase, a prominent feature is the uniform charge density distribution.
In Fig. S1, we show the charge density profiles for more parameters in the d-wave SC phase.
Although the charge densities $n_x$ have amplitude oscillation near the open boundaries due to the open edge effect, the densities in the bulk are uniform.
Therefore, the uniform charge density is a universal property in the d-wave SC phase, which is independent of the doping ratio.
Furthermore, the uniform charge densities in the bulk are also close to the expected value $n_x = 1 - \delta$, which are shown in Fig. S1 as dashed lines. The small deviations are owing to the finite-size boundary effect.
At the doping ratio $\delta = 1/6$, we also confirm the uniform charge densities without static charge order in the infinite DMRG calculation (not shown here), where $n_x = 1 - \delta$ is exactly satisfied due to small finite-size effect.

In the coexistent phase, we have shown the charge density profile for $t_2 / t_1 = 0.22, \delta = 1/12$ in the main text, which contains two holes in each stripe, i.e. $n^{h}_{str} = 2$. Here, we present more results in Fig. S2.
For $t_2 / t_1 = 0.22, \delta = 1/24$, the charge density shows a space modulation with the wavelength $\lambda = 8$, which also agrees with $n^{h}_{str} = 2$.
For $t_2 / t_1 = 0.32, \delta = 1/12$, the density profile has the same modulation wavelength $\lambda = 4$ as that for $t_2 / t_1 = 0.22, \delta = 1/12$.
Therefore, we conclude that in the coexistent phase, the charge densities show the wave modulation with $n^{h}_{str} = 2$ and $\lambda =1 / 3\delta$.
In Fig. S2, we also fit the density profiles using the formula
\begin{equation}
n(x) = n_0(x) + A_0 \cos(Qx + \phi) [x^{-K_c / 2} + (L_x + 1 - x)^{-K_c / 2}].
\label{eq:density_fit}
\end{equation}
In both fittings, we find that the wave vectors $Q$ are close to $2\pi / \lambda$ as expected.
For $t_2 / t_1 = 0.22, \delta = 1/24$ we find $K_c \simeq 2.06$, and for $t_2 / t_1 = 0.32, \delta = 1/12$ we get $K_c \simeq 1.38$.

In Fig. S3, we demonstrate the charge density profiles in the CDW phase at different doping ratios $\delta$ with $t_2 / t_1 = 0.0$.
For $\delta = 1/24, 1/12, 1/6$, the density profiles have modulations with $\lambda = 16, 8, 4$ respectively, which are all consistent with $n^{h}_{str} = 4$ and $\lambda = 2 / 3\delta$.
However, this charge density distribution $n^{h}_{str} = 4$ is incommensurate with $\delta = 1/8$, since it leads to the modulation wavelength $\lambda = 16 / 3$.
Indeed, as shown in Fig. S3(c), the charge density does not show a well defined period.
If we simply take a magnitude modulation as a period, we find that the wavelength is indeed $16 / 3$, consistent with the condition $n^{h}_{str} = 4$.
In the figure, we also show the fittings of the DMRG data using Eq. (S1), which lead to the correct wave vector $Q$ and give the small power-law exponents $K_c = 0.35 - 0.6$.
Compared with the power exponents in the coexistent phase (see Fig. S2), here the much smaller exponents indicate the strong CDW ordering. 
With growing $t_2 / t_1$ in the CDW phase, $K_c$ slightly increases and indicates the weakened CDW order. Two examples are shown below in Fig. S16(a-b).

Here we stress that, Eq. (S1) only applies to the CDW phase and the coexistent phase, but not in the SC phase because the local CDW order parameter is vanished. Hence, in the main text, when we compare the decay behaviors of correlation functions, we used the power-law exponent from the density-density correlation functions, to compare with that from pairing correlations. In this regard, our comparison of power exponents in the main text is on the equal footing.

\section{2. Superconducting pairing correlation function and density-density correlation function}

In the main text, we have shown the behaviors of SC pairing correlation function and density-density correlation function for $t_2 / t_1 = 0.12, \delta = 1/12$ in the d-wave SC phase.
In this section, we show the similar data for $t_2 / t_1 = 0.22, \delta = 1/8$ in the d-wave SC phase, as well as $t_2 / t_1 = 0.2, 0.32, \delta = 1/12$ in the coexistent phase.
Because of  the d-wave symmetry of the SC pairing correlations in the two phases, we choose $P_{y,y}$ as the example to demonstrate the pairing correlation, which is defined for the vertical bonds as
\begin{equation}
P_{y,y}(r) = \frac{1}{L_y} \sum^{L_y}_{y_0=1} \langle \hat{\Delta}^{\dagger}_y (x_0, y_0) \hat{\Delta}_y (x_0 + r, y_0) \rangle,
\end{equation}
where 
\begin{equation}
\hat{\Delta}_y (x_0, y_0) = \frac{1}{\sqrt{2}}(\hat{c}_{(x_0,y_0), \uparrow} \hat{c}_{(x_0,y_0+\hat{e}_y), \downarrow} - \hat{c}_{(x_0,y_0), \downarrow} \hat{c}_{(x_0,y_0+\hat{e}_y), \uparrow}).
\end{equation}
Following the main text, we define the density-density correlation function $D(r)$ as
\begin{equation}
D(r) = \frac{1}{L_y} \sum^{L_y}_{y_0 = 1} [ \langle \hat{n}_{x_0, y_0} \hat{n}_{x_0+r, y_0} \rangle - \langle \hat{n}_{x_0, y_0} \rangle \langle \hat{n}_{x_0+r, y_0} \rangle ].
\end{equation}
For demonstrating both correlations in real space, we choose the reference position at $x_0 = 10$ and present its correlations with other sites cross the bulk of the system.
Very similar results are obtained for other reference sites.

The results for $t_2 / t_1 = 0.22, \delta = 1/8$ in the d-wave SC phase are shown in Fig. S4. Parallel to the discussion in the main text, we have used two different ways to inspect the long-ranged behavior of pairing correlations.
In the semi-logarithmic plot Fig. S4(a), the pairing correlation length $\xi_{sc}$, fitted by the exponential decay functions $P_{y,y}(r) \sim e^{-r/\xi_{sc}}$ for each bond dimension $D$, follows the power-law increase $\xi_{sc} \sim D^{\alpha}$ with bond dimension upto $D = 20000$, supporting a quasi-long-range pairing correlation in the infinite-$D$ limit.
In the double-logarithmic plot Fig. S4(b), we extrapolate the pairing correlations (for each distance $r$) to the $D \rightarrow \infty$ limit by using the polynomial function up to the second order of $1/D$ (see the details in Section 6).
The obtained results for $r \simeq 6 - 17$ fit the power-law behavior quite well, giving a small power exponent $K_{sc} \simeq 0.43$.
In Fig. S4(c), we compare the $K_{sc}$ and we find that $K_{sc}$ significantly decreases with growing circumference $L_y$ from $K_{sc} (L_y=4)=1.17$ to $K_{sc} (L_y=6)=0.43$, which strongly suggests long-range pairing correlation on wider systems.
For density correlations, we first use the exponential fitting $D(r) \sim e^{-r/\xi_{cdw}}$ as shown in Fig. S4(d).
One can find that the density correlations decay slower than the exponential behavior, which makes it impossible to fit the whole curves using exponential function.
Thus, we estimate the correlation length $\xi_{cdw}$ by fitting the data with the best exponential behavior in a range of distance.
The fitted $\xi_{cdw}$ seems also to grow with bond dimension $D$ following a power law.
We further plot the double-logarithmic plot of the density correlations including the extrapolated results in the infinite-$D$ limit, as shown in Fig. S4(e).
Here we also use the polynomial function up to the second order of $1/D$ to fit the data.
Clearly, the extrapolated results fit the power-law behavior quite well in a large region of $r \sim 4 - 20$, giving a large power exponent $K_{cdw} \simeq 2.1$.
Importantly, Fig. S4(f) shows that the density correlations decay faster on the larger size, which is opposite to the behavior of pairing correlations. In short, all of these features are in line with those at the parameter $t_2 / t_1 = 0.12, \delta = 1/12$ shown in the main text.  Thus, we believe these findings are quite robust in the whole d-wave SC phase, which strongly indicate that the pairing correlation is dominate over the density correlation in the d-wave SC phase.

For $t_2 / t_1 = 0.2, \delta = 1/12$ in the coexistent phase, we show the results in Fig. S5.
As shown in Fig. S5(a), the correlation length $\xi_{sc}$ fitted by the exponential decay of the pairing correlations $P_{y,y}$ also follows the power-law behavior $\xi_{sc} \sim D^{\alpha}$, which suggests a quasi-long-range decay in the infinite-$D$ limit.
By extrapolating the pairing correlations to the $D \rightarrow \infty$ limit, the results with the distance $r \simeq 4 - 20$ fit the power-law behavior quite well, giving a small power exponent $K_{sc} \simeq 0.36$.
With growing circumference from $L_y = 4$ to $6$, $K_{sc}$ also quickly reduces from $0.91$ to $0.36$, strongly indicating long-range SC correlation on larger system sizes.
For density correlations, we can find that the curves also decay slower than exponential behavior in Fig. S5(d).
We further plot the density correlations including the extrapolated results in the infinite-$D$ limit in the double-logarithmic way, as shown in Fig. S5(e).
The extrapolated data with $r \simeq 3 - 20$ fit the power-law behavior, giving a power exponent $K_{cdw} \simeq 1.6$.
With growing circumference from $L_y = 4$ to $6$, the density correlations slightly decay faster.

We have also carefully examined  the point $t_2 / t_1 = 0.32, \delta = 1/12$ in the coexistent phase as shown in Fig. S6.
The overall features are similar to $t_2 / t_1 = 0.2, \delta = 1/12$ in Fig. S5. Via the fitting and extrapolation to infinite-bond-dimension limit, we get $K_{sc}\sim 1.0$, compared with $K_{cdw} \sim1.6$. To sum up, in the coexistent phase, the pairing correlation is also quasi-long-ranged with $K_{sc} < K_{cdw}$. However, the density correlations become stronger, with $K_{cdw} < 2$. This is the main difference between the d-wave SC phase and the SC + CDW coexistent phase. 
In this coexistent phase, the power exponent $K_{cdw}$ obtained by fitting the density correlation function also agrees with the exponent $K_{c}$ that describes the power-law behavior of the CDW order parameter. For $t_2 / t_1 = 0.22$ and $0.32$, we find $K_c \simeq 1.57$ (see Fig. 2(c) in the main text) and $1.38$ (see Fig. S2(b)) respectively, which are consistent with $K_{cdw} \simeq 1.6$ found here.  

We also would like to point out that, in the direct comparison of pairing correlations and density correlations in the SC phase and the SC + CDW coexistent phase, the long-distance amplitudes of pairing correlations are always larger than those of the density correlations on the $L_y = 6$ systems (especially nearly one order larger in the d-wave SC phase), as shown in Fig. S4-Fig. S6 (notice that the scales in the left column and right column are different).
This serves as another direct evidence that the SC pairing correlation is dominate over the charge correlation.

To confirm the discussed results, we have also compared the obtained correlation functions on the systems with different lengths.
As shown in Fig. S7, the pairing  and density correlations are carefully compared on the $L_x = 48$ and $L_x = 64$ cylinders, by keeping the bond dimensions $D = 12000$ and $20000$.
For each given bond dimension, the correlations are highly consistent. However, near the boundary of the $L_x = 48$ system, we find that the correlations are slightly enhanced on the $L_x = 64$ system, indicating a reduced boundary effect.
The power-law growth of the correlation length with $D$ is robust for the larger systems, suggesting  a strong quasi-long-range order for SC at large $N_x$ limit.

\section{3. Single-particle correlation function and electron densities in momentum space}

In Fig. S8, we show the single-particle correlations for different parameters in the d-wave SC phase.
The results for $t_2 / t_1 = 0.12, \delta = 1/12$ are shown for different bond dimensions $D = 8000 - 20000$, which show good convergence. 
On the $L_y = 6$ cylinder, $|G(r)|$ decays exponentially with short correlation lengths $\xi_{G} \sim 2 - 3$.
Compared with the results on the $L_y = 4$ cylinder, the single-particle correlations also decay faster with growing circumference, which suggests short single-particle correlation lengths on wider systems  in the d-wave SC phase.

In the coexistent phase as shown in Fig. S9, the single-particle correlations are also suppressed with growing circumference.
However, different from the d-wave SC phase, the correlations on the $L_y = 6$ cylinder clearly enhance with raising $t_2 / t_1$ in the SC + CDW coexistent phase.
In the semi-logarithmic plot, the exponential fitting gives that $\xi_{G}$ increases from $2.5$ to $6.9$ with $t_2 / t_1$ from $0.2$ to $0.32$ (see Fig. S9(a-c)).
Interestingly, for $t_2 / t_1 = 0.32$, the correlations can also be fitted well by using the power-law behavior $|G(r)| \sim r^{-K_{G}}$ with a power exponent $K_{G} \simeq 1.74$. It demonstrates that the single-particle correlations are gradually enhanced by increasing $t_2 / t_1$ in the SC + CDW coexistent phase.

In Fig. S10, we show the electron densities in the momentum space $n(\bf k)$ for more parameters.
The overall feature is very similar to the cases we show in the main text, i.e. 
a large (small) Fermi surface is identified in the non-SC (SC) phase. This behavior seems quite robust, independent of the specific doping level or the coupling ratio $t_2 / t_1$. 
The scans of the $n(\bf k)$ data for each given $k_y$ are also shown in Fig. S11 for the doping level $\delta = 1/12$, which clearly show the change of electron occupation near ${\bf k} = (0, \pi)$ and $(\pi, 0)$ with growing $t_2 / t_1$. Thus, we believe that the change of Fermi surface topology may be closely related to the transition between the normal CDW and the SC phases.

\section{4. Spin correlation function and spin structure factor}

In this section, we further discuss spin correlation functions.
In the d-wave SC phase, we show the results for different couplings and doping ratios in Fig. S12.
By comparing the spin correlations for different bond dimensions $D = 8000 - 20000$ in Fig. S12(a), one can find that spin correlations quickly converge with growing bond dimensions, which ensures the good convergence.
On the $L_y = 6$ cylinder, the spin correlations show a good exponential decay with short correlation lengths $\xi_{s} \sim 2.3 - 3.0$. 
Compared with the results on the $L_y = 4$ cylinder ($\xi_{s} \sim 3.0 - 4.0$), spin correlations are clearly suppressed with growing circumference, which suggests very short spin correlation lengths on larger system size in the d-wave SC phase.

In the coexistent phase, spin correlations show different behaviors from those in the d-wave SC phase, as shown in Fig. S13.
In Fig. S13(a), we show that the spin correlations with growing bond dimension also quickly converge, confirming the good convergence.
A prominent feature is that for both $L_y = 4$ and $6$ the spin correlations exhibit the N\'eel-type oscillation.
Different from the d-wave SC phase, here the antiferromagnetic spin correlations enhance with growing circumference.
In particular, on the $L_y = 6$ cylinder $\xi_{s}$ grows rapidly with increased $t_2 / t_1$, which reaches $\xi_{s} \simeq 6.48$ for $t_2 / t_1 = 0.32, \delta = 1/12$, showing an enhanced spin correlation.
In Fig. S13(d), we further compare the correlations for $t_2 / t_1 = 0.32, \delta = 1/12$ and the spin-$1/2$ $J_1 - J_2$ square-lattice Heisenberg model with $J_2 / J_1 = 0.1$, which has been identified in the N\'eel antiferromagnetic phase. 
Although spin correlation is suppressed by the doped holes, it is still interesting to study these correlations for larger range of $t_2 / t_1$ and for wider systems to identify possible intertwined magnetic ordering in the coexistent phase. 

In Fig. S14, we also demonstrate the spin structure factor $S(\bf k)$ in the different phases. 
Overall, the structure factor peak enhances with decreased doping ratio.
In the CDW phase, $S(\bf k)$ shows round peaks near ${\bf k} = (\pi, \pi)$, and this peak splitting continues to grow with doping. This can be understood by the quasi-periodic oscillations of the spin correlations in Fig. S15(d). The quasi-period is largely consistent with period of charge density, hence we speculate that the quasi-period of spin correlation is due to the influence of charge density wave order.
With further growing $t_2 / t_1$, the two peaks gradually move towards ${\bf k} = (\pi, \pi)$ and change slowly in the d-wave SC phase, which is consistent with the short spin correlation length $\xi_{s}$ shown in Fig. S12.
In the coexistent phase, $S(\bf k)$ shows an enhanced peak at ${\bf k} = (\pi, \pi)$, agreeing with the N\'eel antiferromagnetic correlations with increased $\xi_{s}$ in Fig. S13.

\section{5. Correlation functions in the charge density wave phase}

In the previous sections, we focus on the correlation functions in the d-wave SC phase and the SC + CDW phase.
In this section, we show the different correlations in the CDW phase.
The results for $t_2 / t_1 = 0$ and $\delta = 1/8, 1/12$ are shown in Fig. S15.
It seems that all the correlation functions decay fast, especially the SC pairing correlation and single-particle correlation have very short correlation lengths.
With the increase of $t_2/t_1$, the SC pairing correlations are found to be enhanced. While the algebraic fitting of pairing correlations gives the power exponent $K_{sc} > 2$ near $t_2 / t_1 = 0$, the exponent $K_{sc}$ reduces quickly crossing over to the uniform d-wave SC phase but the CDW charge distribution remains robust, as shown in Fig. S16. The exact dependence of SC order on doping level and $t_2 / t_1$ in the CDW phase needs more comprehensive calculations, which we leave to future study.

\section{6. Extrapolation of correlation functions with growing bond dimension}

In the DMRG simulations for wider systems, it inevitably has  the finite bond-dimension effect.
To eliminate this effect and extract the intrinsic physics, the extrapolated pairing and density correlations are shown in the main text. Here we explain the extrapolation process in  more details. 

We perform polynomial extrapolations to best fit the data for a range of bond dimensions up to the largest $SU(2)$ bond dimension $D = 20000$ (equivalent to $\sim 60000$ $U(1)$ states), which is the largest bond dimension that has been achieved in the $t-J$ model. One typical example of the data extrapolation is shown in Fig. S16. 
For each given distance $r$, the correlations obtained by at least five different bond dimensions are extrapolated by the polynomial function $C(1/D) = C(0) + a / D + b / D^2$. We have ensured that the cubic extrapolation leads to the similar results, which does not change the conclusions made in the main text.

\section{7. Technical details on implementing spin-rotational $SU(2)$ symmetry on the $t-J$ model}

As we have emphasized in the main text, one advantage of our work is the application of the full $U(1) \times SU(2)$ symmetry on the $t-J$ model (the $U(1)$ symmetry denotes charge conservation), which allows us to achieve much larger bond dimensions in the simulation and obtain more accurate results compared with the usual $U(1) \times U(1)$ algorithm. 
Here we briefly describe the technical details on how to implement the spin $SU(2)$ symmetry.
We first recall the Hamiltonian of the $t - J$ model 
\begin{equation}\label{Hamiltonian}
H = -\sum_{\{ij\},\sigma} t_{ij} (\hat{c}^{\dagger}_{i,\sigma}\hat{c}_{j,\sigma} + h.c.)  + \sum_{\{ij\}} J_{ij} ({\bf S}_i \cdot {\bf S}_j - \frac{1}{4} \hat{n}_i \hat{n}_j),
\end{equation}
where $\hat{c}^{\dagger}_{i,\sigma}$ and $\hat{c}_{i,\sigma}$ are the creation and annihilation operators for the electron at the site $i$ with spin magnitude $\sigma$ ($\sigma = \pm 1/2$), ${\bf S}_{i}$ is the spin-$1/2$ operator, and $\hat{n}_i$ is the particle number operator $\hat{n}_i \equiv \sum_{\sigma} \hat{c}^{\dagger}_{i,\sigma} \hat{c}_{i,\sigma}$.

To use the $SU(2)$ symmetry, we define the rank-$1/2$ irreducible tensor operators $T^{(\frac{1}{2})}_{i, q}$ and $T^{\dagger(\frac{1}{2})}_{i, q}$ ($q = \pm 1/2$) for the electron creation and annihilation operators at each site: 
\begin{align}
T^{(\frac{1}{2})}_{i,-\frac{1}{2}} = \hat{c}_{i,\uparrow}, \quad  T^{(\frac{1}{2})}_{i,\frac{1}{2}} = -\hat{c}_{i,\downarrow}, \quad
T^{\dagger(\frac{1}{2})}_{i,-\frac{1}{2}} = \hat{c}^{\dagger}_{i,\downarrow}, \quad T^{\dagger(\frac{1}{2})}_{i,\frac{1}{2}} = \hat{c}^{\dagger}_{i,\uparrow},
\end{align}
and the rank-$1$ irreducible tensor operators $T^{(1)}_{i, q}$ $(q = +1, 0, -1)$ to describe the spin operators as
\begin{align}
T^{(1)}_{i,1} = -\frac{1}{\sqrt{2}} \hat{S}^{+}_i, \quad  T^{(1)}_{i,0} = \hat{S}^{z}_i, \quad
T^{(1)}_{i,-1} = \frac{1}{\sqrt{2}} \hat{S}^{-}_i.
\end{align}
These irreducible tensor operators can be combined to obtain the rank-$0$ irreducible tensor operators such as
\begin{align}
H^{(0)}_{0,ij} = \frac{1}{\sqrt{2}} (T^{\dagger (\frac{1}{2})}_{i,\frac{1}{2}} T^{(\frac{1}{2})}_{j,-\frac{1}{2}} - T^{\dagger (\frac{1}{2})}_{i,-\frac{1}{2}} T^{(\frac{1}{2})}_{j,\frac{1}{2}}) = \frac{1}{\sqrt{2}} \sum_{\sigma} \hat{c}^{\dagger}_{i,\sigma}\hat{c}_{j,\sigma},
\end{align}
\begin{align}
H^{(0)}_{0,ii} = \frac{1}{\sqrt{2}} (T^{\dagger (\frac{1}{2})}_{i,\frac{1}{2}} T^{(\frac{1}{2})}_{i,-\frac{1}{2}} - T^{\dagger (\frac{1}{2})}_{i,-\frac{1}{2}} T^{(\frac{1}{2})}_{i,\frac{1}{2}}) = \frac{1}{\sqrt{2}} \sum_{\sigma} \hat{c}^{\dagger}_{i,\sigma}\hat{c}_{i,\sigma} =  \frac{1}{\sqrt{2}} \hat{n}_i,
\end{align}
\begin{align}
S^{(0)}_{0, ij} = \frac{1}{\sqrt{3}} ( T^{(1)}_{i, 1} T^{(1)}_{j, -1} + T^{(1)}_{i, -1} T^{(1)}_{j, 1} - T^{(1)}_{i, 0} T^{(1)}_{j, 0} ),
\end{align}
where we denote $H^{(0)}_{0,ij}$ and $S^{(0)}_{0, ij}$ as the rank-$0$ irreducible tensor operators coupled by the rank-$1/2$ and rank-$1$ operators, respectively.

With the help of these irreducible tensor operators, the different parts of the Hamiltonian can be expressed as
\begin{eqnarray}
H&=& H_t + H_J +H_n, \\
H_{t} &=& -\sum_{\{i,j\}, \sigma} t_{ij} (\hat{c}^{\dagger}_{i,\sigma} \hat{c}_{j,\sigma} + \hat{c}^{\dagger}_{j,\sigma} \hat{c}_{i,\sigma}) = -\sqrt{2} \sum_{\{i,j\}} t_{ij} (H^{(0)}_{0, ij} + H^{(0)}_{0, ji}), \\
H_J &=& \sum_{\{i,j\}} J_{ij} ( \frac{1}{2} \hat{S}^{+}_i \hat{S}^{-}_j + \frac{1}{2} \hat{S}^{-}_i \hat{S}^{+}_j + \hat{S}^{z}_i \hat{S}^{z}_j) = -\sqrt{3} \sum_{\{i,j\}} J_{ij} S^{(0)}_{0, ij} \\
H_n &=&  -\frac{1}{4} \sum_{\{i,j\}} J_{ij} \hat{n}_i \hat{n}_j = -\frac{1}{4} \sum_{\{i,j\}} J_{ij} 2 H^{(0)}_{0,ii} H^{(0)}_{0,jj}. 
\end{eqnarray}
Furthermore, we can compute the matrix elements of the original operators by using the Wigner-Eckart theorem, which expresses the element as the product of a Clebsch-Gordan coefficient and the reduced matrix element of the corresponding irreducible tensor operator.
Therefore, in the DMRG simulation we only need to deal with the reduced matrix elements of the different irreducible tensor operators, which only carry the total electron number and total angular momentum quantum numbers.
Since the double occupation is excluded in the $t-J$ model, encoding in total angular momentum quantum number can largely reduce the matrix dimension by a factor of $3$, i.e. the bond dimension $D$ using in $SU(2)$ simulation is approximately equivalent to $3D$ using in the $U(1)$ DMRG simulation. This is the key to accelerate the calculation and access larger bond dimensions.  

In Fig. S17, we demonstrate the obtained total energy per site $E/N$ versus the DMRG truncation error.
With the $SU(2)$ bond dimension up to $D = 20000$, the truncation errors are reduced to about $3\times 10^{-6}$.
The extrapolated energy is also very close to the lowest energy that we obtain with $D = 20000$.
These results clearly indicate the good convergence of our calculations.

\clearpage

\begin{figure}[htp]
\includegraphics[width=0.8\linewidth]{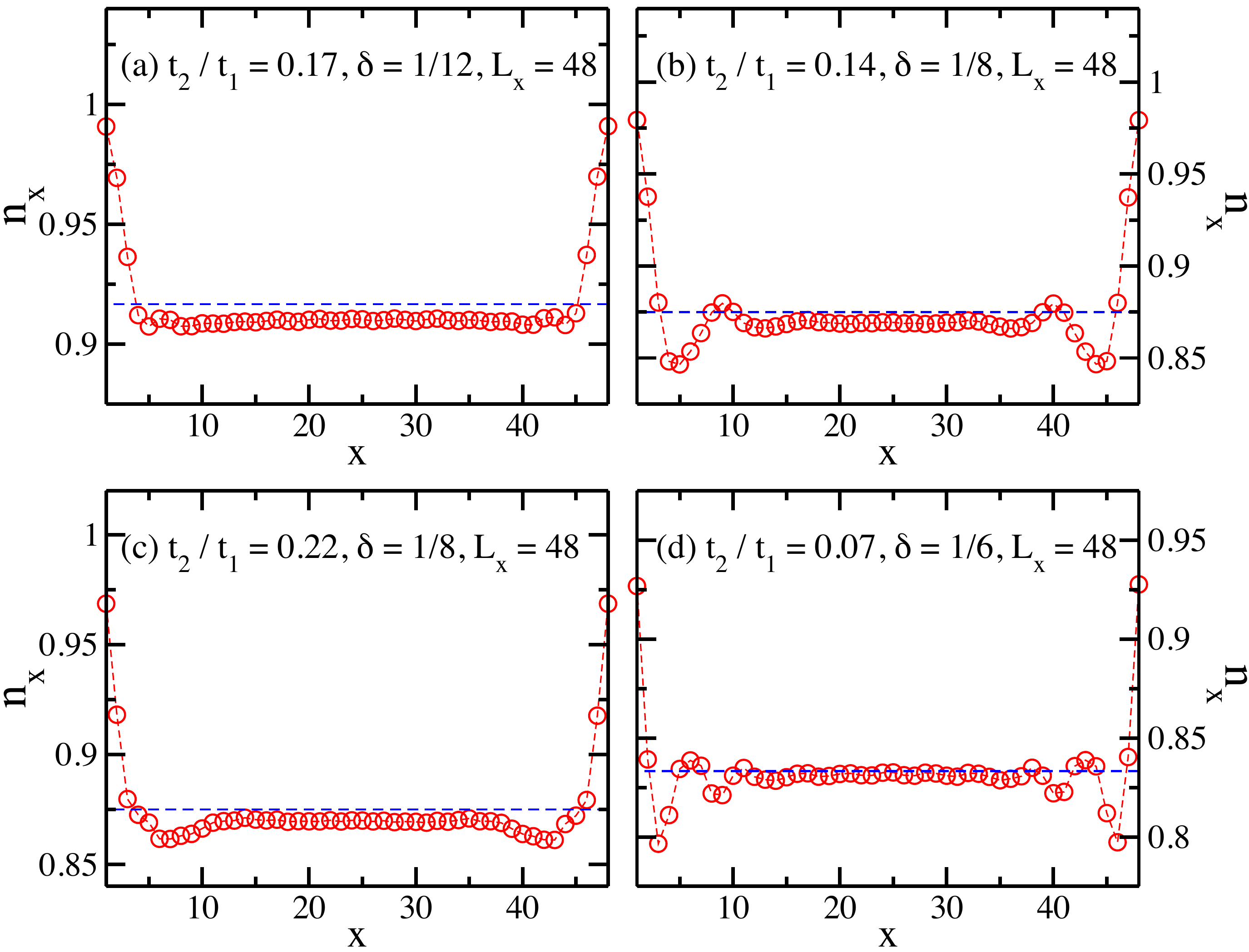}
\caption{Charge density profile $n_{x}$ in the d-wave superconducting phase.
The results for different doping ratios $\delta$ are shown on the $L_y = 6, L_x = 48$ cylinder.
The dashed blue lines denote $n_{x} = 1 - \delta$.
}
\label{supfig:cdw_sc}
\end{figure}

\begin{figure}[htp]
\includegraphics[width=0.8\linewidth]{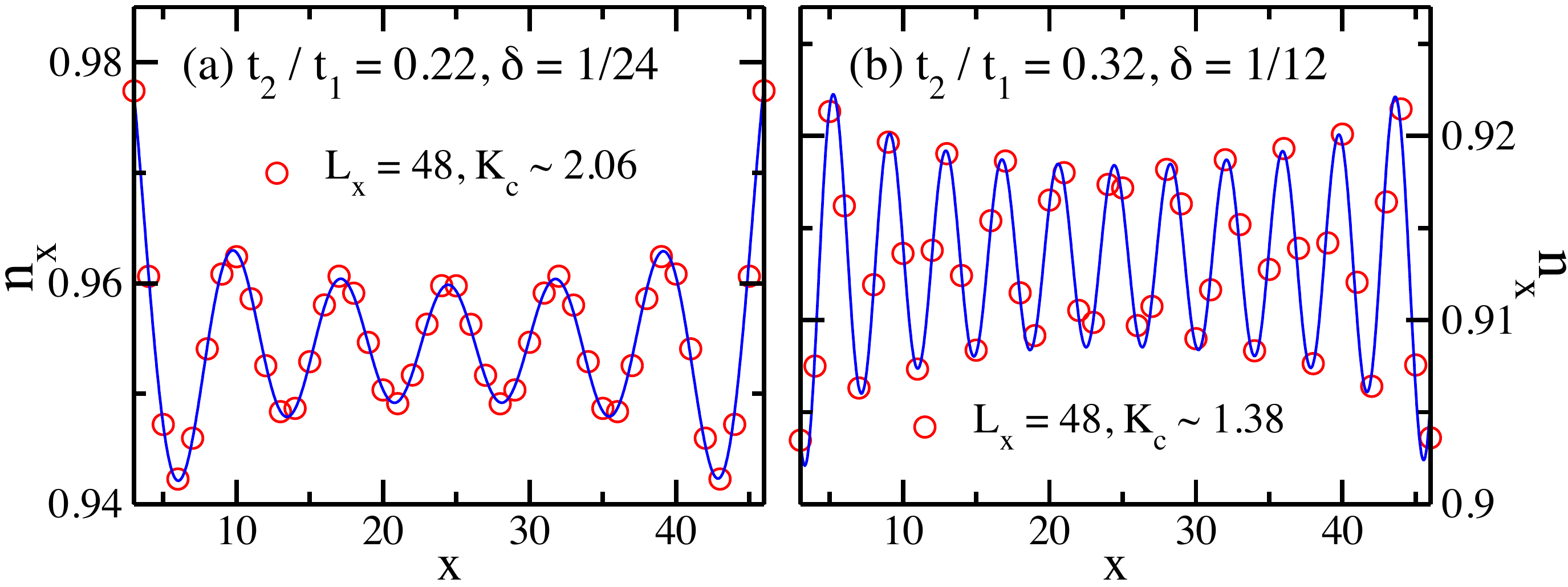}
\caption{Charge density profile $n_{x}$ in the coexistent phase.
The results for (a) $t_2 / t_1 = 0.22, \delta = 1/24$ and (b) $t_2 / t_1 = 0.32, \delta = 1/12$ on the $L_y = 6, L_x = 48$ cylinder.
The DMRG data are shown as the red circles, which are fitted by the formula $n(x) = n_0(x) + A_0 \cos(Qx + \phi) [x^{-K_c / 2} + (L_x + 1 - x)^{-K_c / 2}]$ (see the blue lines). The fittings give $K_c \simeq 2.06$ and $1.38$ for the two systems, respectively.
}
\label{supfig:cdw_cdw_sc}
\end{figure}

\begin{figure}[htp]
\includegraphics[width=0.8\linewidth]{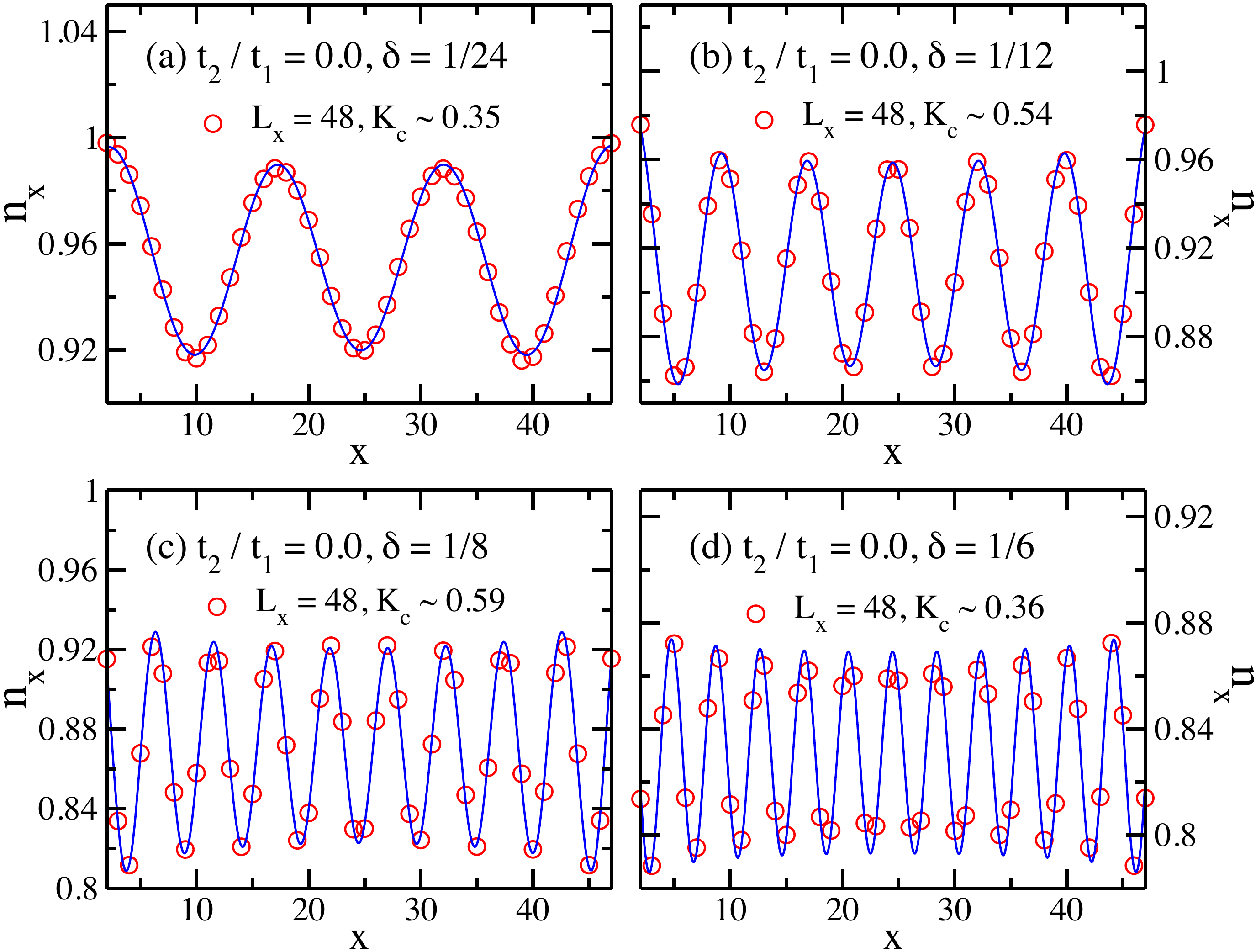}
\caption{Charge density profile $n_{x}$ in the charge density wave phase.
The results for $t_2 / t_1 = 0.0$ and different doping ratios are shown on the $L_y = 6, L_x = 48$ cylinder.
The DMRG data are shown as the red circles, which are also fitted by the formula $n(x) = n_0(x) + A_0 \cos(Qx + \phi) [x^{-K_c / 2} + (L_x + 1 - x)^{-K_c / 2}]$ (see the blue lines). The good fittings give the small power exponents $K_c = 0.35 - 0.6$ for these systems.
}
\label{supfig:cdw_cdw}
\end{figure}

\begin{figure}[htp]
\includegraphics[width=0.8\linewidth]{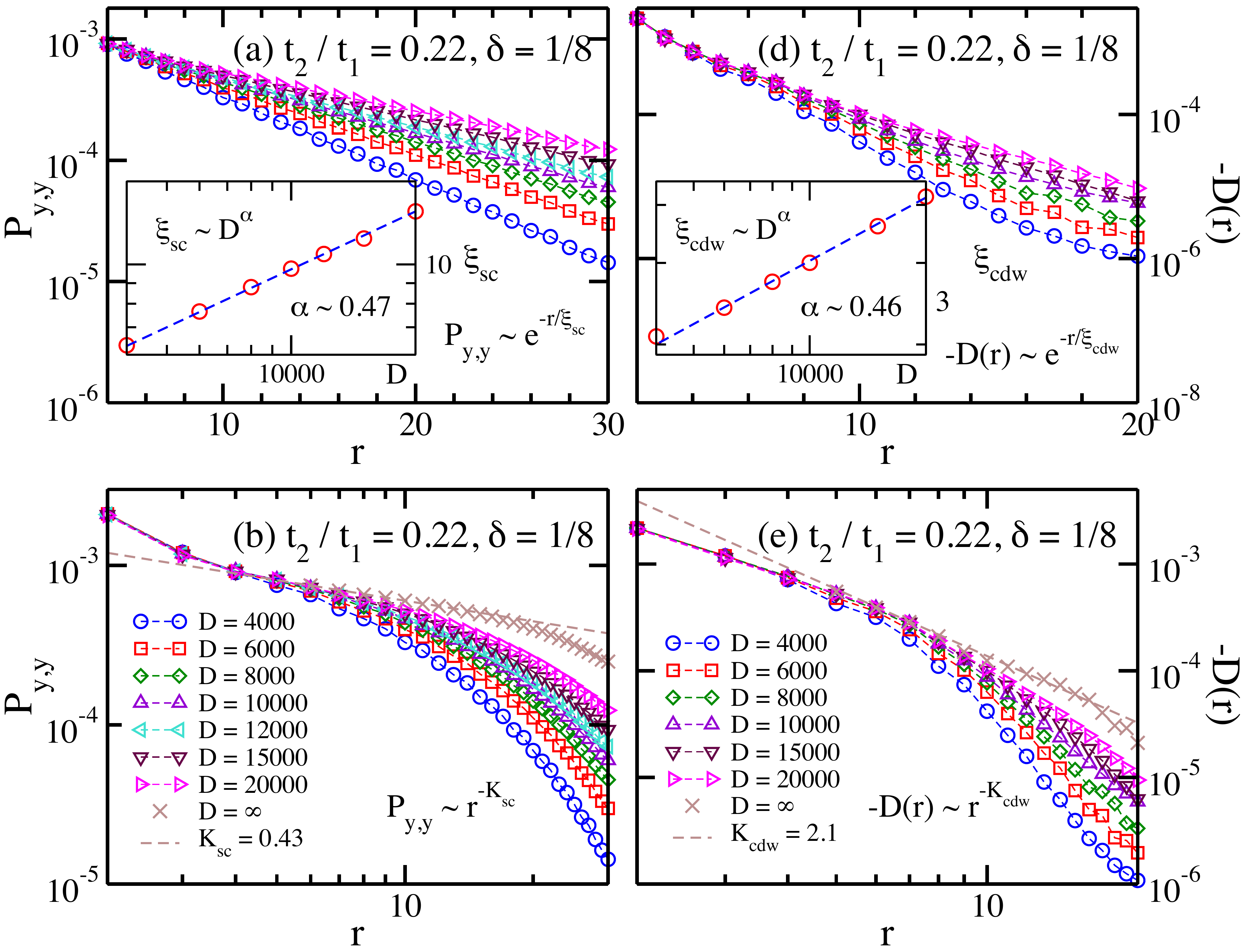}
\includegraphics[width=0.8\linewidth]{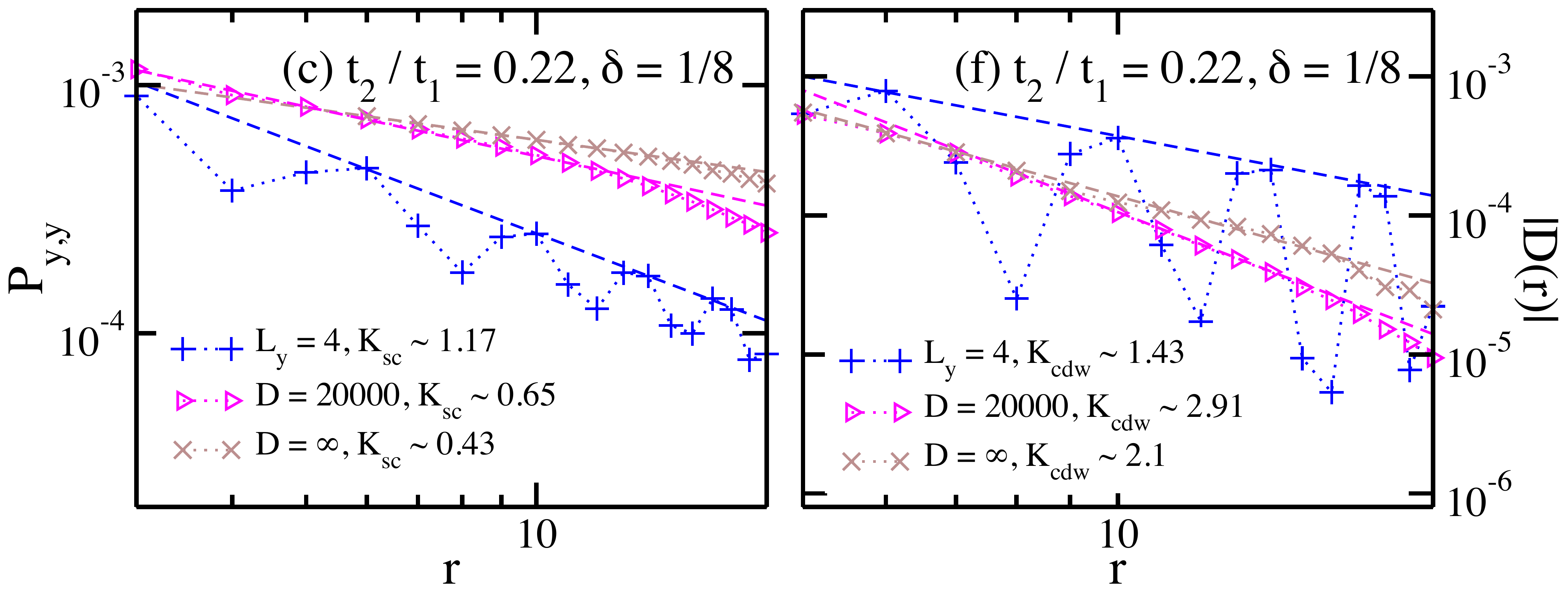}
\caption{Analyses of the SC pairing correlation and density-density correlation for $t_2 / t_1 = 0.22, \delta = 1/8$ in the d-wave SC phase.
(a) Semi-logarithmic plot of the SC pairing correlation $P_{y,y}(r)$ obtained by different $SU(2)$ bond dimensions $D = 4000 - 20000$. The inset shows the dependence of correlation length $\xi_{sc}$ on $D$, where $\xi_{sc}$ is obtained by fitting  $P_{y,y} \sim \exp(-r/\xi_{sc})$.  
In the range of $D = 4000 - 20000$ $SU(2)$ states (equivalent to $12000 - 60000$ $U(1)$ states), $\xi_{sc}$ fits to $\xi_{sc} \sim D^\alpha$ with $\alpha = 0.47$. (b) Double-logarithmic plot of the correlation $P_{y,y}(r)$ (same data in the subfigure (a)). The dashed crossed line denotes the power-law fitting of the extrapolated $D \rightarrow \infty$ results.
(c) Comparison of the power exponents $K_{sc}$ of the pairing correlations on the $L_y = 4$ and $L_y = 6$ cylinders.
(d) Semi-logarithmic plot of the density correlation $D(r)$. The inset shows the dependence of correlation length $\xi_{cdw}$ on $D$, where $\xi_{cdw}$ is obtained by fitting $D(r) \sim \exp(-r/\xi_{cdw})$.  
In the range of $D = 8000 - 20000$, $\xi_{cdw}$ fits to $\xi_{cdw} \sim D^\alpha$ with $\alpha = 0.46$. (e) Double-logarithmic plot of the density correlation $D(r)$ (same data in the subfigure (d)). The dashed crossed line denotes the power-law fitting of the extrapolated $D \rightarrow \infty$ results.
(f) Comparison of the power exponents $K_{cdw}$ of the density correlations on the $L_y = 4$ and $L_y = 6$ cylinders.
}
\label{supfig:pairing_density_022_8}
\end{figure}

\begin{figure}[htp]
\includegraphics[width=0.8\linewidth]{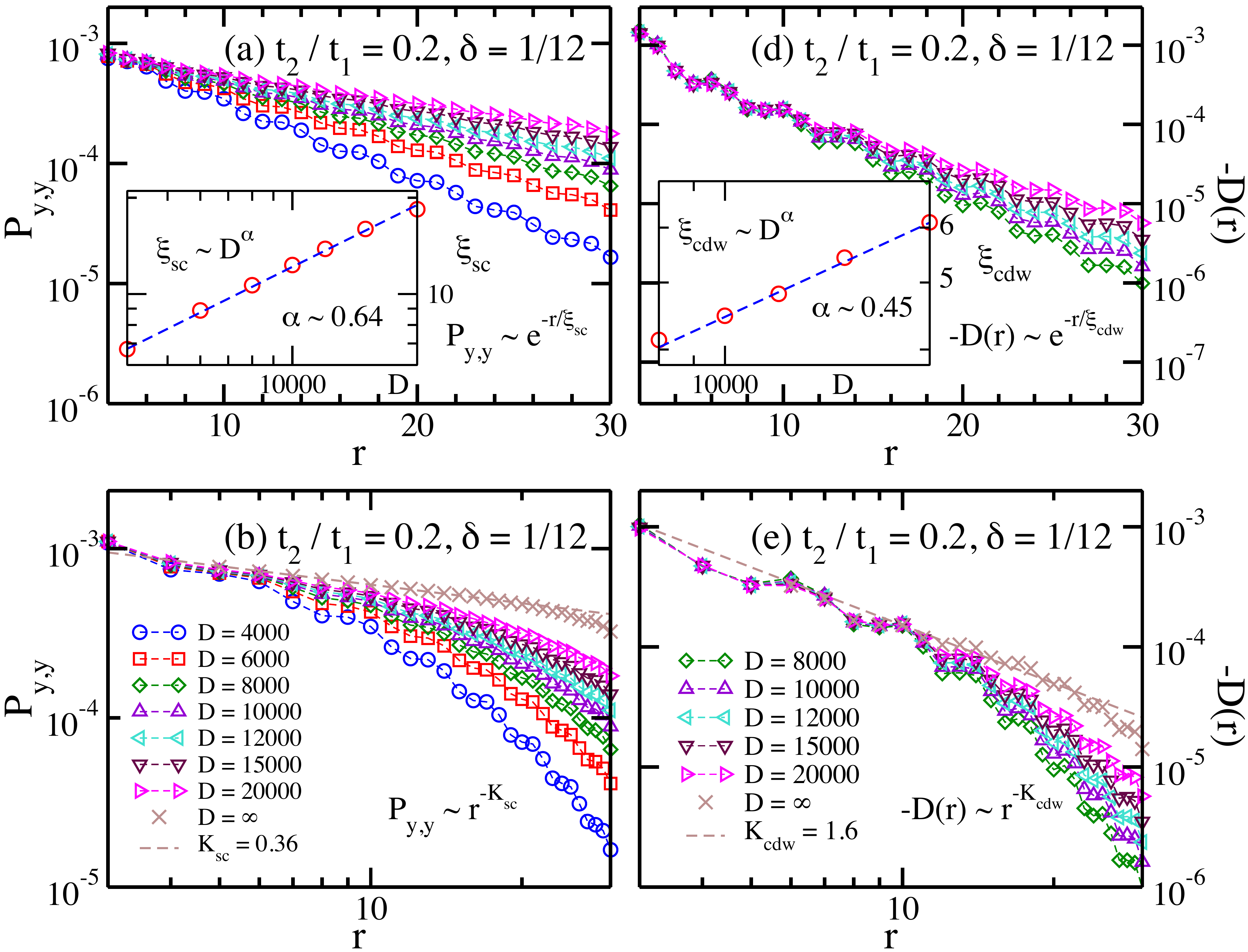}
\includegraphics[width=0.8\linewidth]{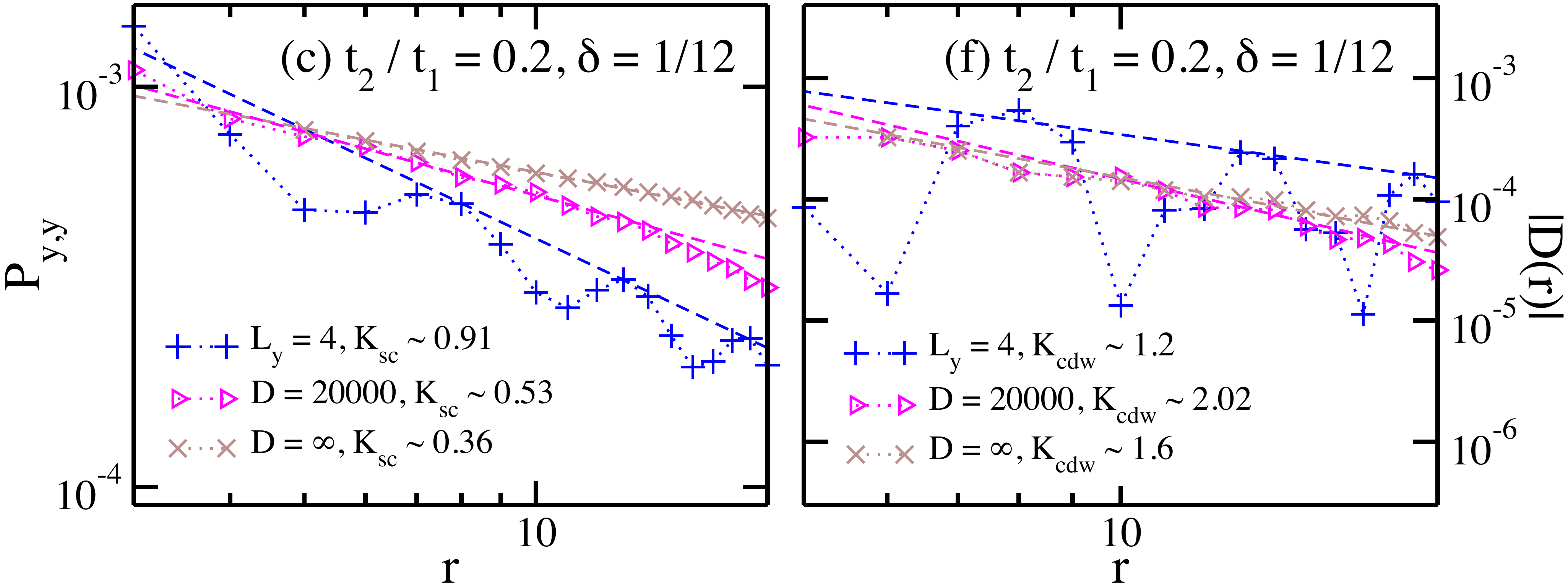}
\caption{Analyses of the SC pairing correlation and density-density correlation for $t_2 / t_1 = 0.2, \delta = 1/12$ in the coexistent phase.
(a) Semi-logarithmic plot of the SC pairing correlation $P_{y,y}(r)$ obtained by different $SU(2)$ bond dimensions $D = 4000 - 20000$. The inset shows the dependence of correlation length $\xi_{sc}$ on $D$, where $\xi_{sc}$ is obtained by fitting  $P_{y,y} \sim \exp(-r/\xi_{sc})$.  
In the range of $D = 4000 - 20000$ $SU(2)$ states (equivalent to $12000 - 60000$ $U(1)$ states), $\xi_{sc}$ fits to $\xi_{sc} \sim D^\alpha$ with $\alpha = 0.64$. (b) Double-logarithmic plot of the correlation $P_{y,y}(r)$ (same data in the subfigure (a)). The dashed crossed line denotes the power-law fitting of the extrapolated $D \rightarrow \infty$ results.
(c) Comparison of the power exponents $K_{sc}$ of the pairing correlations on the $L_y = 4$ and $L_y = 6$ cylinders.
(d) Semi-logarithmic plot of the density correlation $D(r)$. The inset shows the dependence of correlation length $\xi_{cdw}$ on $D$, where $\xi_{cdw}$ is obtained by fitting $D(r) \sim \exp(-r/\xi_{cdw})$.  
In the range of $D = 4000 - 20000$, $\xi_{cdw}$ fits to $\xi_{cdw} \sim D^\alpha$ with $\alpha = 0.45$. (e) Double-logarithmic plot of the density correlation $D(r)$ (same data in the subfigure (d)). The dashed crossed line denotes the power-law fitting of the extrapolated $D \rightarrow \infty$ results.
(f) Comparison of the power exponents $K_{cdw}$ of the density correlations on the $L_y = 4$ and $L_y = 6$ cylinders.
}
\label{supfig:pairing_density_02_12}
\end{figure}

\begin{figure}[htp]
\includegraphics[width=0.8\linewidth]{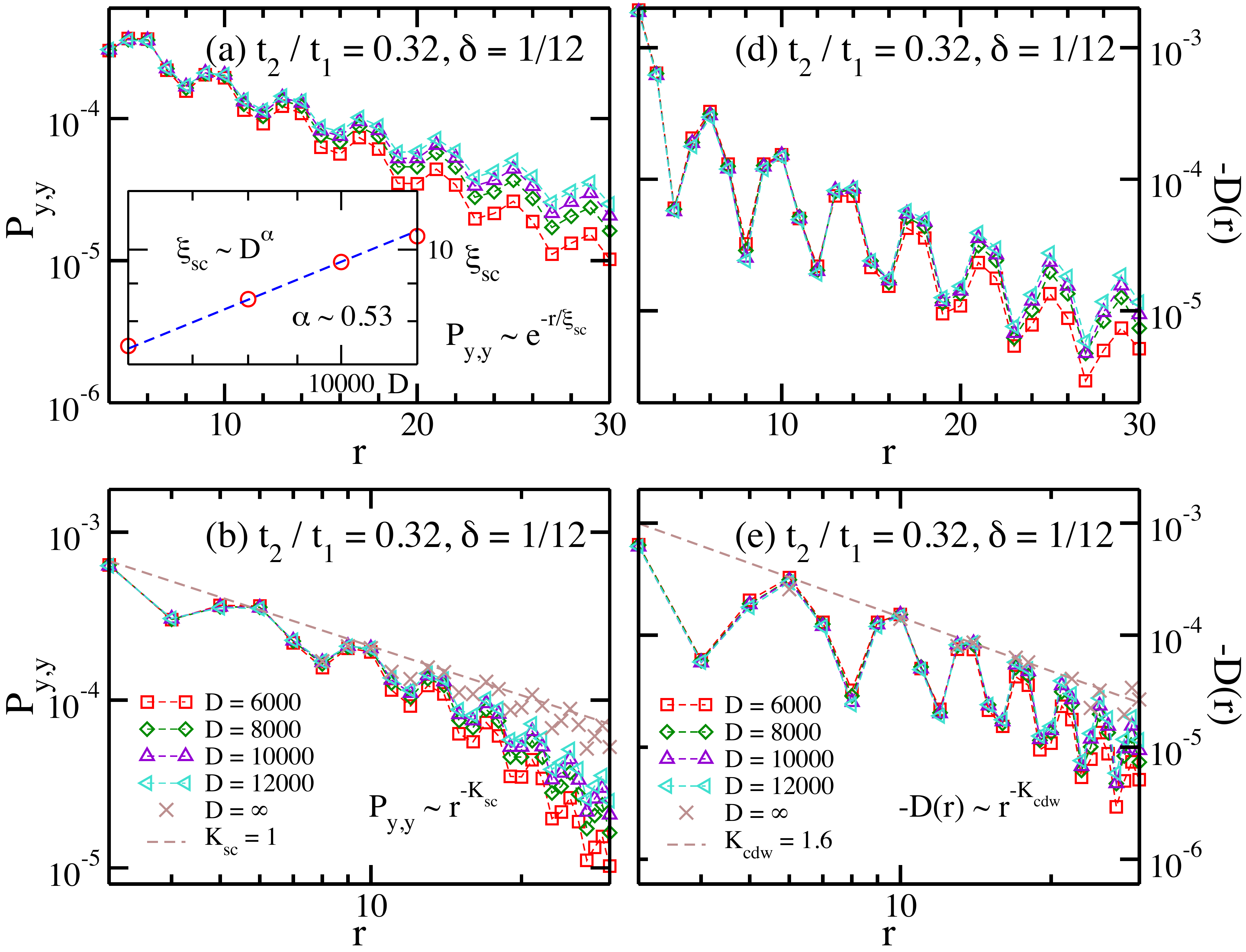}
\includegraphics[width=0.8\linewidth]{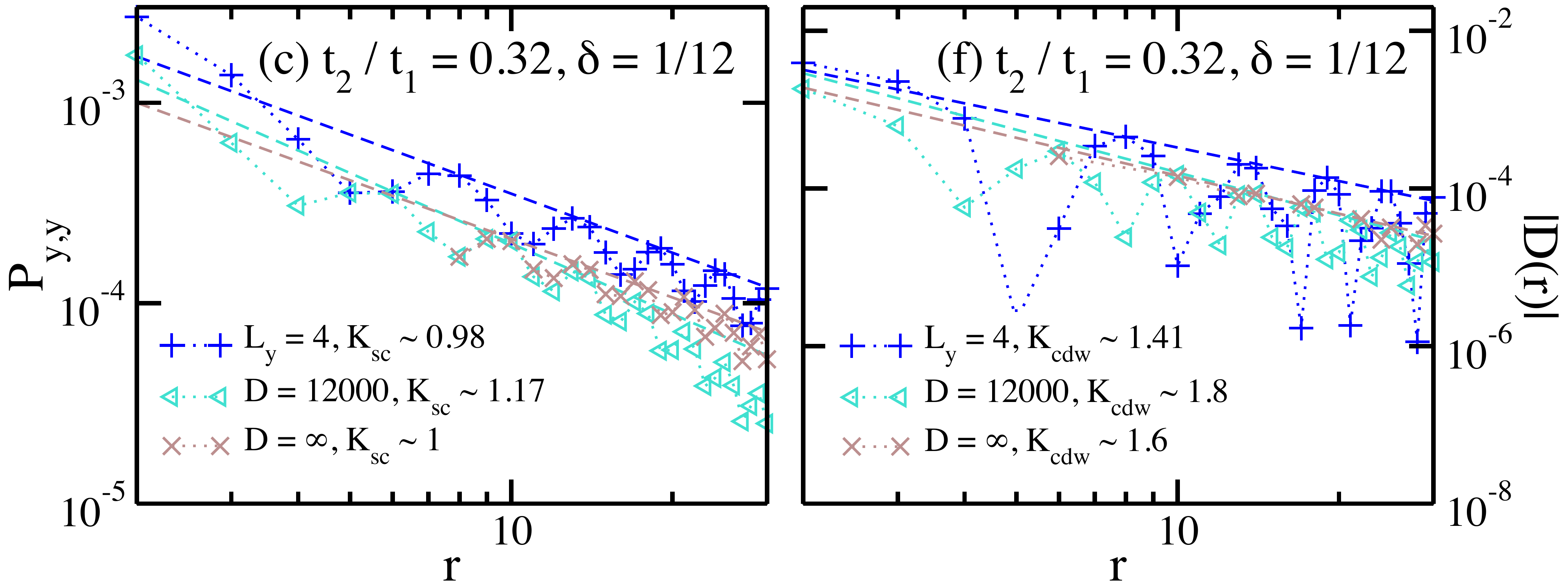}
\caption{Analyses of the SC pairing correlation and density-density correlation for $t_2 / t_1 = 0.32, \delta = 1/12$ in the coexistent phase.
(a) Semi-logarithmic plot of the SC pairing correlation $P_{y,y}(r)$ obtained by different $SU(2)$ bond dimensions $D = 6000 - 12000$. The inset shows the dependence of correlation length $\xi_{sc}$ on $D$, where $\xi_{sc}$ is obtained by fitting  $P_{y,y} \sim \exp(-r/\xi_{sc})$.  
In the range of $D = 6000 - 12000$ $SU(2)$ states (equivalent to $18000 - 36000$ $U(1)$ states), $\xi_{sc}$ fits to $\xi_{sc} \sim D^\alpha$ with $\alpha = 0.53$. (b) Double-logarithmic plot of the correlation $P_{y,y}(r)$ (same data in the subfigure (a)). The dashed crossed line denotes the power-law fitting of the extrapolated $D \rightarrow \infty$ results.
(c) Comparison of the power exponents $K_{sc}$ of the pairing correlations on the $L_y = 4$ and $L_y = 6$ cylinders.
(d) Semi-logarithmic plot of the density correlation $D(r)$. The density correlations clearly decay slower than the exponential behavior. (e) Double-logarithmic plot of the density correlation $D(r)$ (same data in the subfigure (d)). The dashed crossed line denotes the power-law fitting of the extrapolated $D \rightarrow \infty$ results.
(f) Comparison of the power exponents $K_{cdw}$ of the density correlations on the $L_y = 4$ and $L_y = 6$ cylinders.}
\label{supfig:pairing_density_032_12}
\end{figure}

\begin{figure}[htp]
\includegraphics[width=0.8\linewidth]{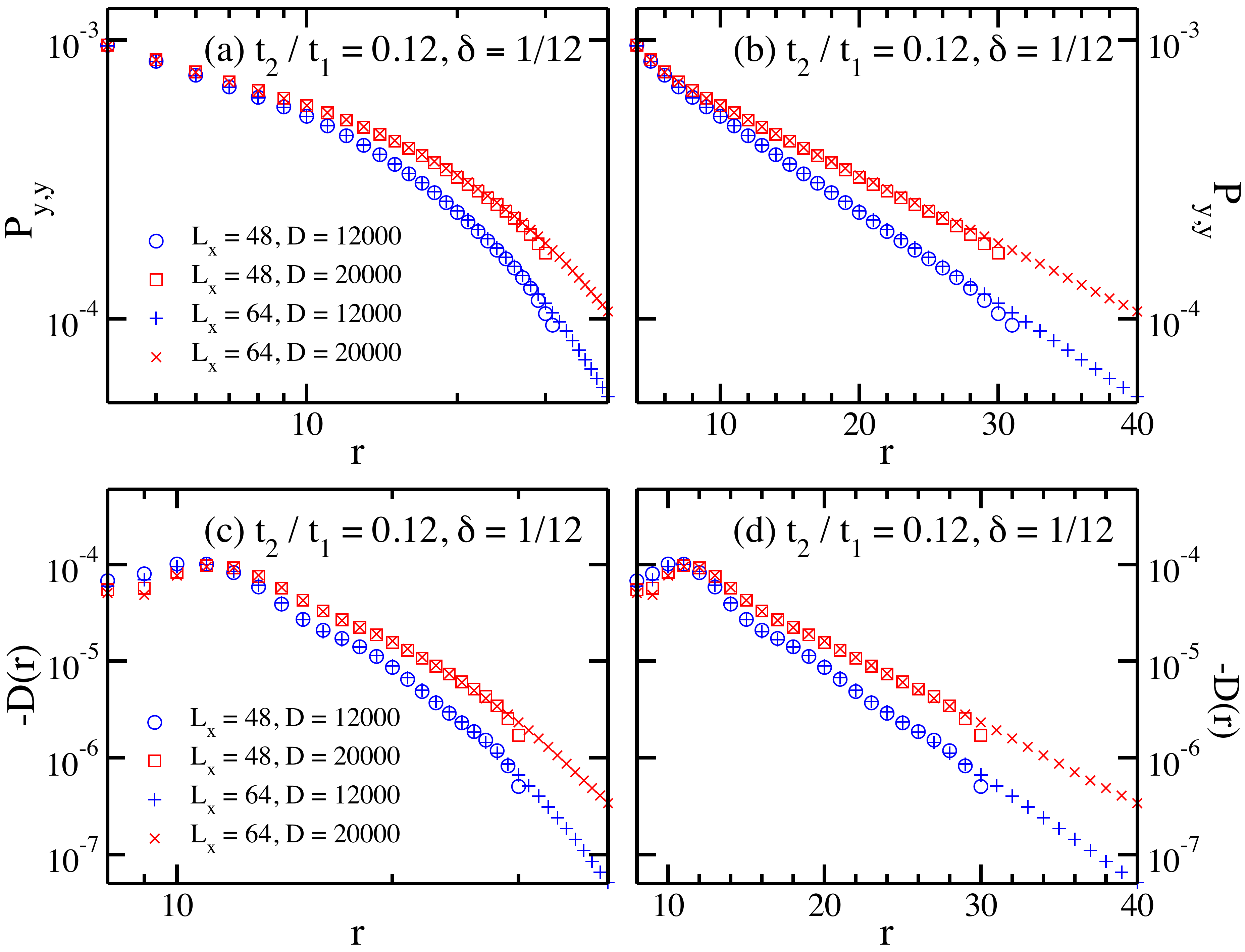}
\caption{Comparing correlation functions on the cylinder systems with different lengths.
(a) The double-logarithmic plot of the SC pairing correlations $P_{y,y}(r)$ for $t_2 / t_1 = 0.12, \delta = 1/12$, which are obtained by the bond dimensions $D = 12000, 20000$ on the $L_x = 48$ and $L_x = 64$ cylinders.
(b) The semi-logarithmic plot of the same data shown in the subfigure (a).
(c) The double-logarithmic plot of the density correlations $-D(r)$ for $t_2 / t_1 = 0.12, \delta = 1/12$, which are obtained by the bond dimensions $D = 12000, 20000$ on the $L_x = 48$ and $L_x = 64$ cylinders.
(d) The semi-logarithmic plot of the same data shown in the subfigure (c).
}
\label{supfig:correlation_length_dependence}
\end{figure}

\begin{figure}[htp]
\includegraphics[width=0.8\linewidth]{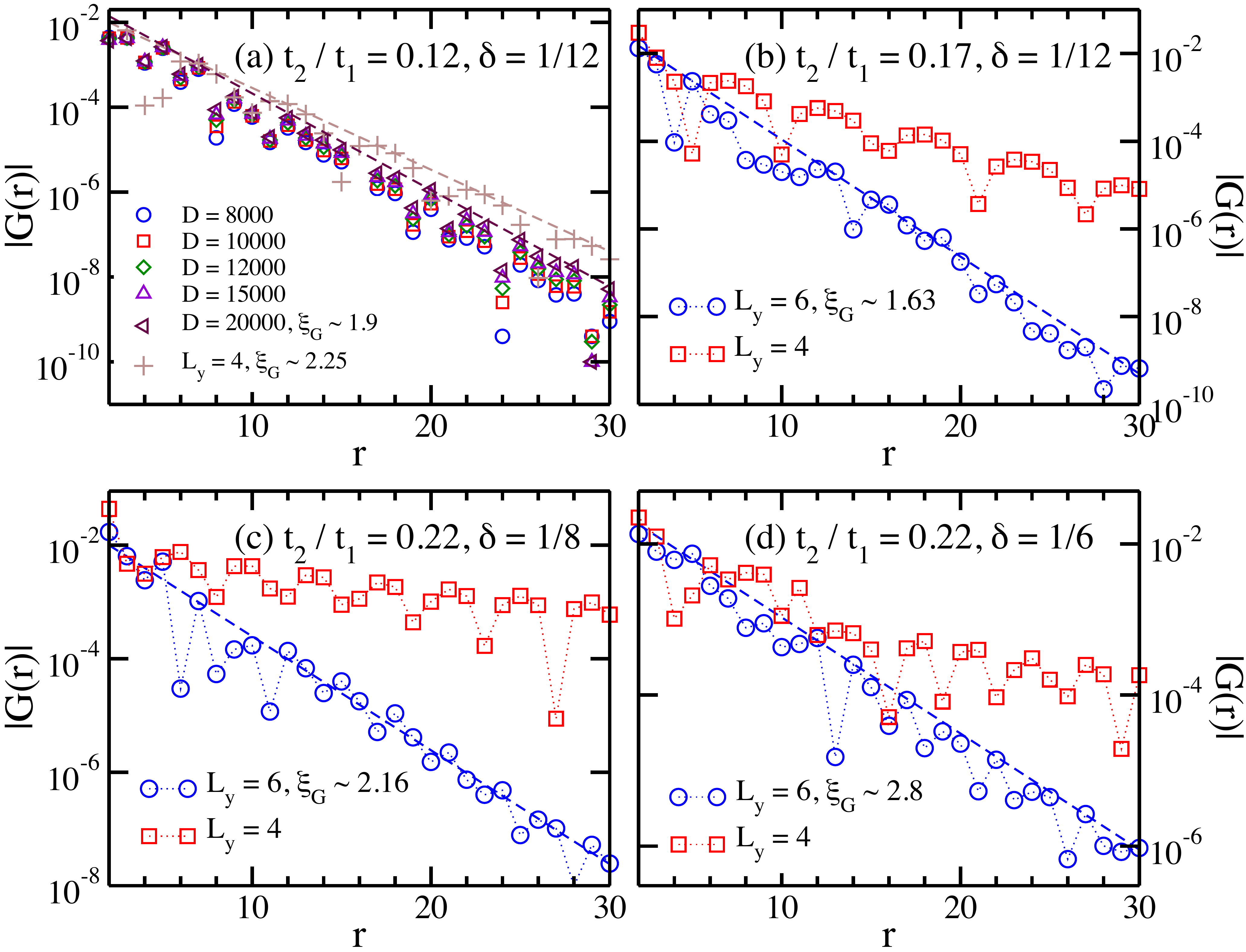}
\caption{Single-particle correlation functions in the d-wave superconducting phase.
(a) Semi-logarithmic plot of the correlations for $t_2 / t_1 = 0.12, \delta = 1/12$. The results for $L_y = 6$ are shown for the bond dimensions $D = 8000 - 20000$. The exponential fitting of $|G(r)| \sim e^{-r/\xi_{G}}$ for $D = 20000$ gives $\xi_{G} \simeq 1.9$.
(b)-(d) show the semi-logarithmic plots for $t_2 / t_1 = 0.17, \delta = 1/12$ ($D = 12000$), $t_2 / t_1 = 0.22, \delta = 1/8$ ($D = 15000$), and $t_2 / t_1 = 0.22, \delta = 1/6$ ($D = 12000$), respectively.
}
\label{supfig:green_dwave}
\end{figure}

\begin{figure}[htp]
\includegraphics[width=0.8\linewidth]{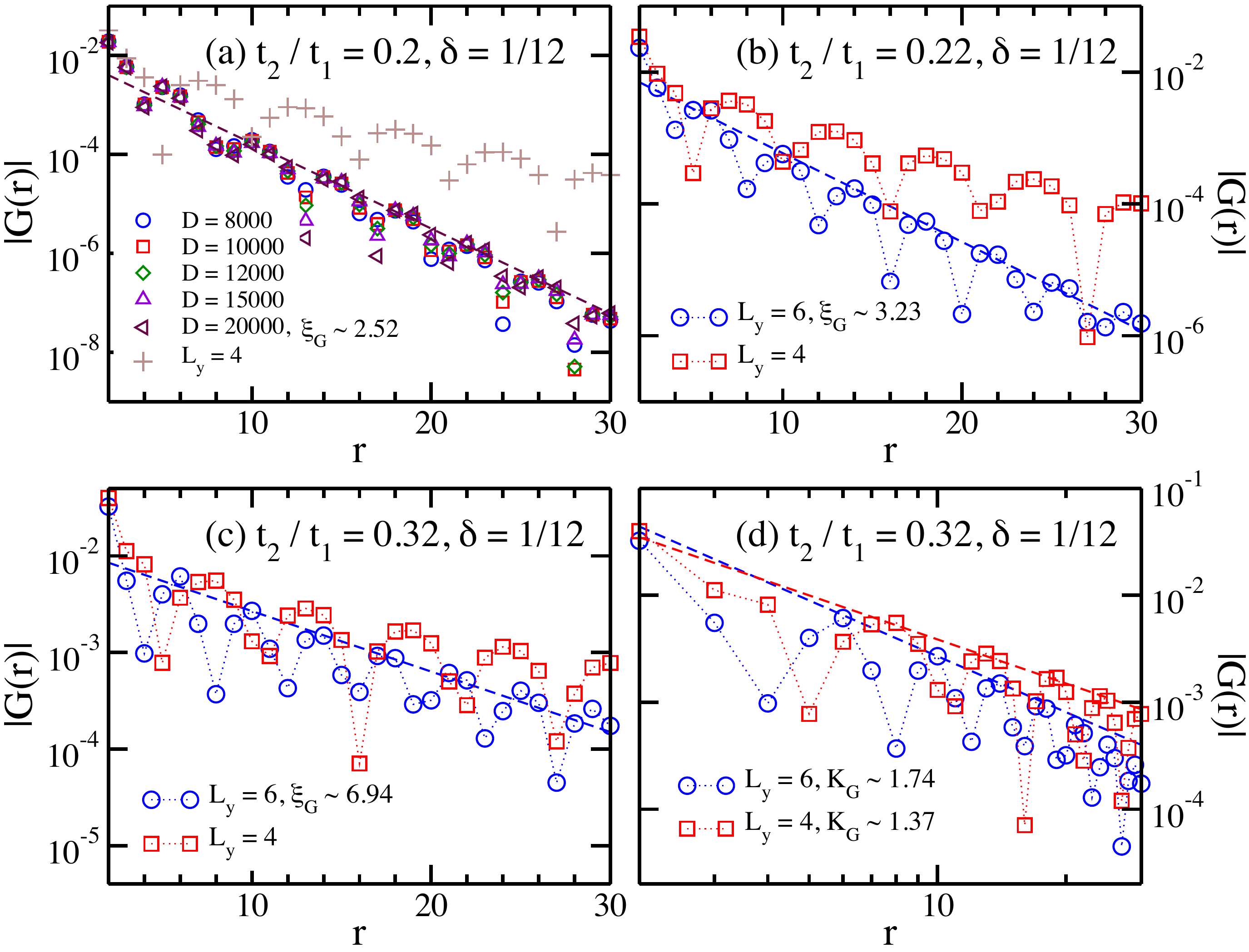}
\caption{Single-particle correlation functions in the coexistent phase.
(a) Semi-logarithmic plot of the correlations for $t_2 / t_1 = 0.2, \delta = 1/12$. The results for $L_y = 6$ are obtained by keeping the bond dimensions $D = 8000 - 20000$. (b) and (c) show the semi-logarithmic results for $t_2 / t_1 = 0.22, \delta = 1/12$ ($D = 15000$) and $t_2 / t_1 = 0.32, \delta = 1/12$ ($D = 12000$), respectively. (d) shows the double-logarithmic plot for $t_2 / t_1 = 0.32, \delta = 1/12$. The data for $L_y = 6$ could also be fitted well by the power-law decay with the power exponent $K_{G} \simeq 1.74$.
}
\label{supfig:green_coexistence}
\end{figure}

\begin{figure}[htp]
\includegraphics[width=0.32\linewidth]{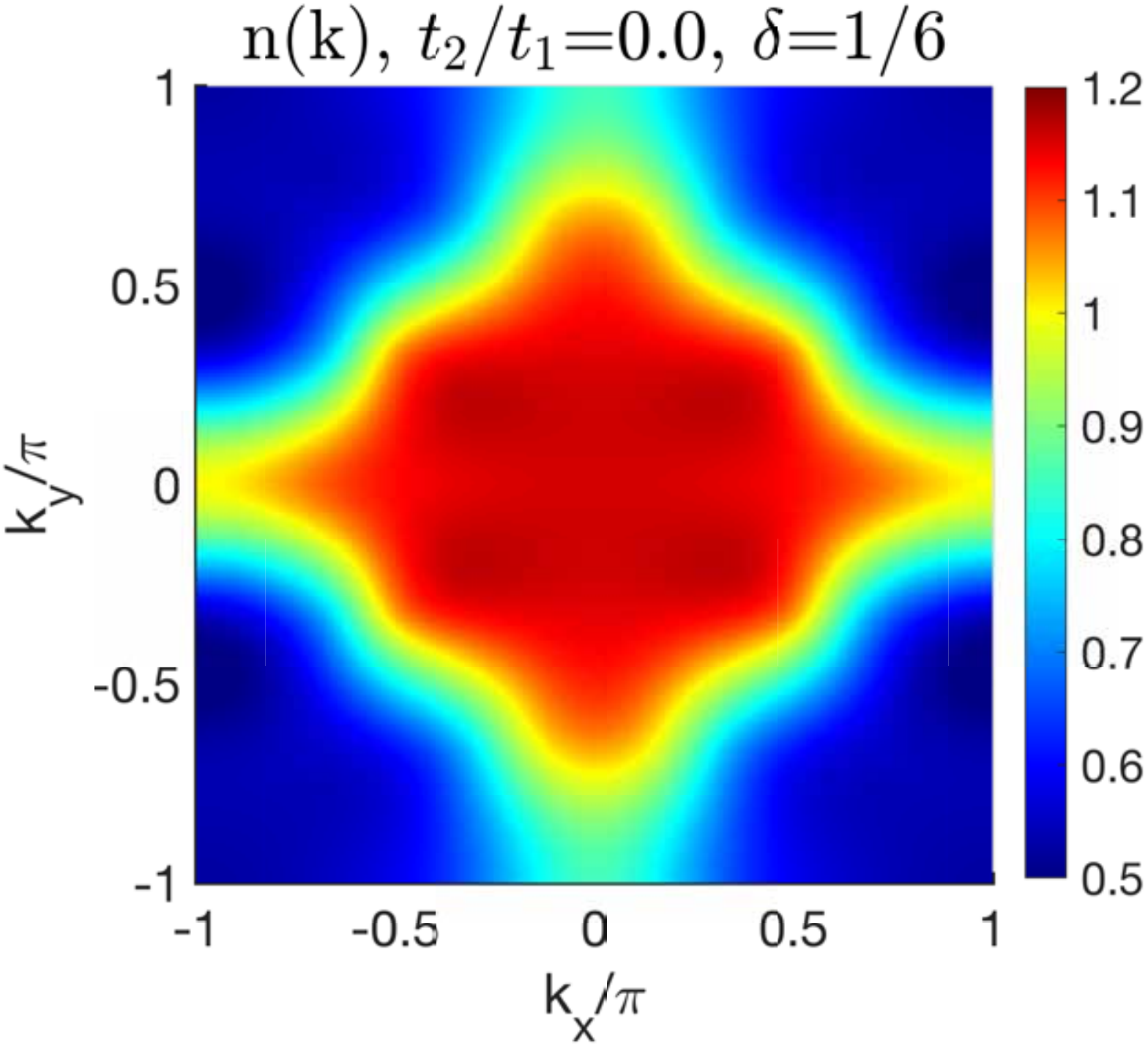}
\includegraphics[width=0.32\linewidth]{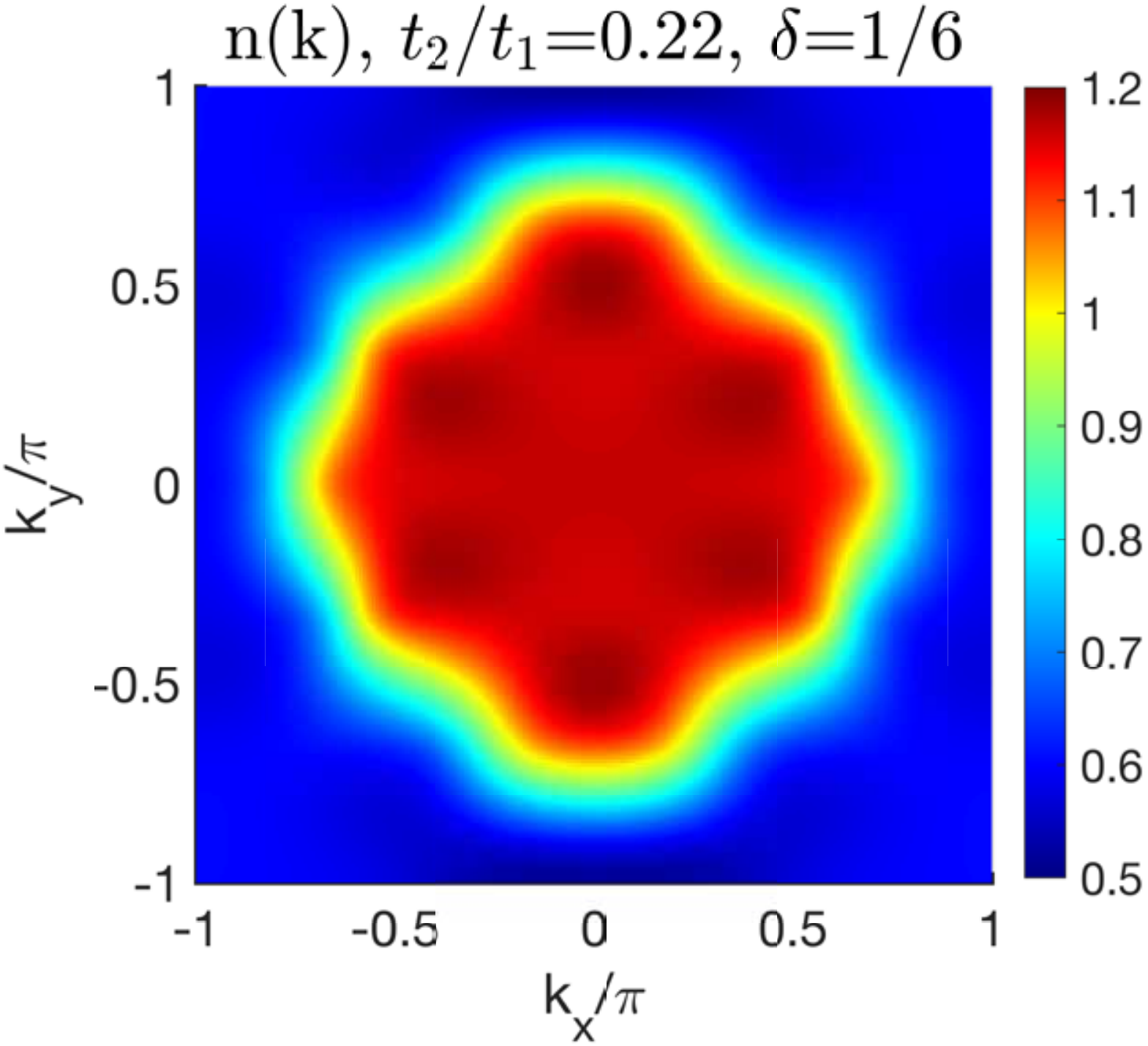}
\includegraphics[width=0.32\linewidth]{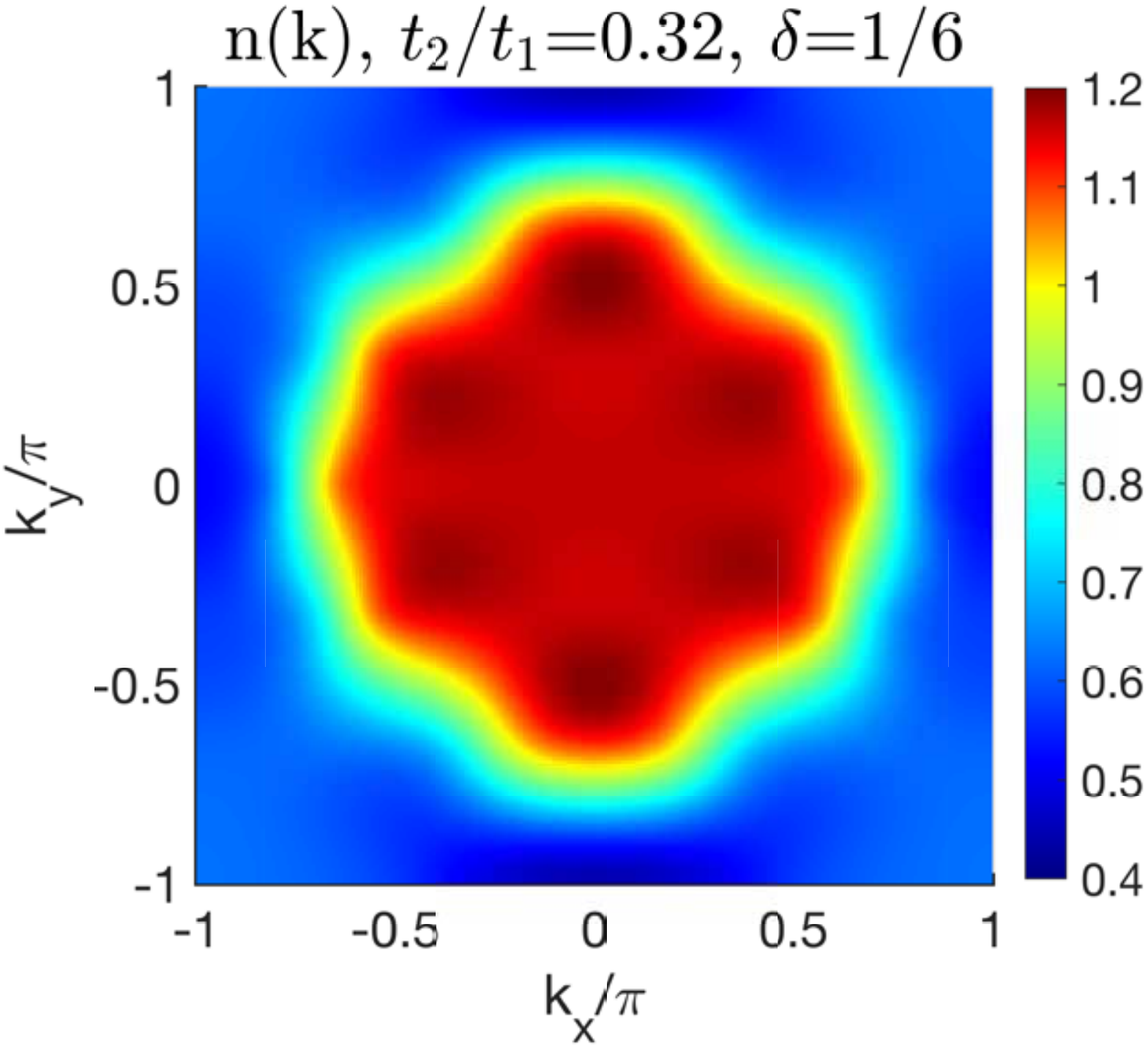}
\includegraphics[width=0.32\linewidth]{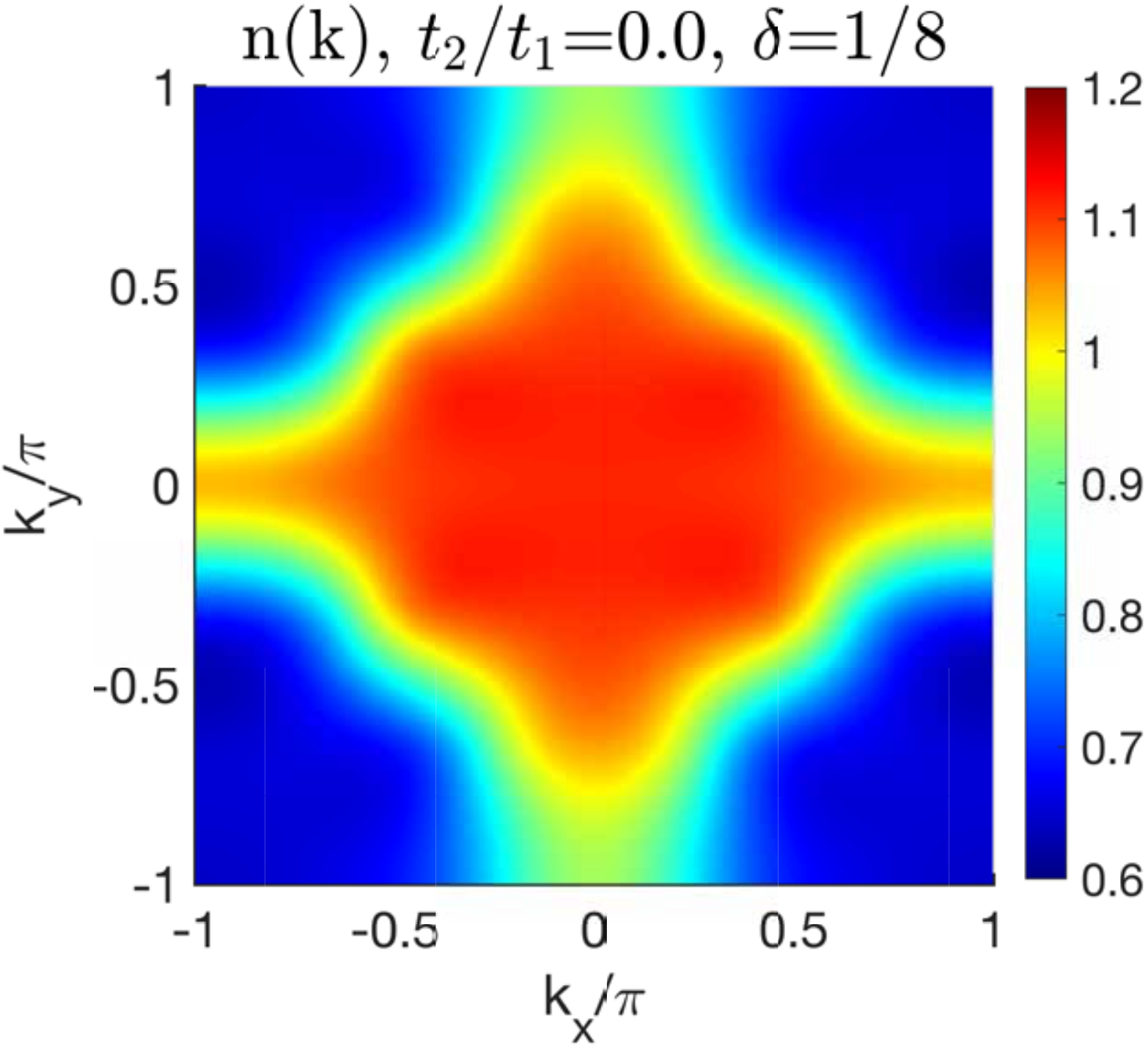}
\includegraphics[width=0.32\linewidth]{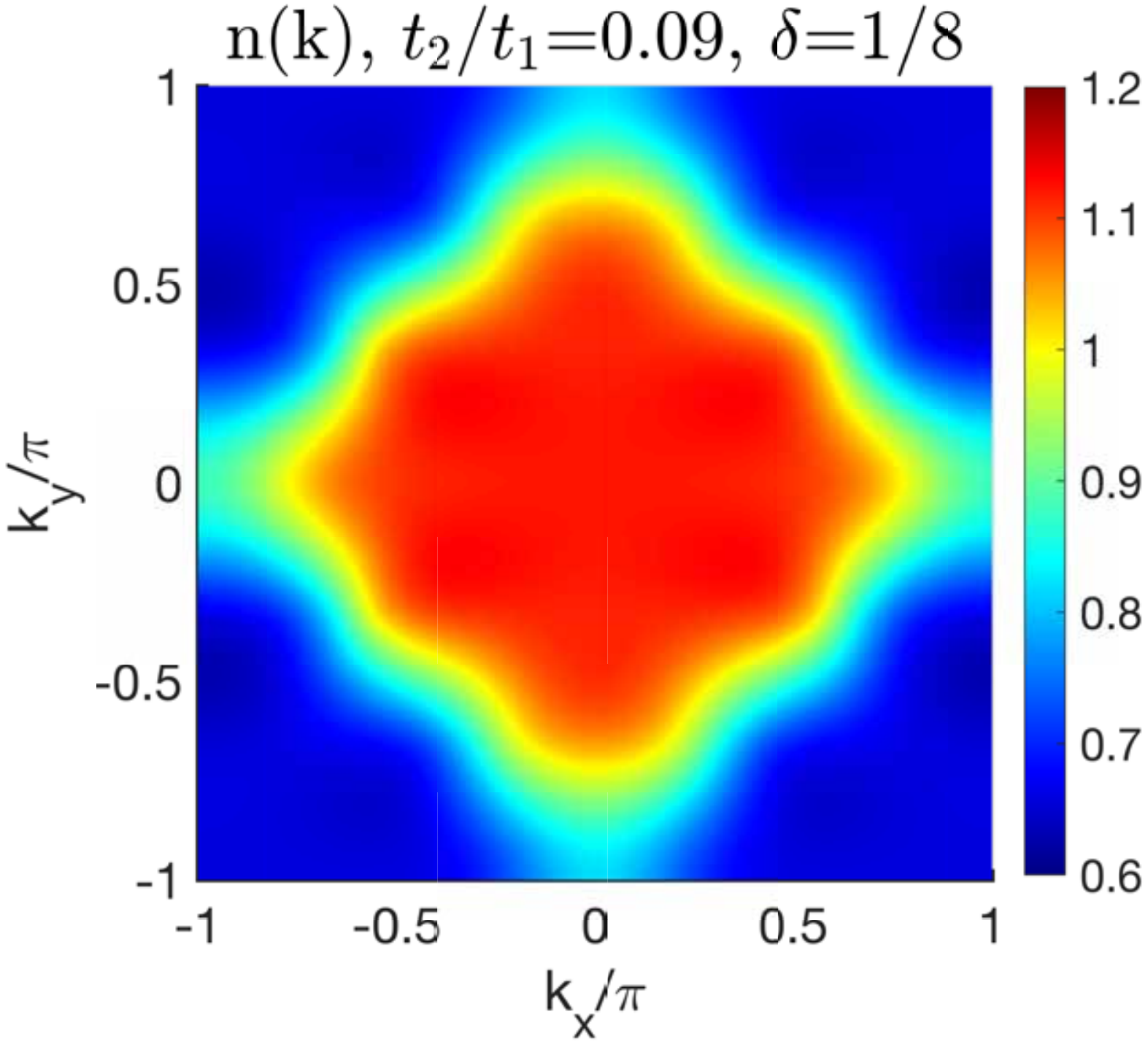}
\includegraphics[width=0.32\linewidth]{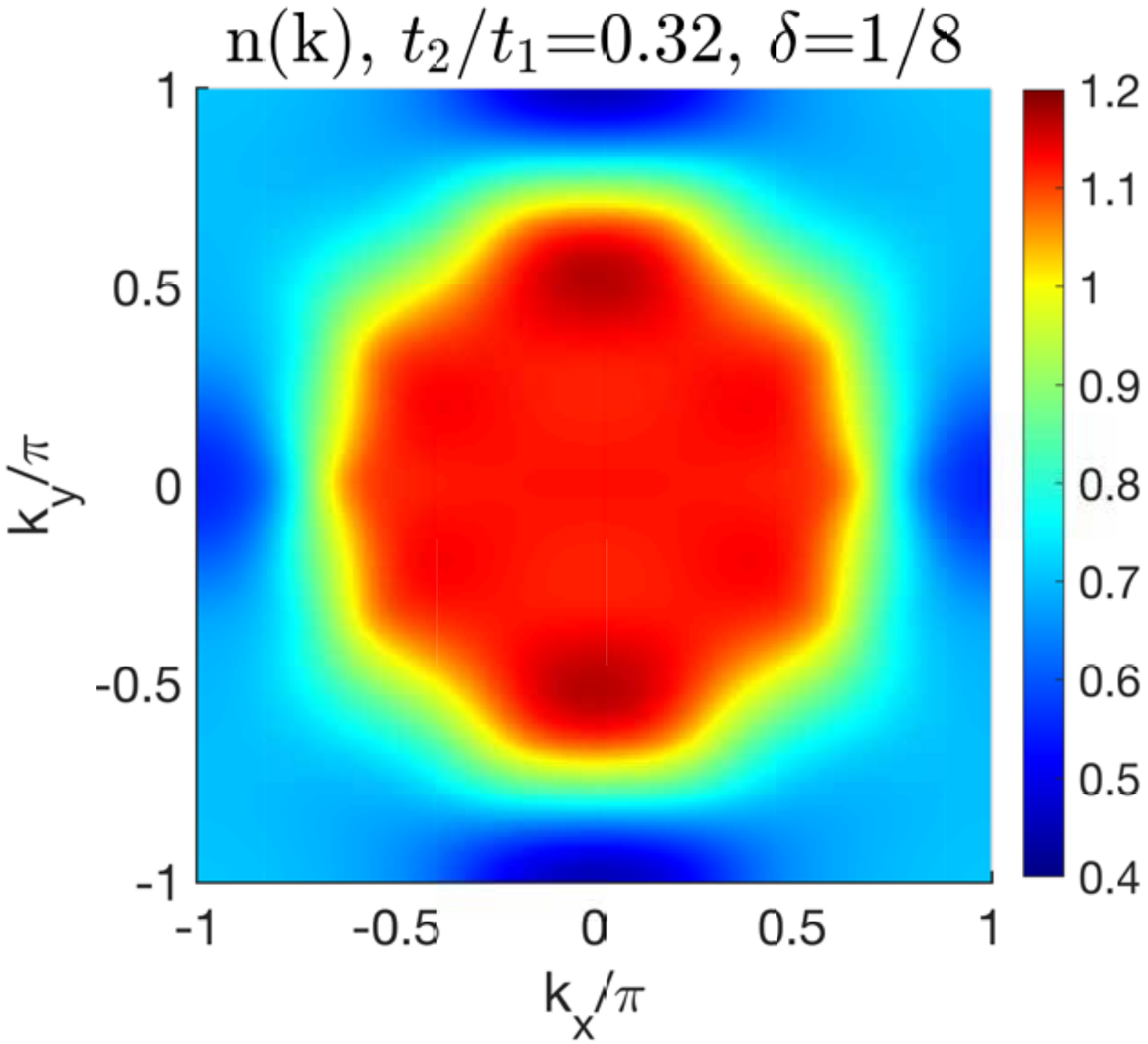}
\includegraphics[width=0.32\linewidth]{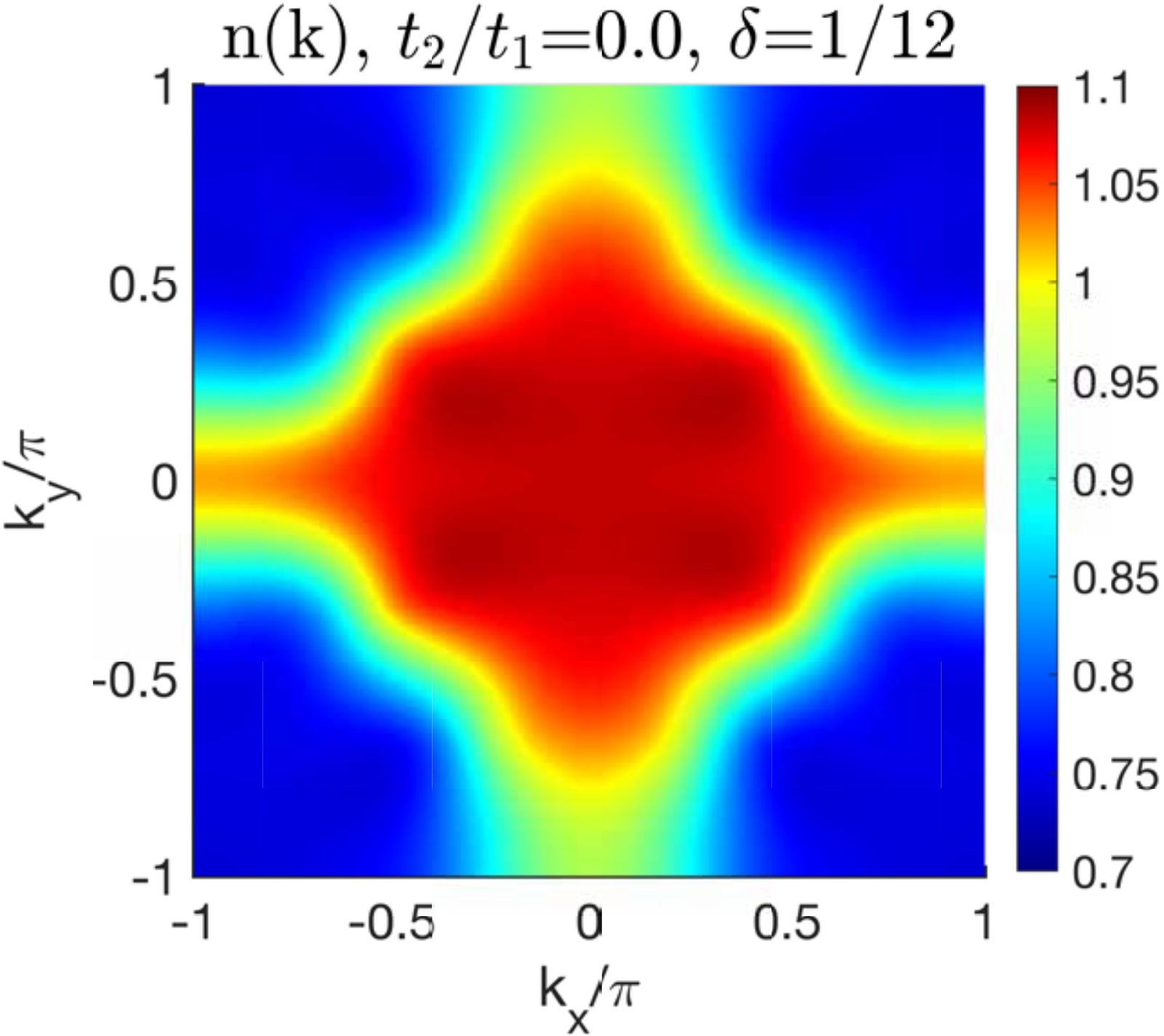}
\includegraphics[width=0.32\linewidth]{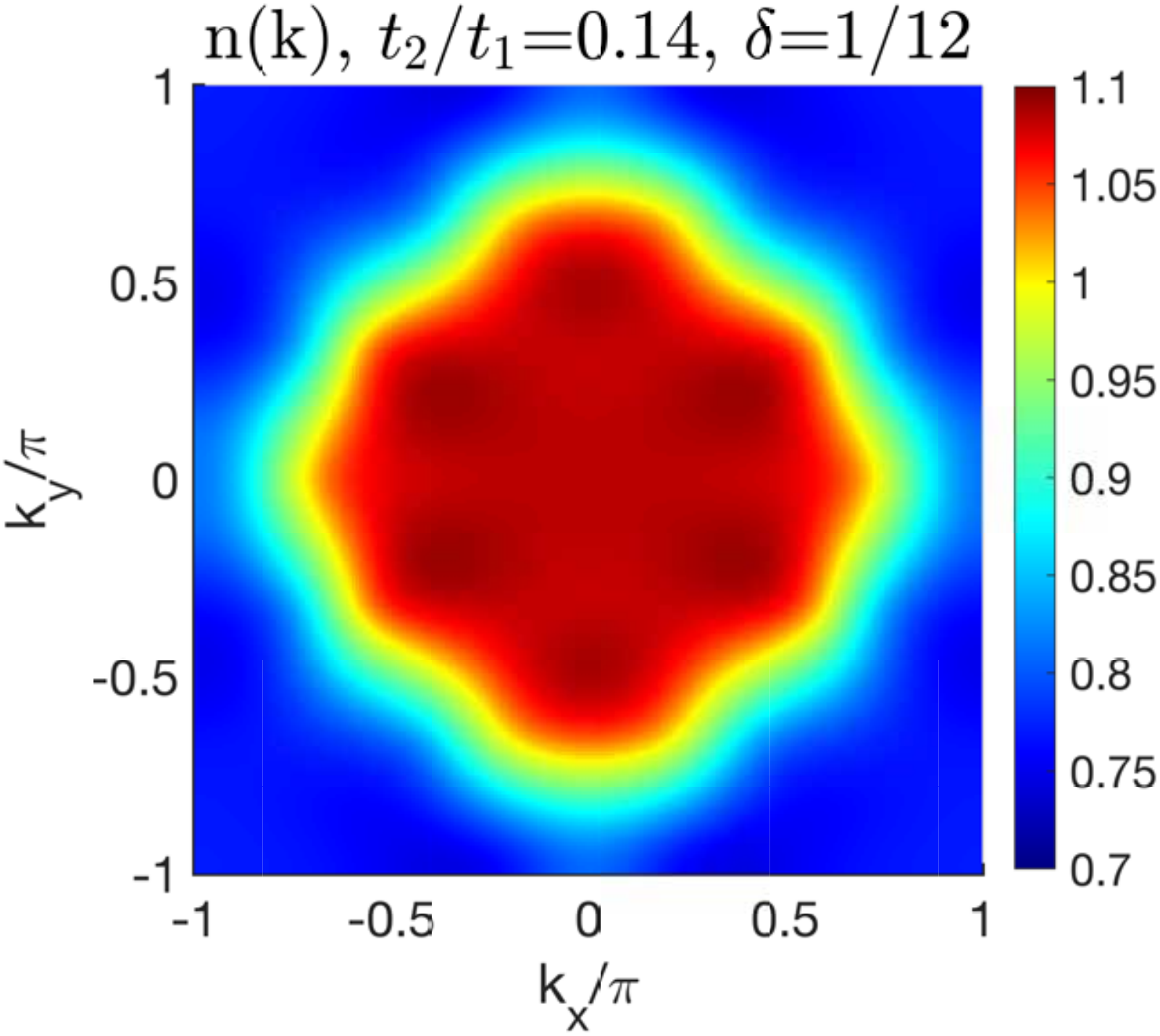}
\includegraphics[width=0.32\linewidth]{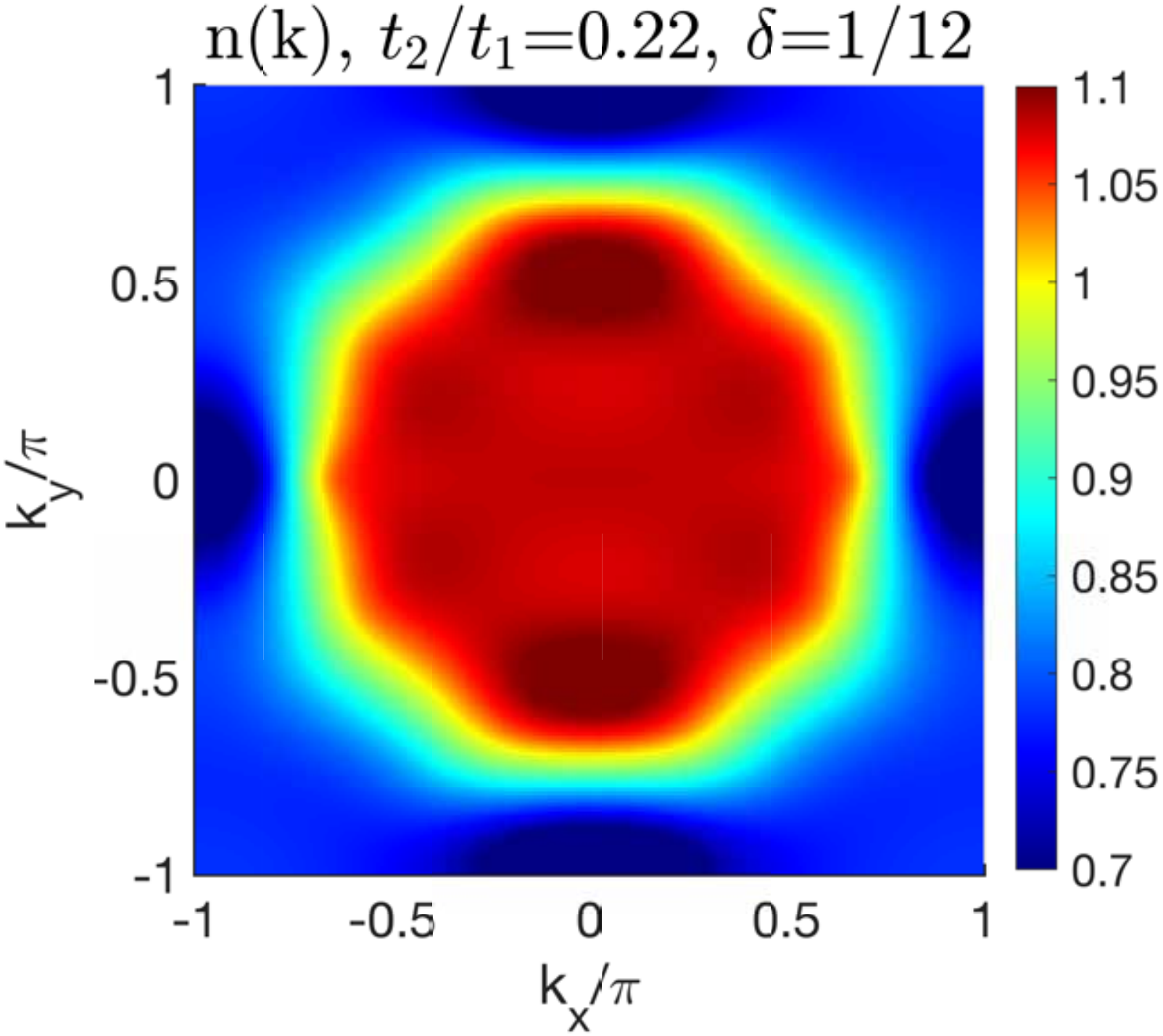}
\caption{Momentum distribution functions $n(\bf k)$ for different $t_2 / t_1$ and doping ratios $\delta$.
$n(\bf k)$ are obtained by taking the Fourier transformation for the single-particle correlations of the middle $6 \times 24$ sites on the $6 \times 48$ cylinder. These results are obtained by keeping the bond dimension $D = 10000$.
}
\label{supfig:nk}
\end{figure}

\begin{figure}[htp]
\includegraphics[width=0.5\linewidth]{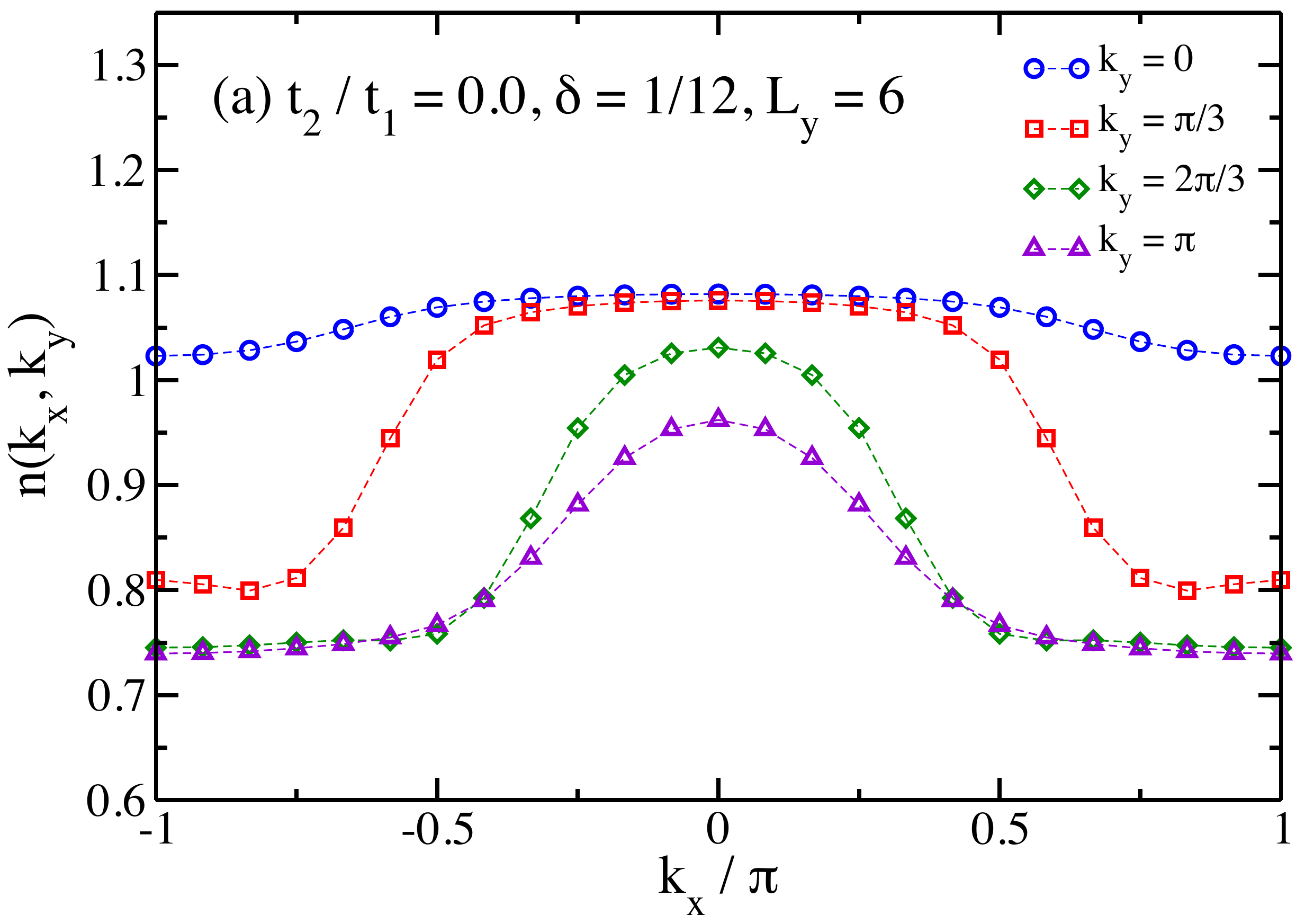}\\
\includegraphics[width=0.5\linewidth]{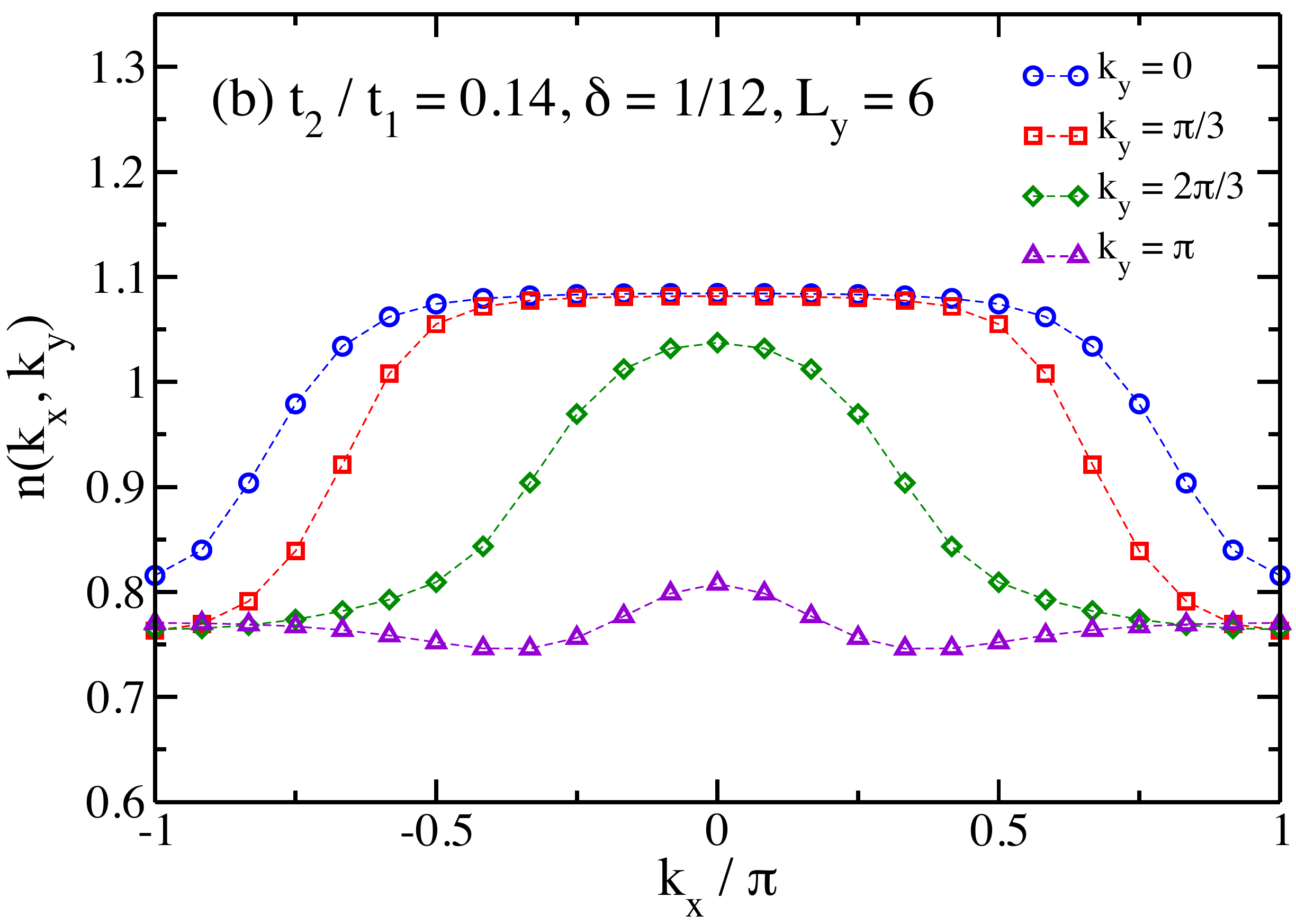}\\
\includegraphics[width=0.5\linewidth]{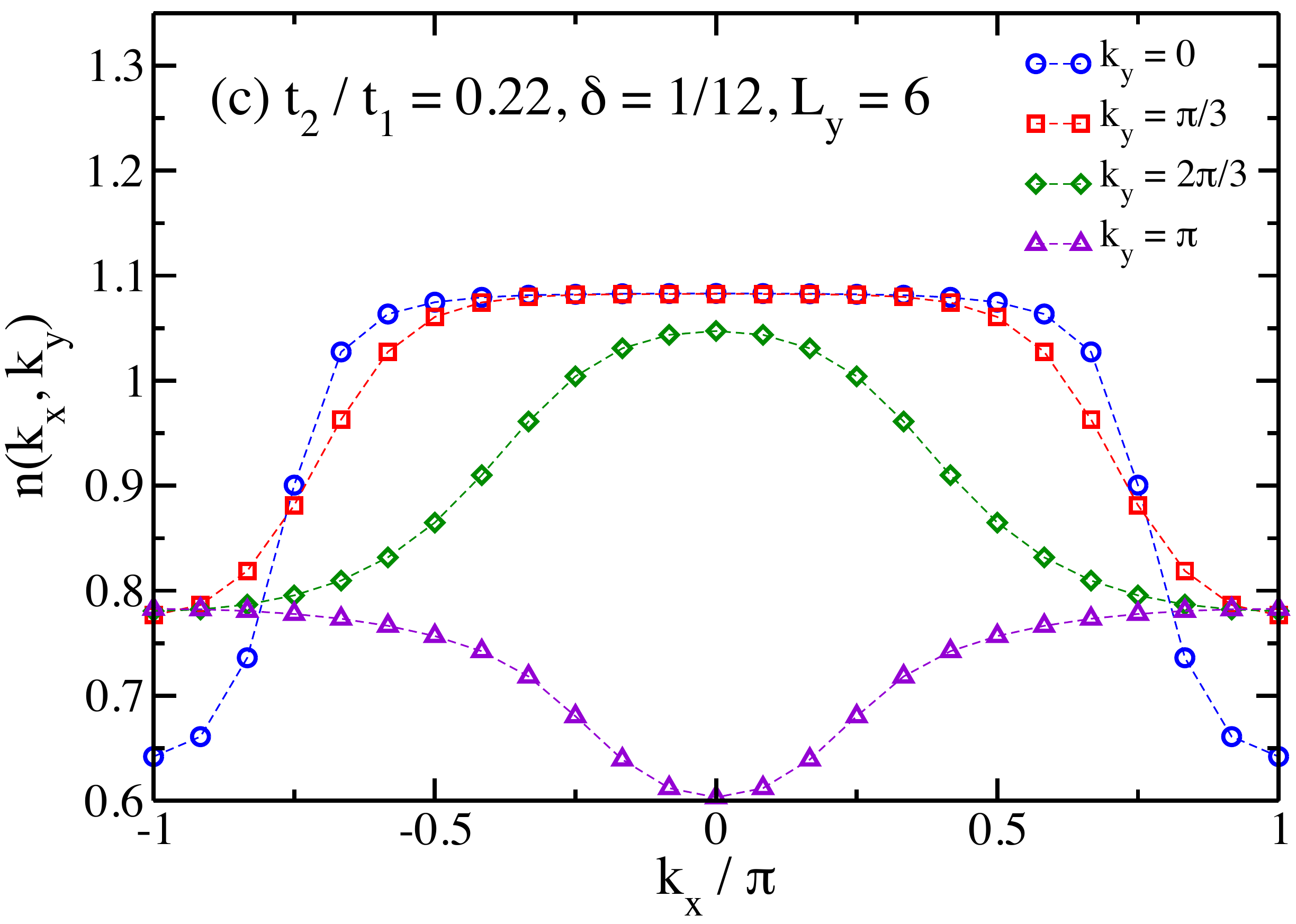}
\caption{Momentum distribution functions $n(\bf k)$ for different $t_2 / t_1$ at the doping level $\delta = 1/12$.
The data are the same as those shown in Fig.~\ref{supfig:nk}. Here the results for $k_y = 0, \pi/3, 2\pi/3$, and $\pi$ are shown separately.
}
\label{supfig:nk_cut}
\end{figure}

\begin{figure}[htp]
\includegraphics[width=0.8\linewidth]{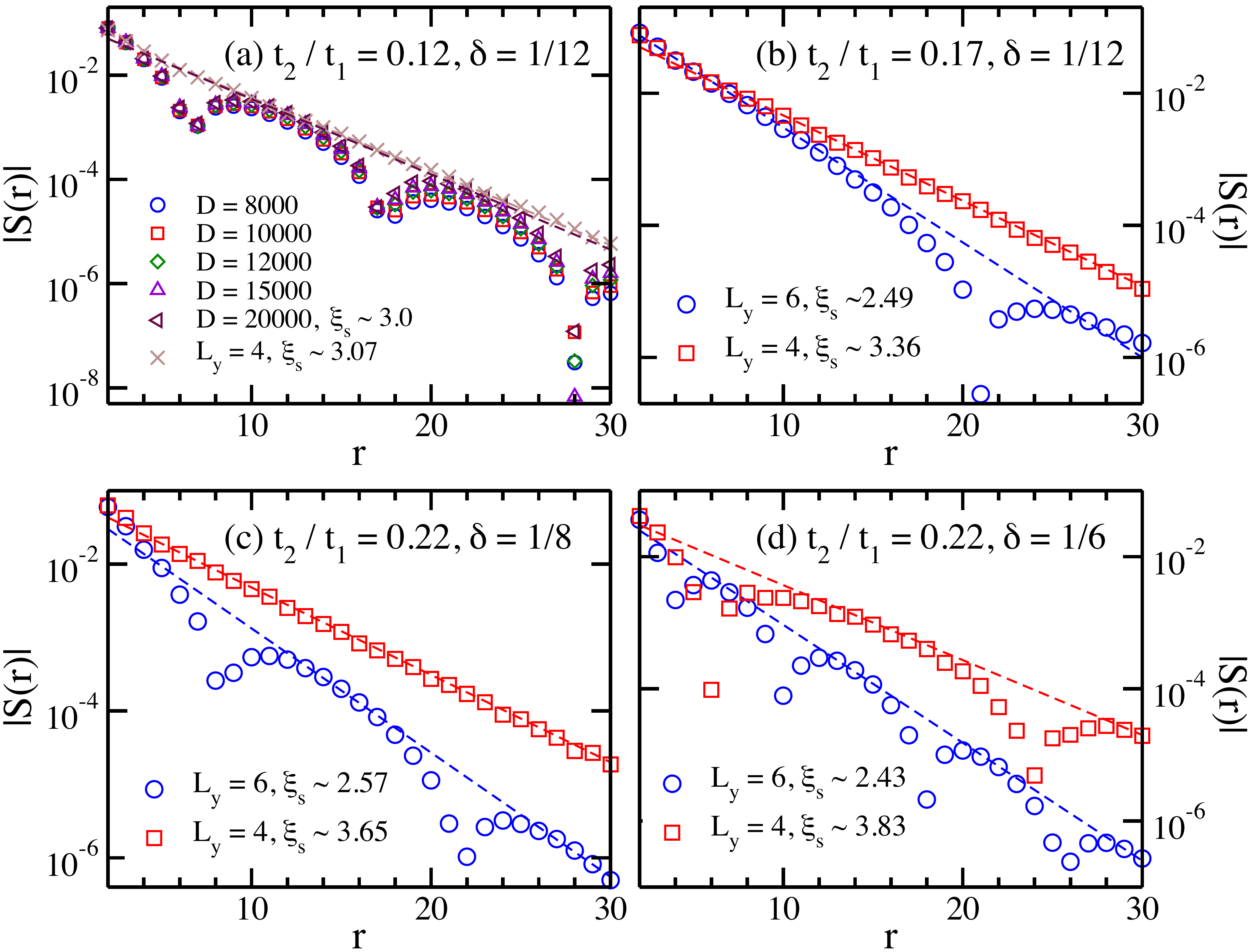}
\caption{Spin correlation functions in the d-wave superconducting phase.
(a) Semi-logarithmic plot of spin correlations $|S(r)|$ for $t_2 / t_1 = 0.12, \delta = 1/12$. The results of the $L_y = 6$ system are shown for the bond dimensions $D = 8000 - 20000$. The exponential fitting of $|S(r)| \sim e^{-r / \xi_{s}}$ for $D = 20000$ gives $\xi_{s} \simeq 3.0$. The spin correlations for $L_y = 4$ have $\xi_{s} \simeq 3.07$. 
(b)-(d) show the semi-logarithmic plots for $t_2 / t_1 = 0.17, \delta = 1/12$ ($D = 12000$),  $t_2 / t_1 = 0.22, \delta = 1/8$ ($D = 15000$), and $t_2 / t_1 = 0.22, \delta = 1/6$ $(D = 12000)$, respectively. Spin correlations decay faster with growing circumference from $L_y = 4$ to $L_y = 6$.}
\label{supfig:spin_dwave}
\end{figure}

\begin{figure}[htp]
\includegraphics[width=0.8\linewidth]{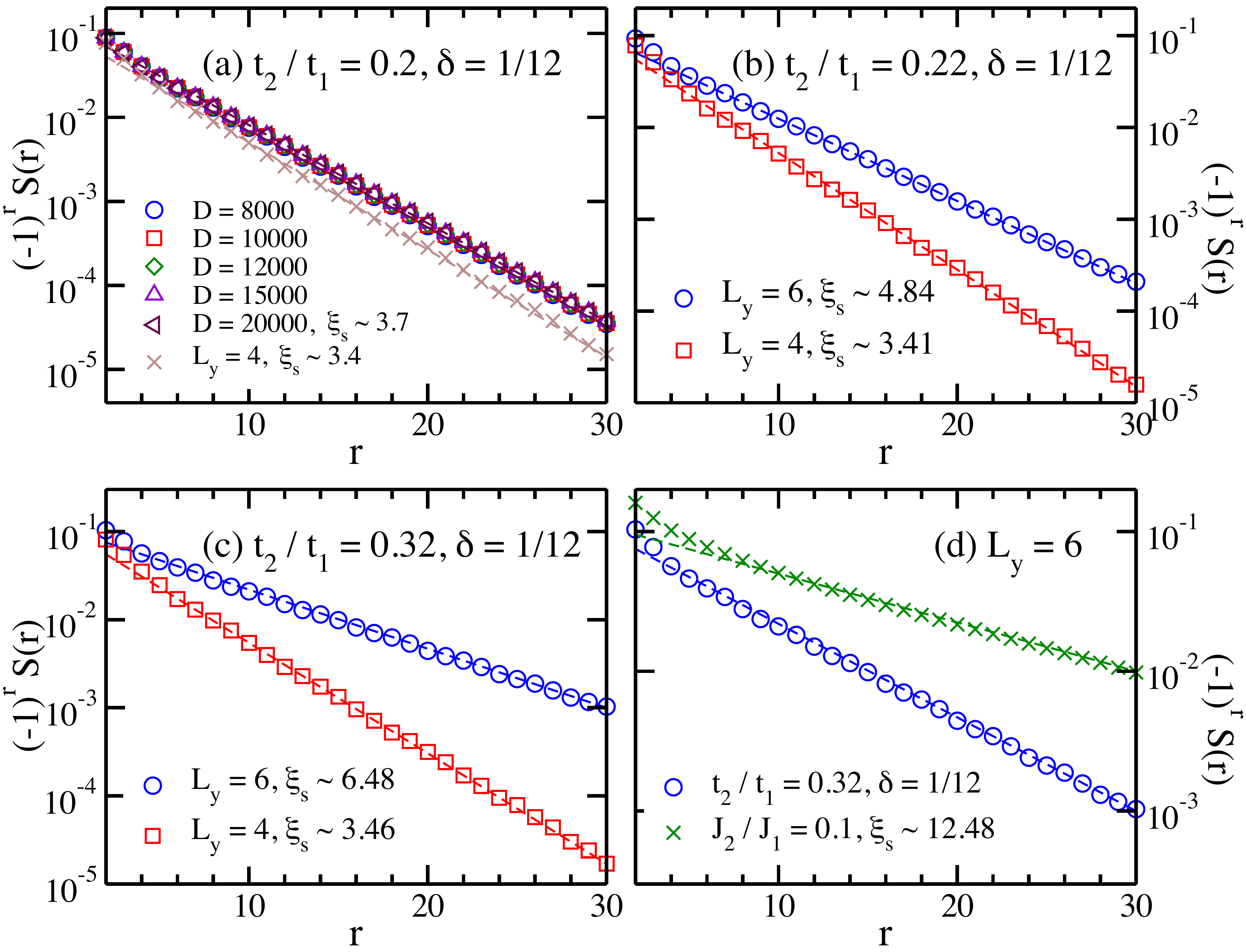}
\caption{Spin correlation functions in the coexistence phase.
(a) Semi-logarithmic plot of spin correlations for $t_2 / t_1 = 0.2, \delta = 1/12$. The results of the $L_y = 6$ system are shown for the bond dimensions $D = 8000 - 20000$. The exponential fitting of $|S(r)| \sim e^{-r / \xi_{s}}$ for $D = 20000$ gives $\xi_{s} \simeq 3.7$. The spin correlations for $L_y = 4$ have $\xi_{s} \simeq 3.4$.
(b) and (c) show the semi-logarithmic results for $t_2 / t_1 = 0.22, \delta = 1/12$ ($D = 15000$) and $t_2 / t_1 = 0.32, \delta = 1/12$ ($D = 12000$), respectively. For both $L_y = 4$ and $6$, spin correlations all have the N\'eel-type oscillation with spin structure factor peak at ${\bf k} = (\pi, \pi)$. For all the cases, spin correlations enhance with growing circumference from $L_y = 4$ to $L_y = 6$. For $L_y = 6$, $\xi_{s}$ quickly increases with growing $t_2$. (d) shows the spin correlations for $t_2 / t_1 = 0.32, \delta = 1/12$ and the $J_1 - J_2$ square Heisenberg model with $J_2 / J_1 = 0.1$ on the $L_y = 6$ cylinder for a comparison. 
}
\label{supfig:spin_coexistence}
\end{figure}

\begin{figure}[htp]
\includegraphics[width=0.245\linewidth]{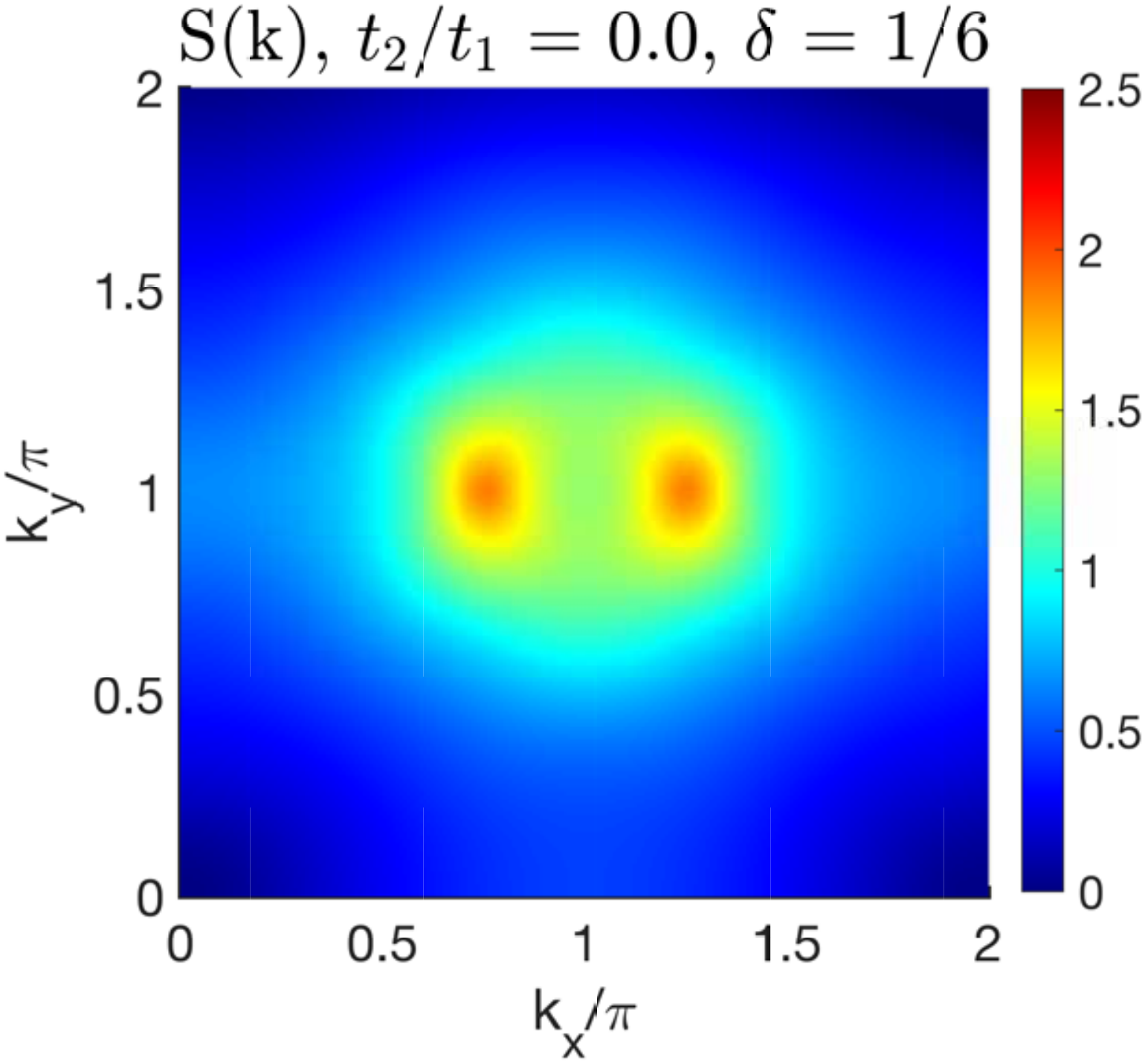}
\includegraphics[width=0.245\linewidth]{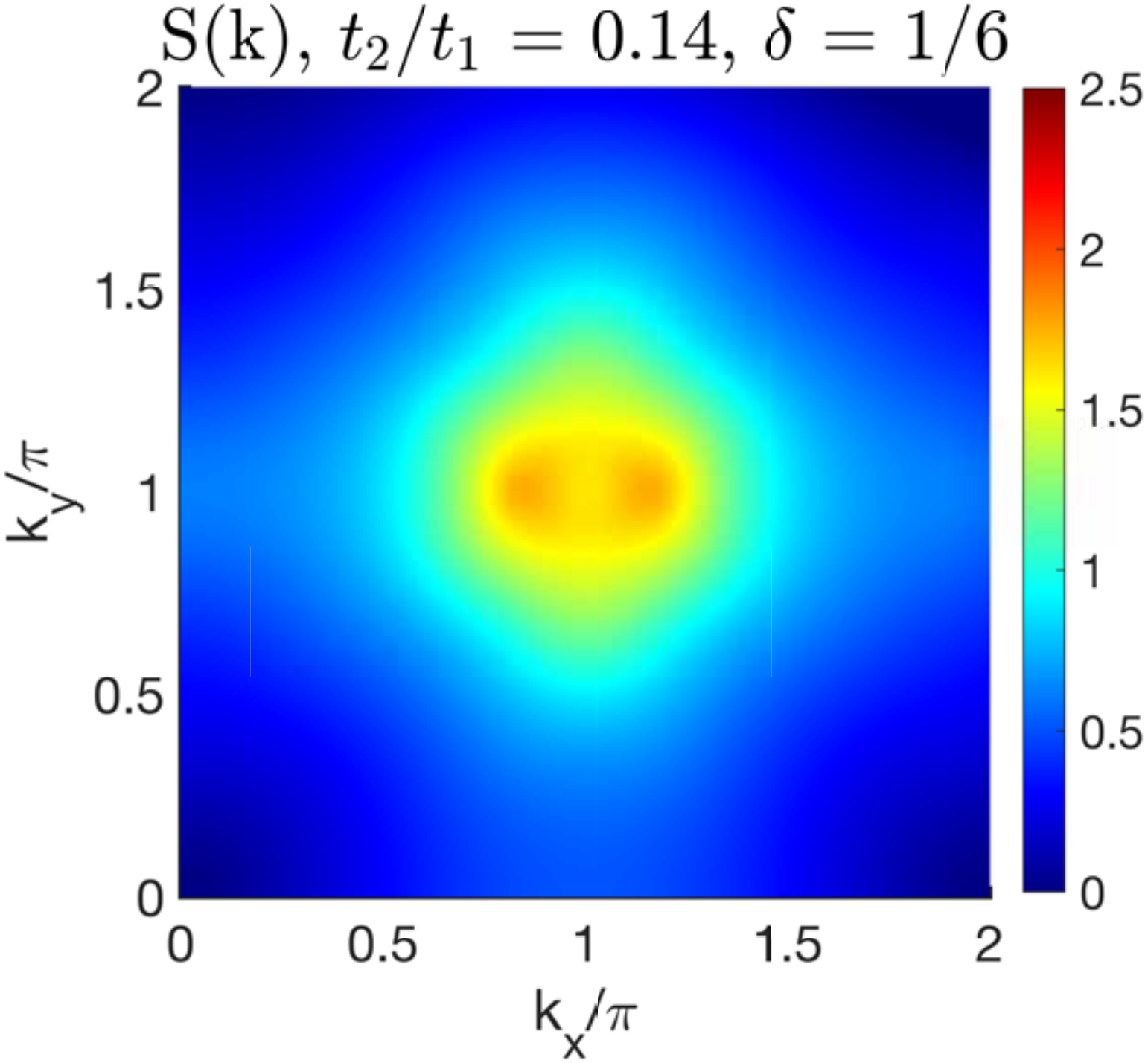}
\includegraphics[width=0.245\linewidth]{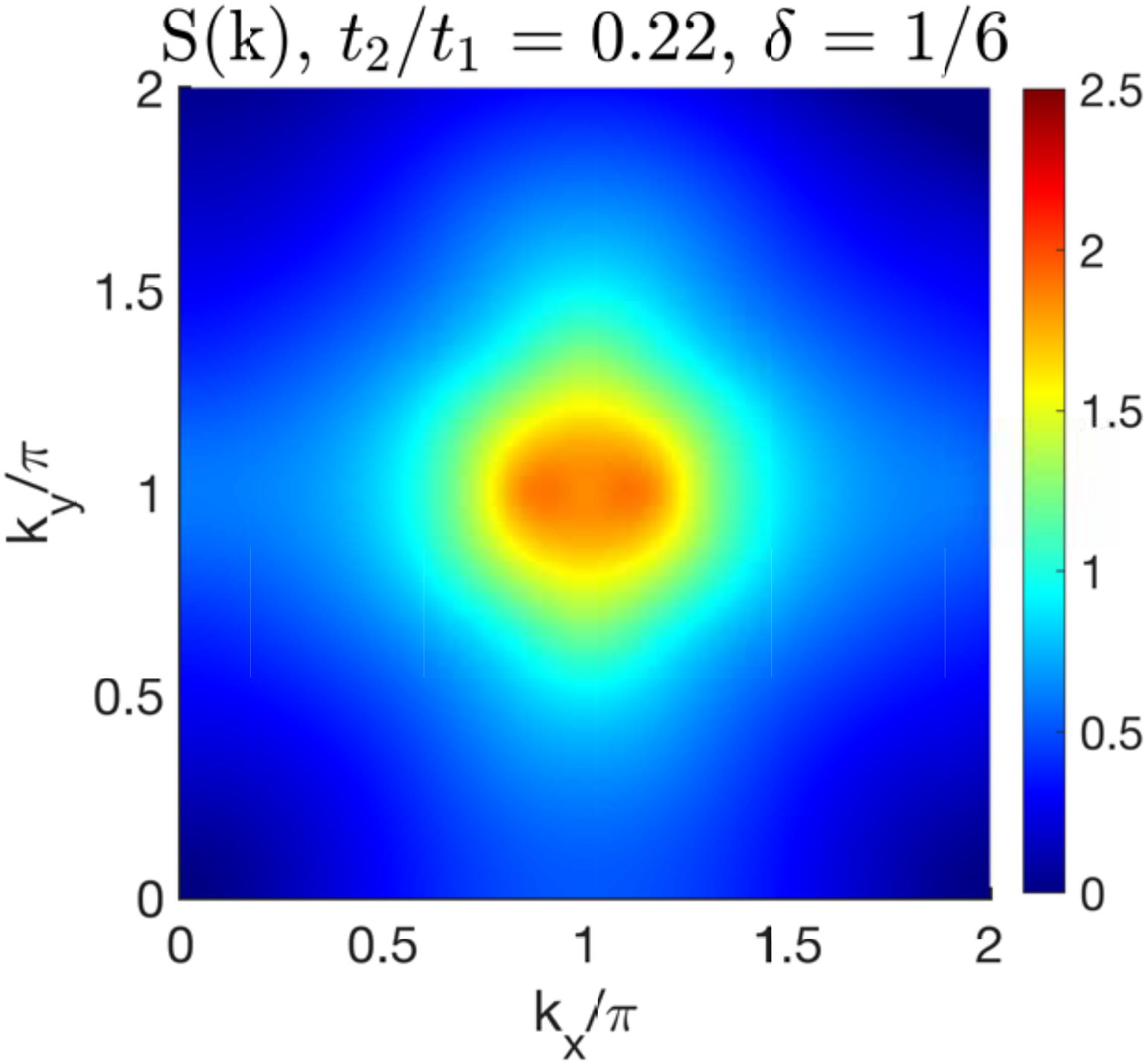}
\includegraphics[width=0.245\linewidth]{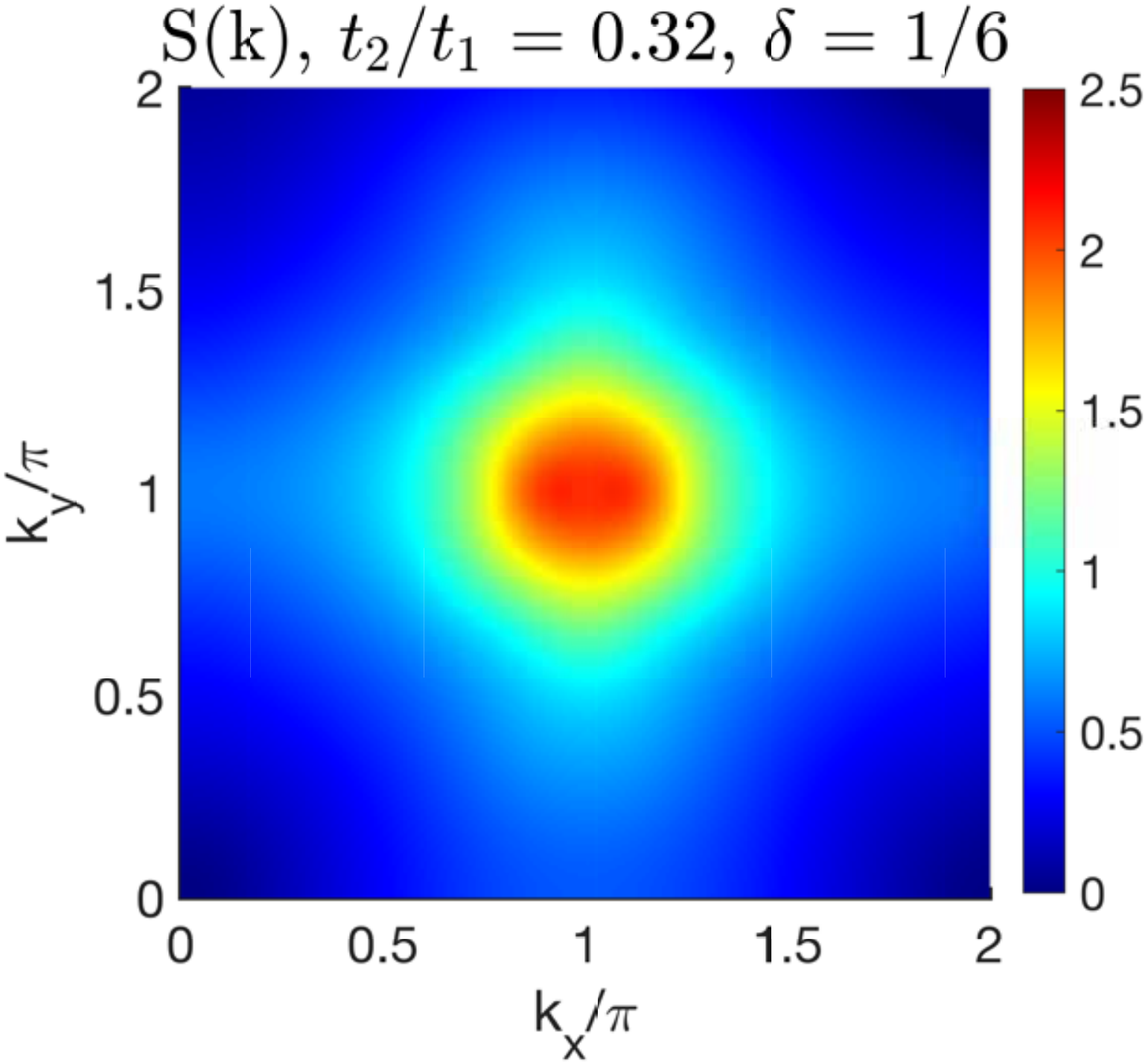}
\includegraphics[width=0.245\linewidth]{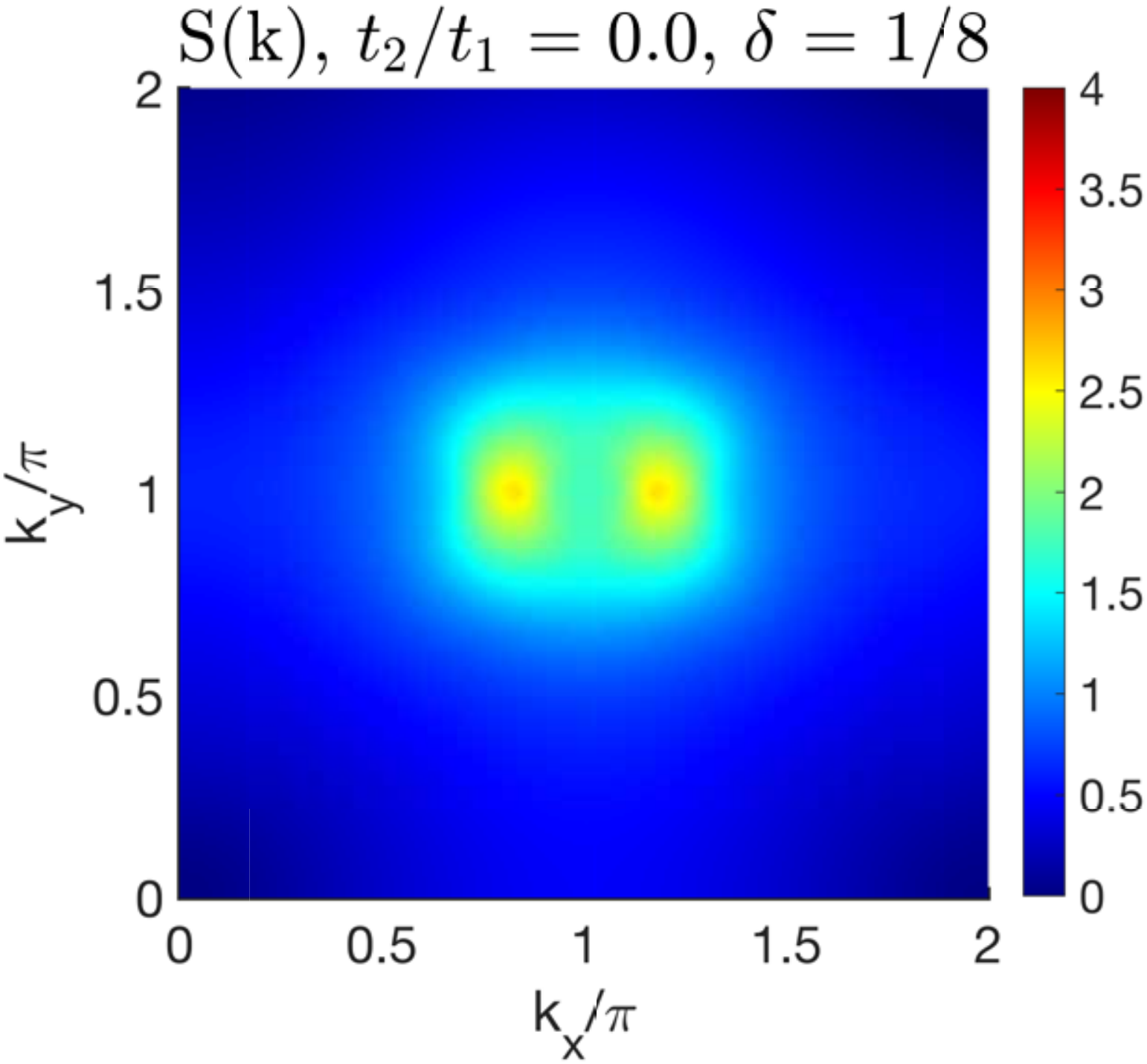}
\includegraphics[width=0.245\linewidth]{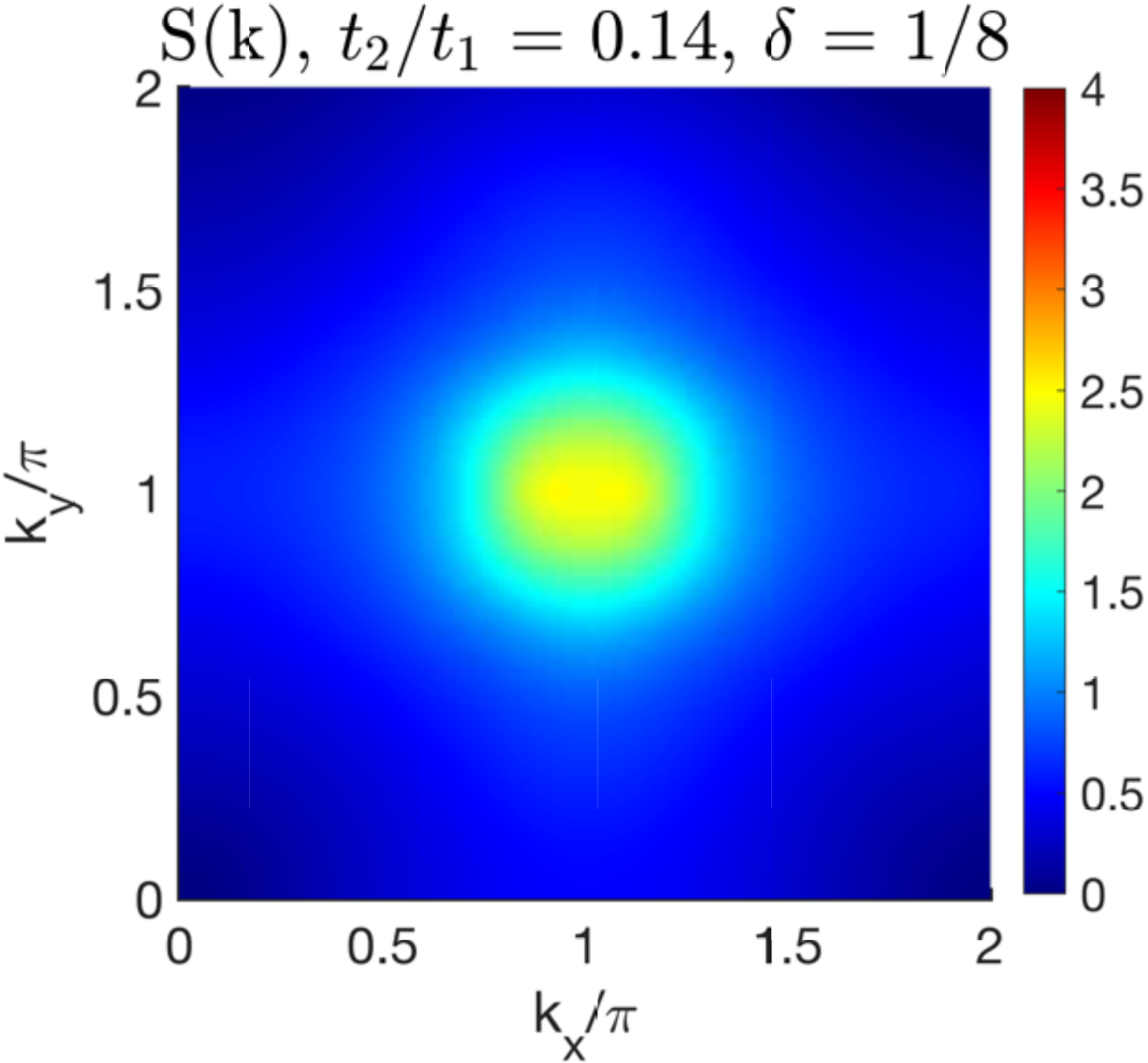}
\includegraphics[width=0.245\linewidth]{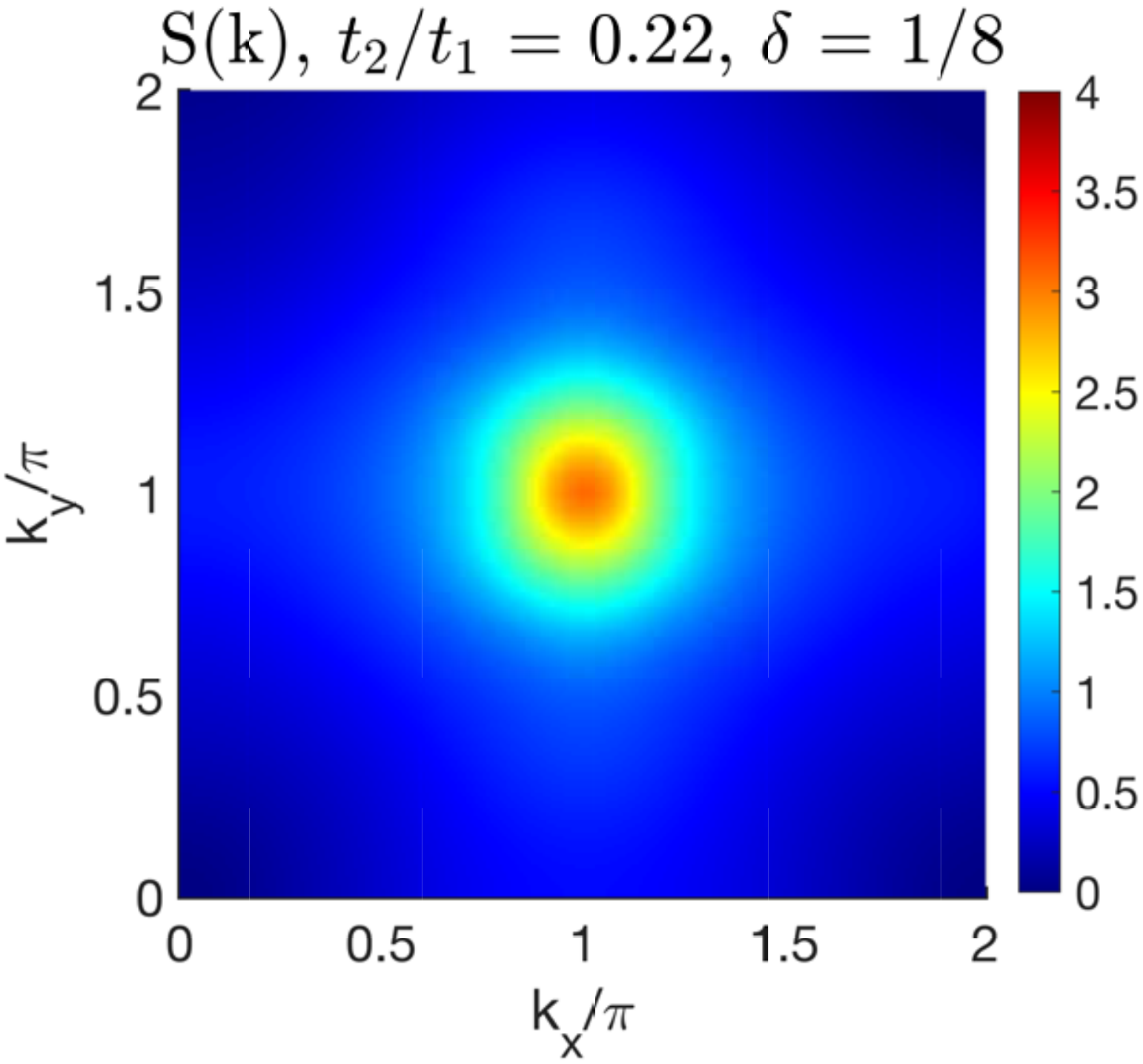}
\includegraphics[width=0.245\linewidth]{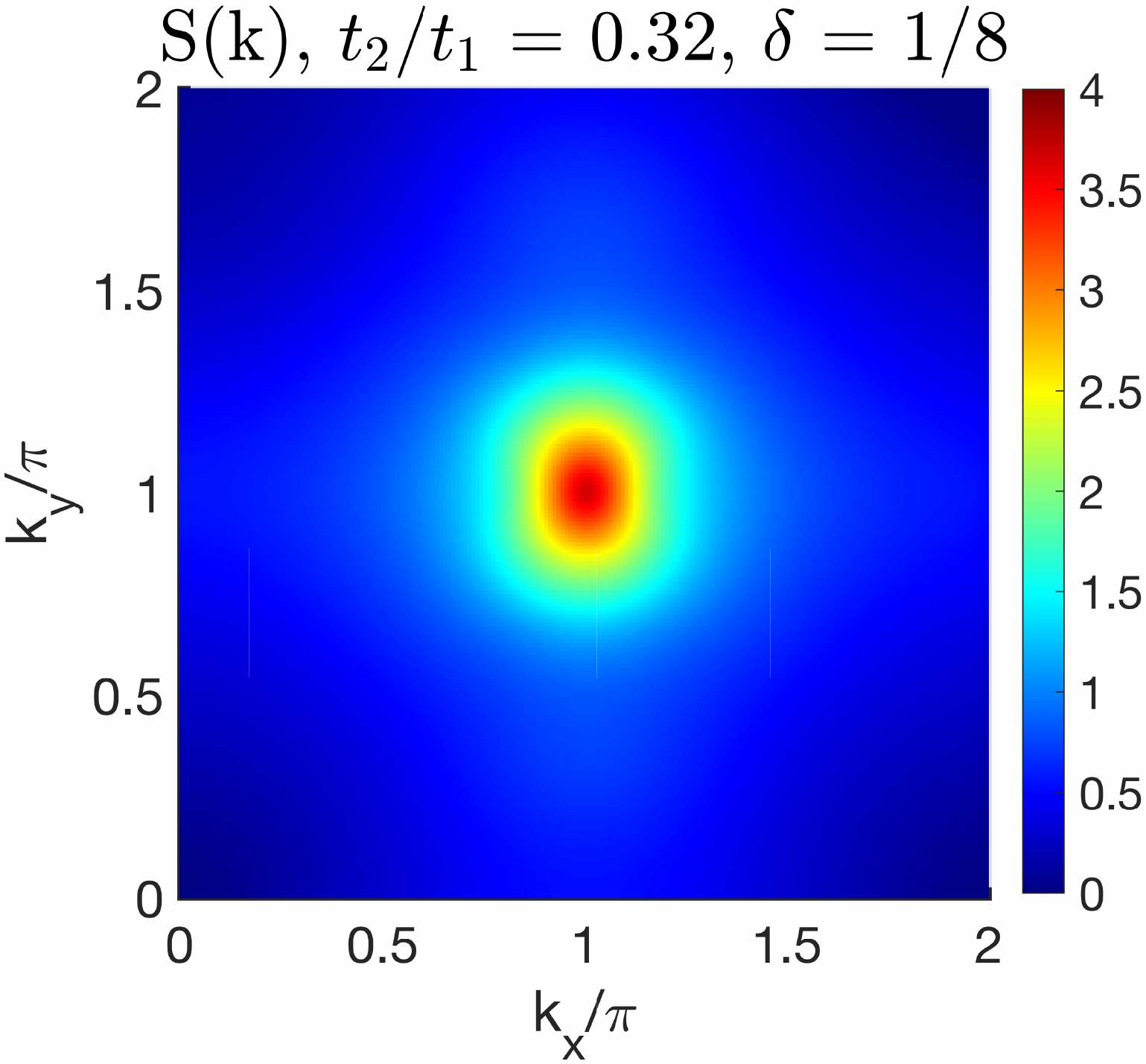}
\includegraphics[width=0.245\linewidth]{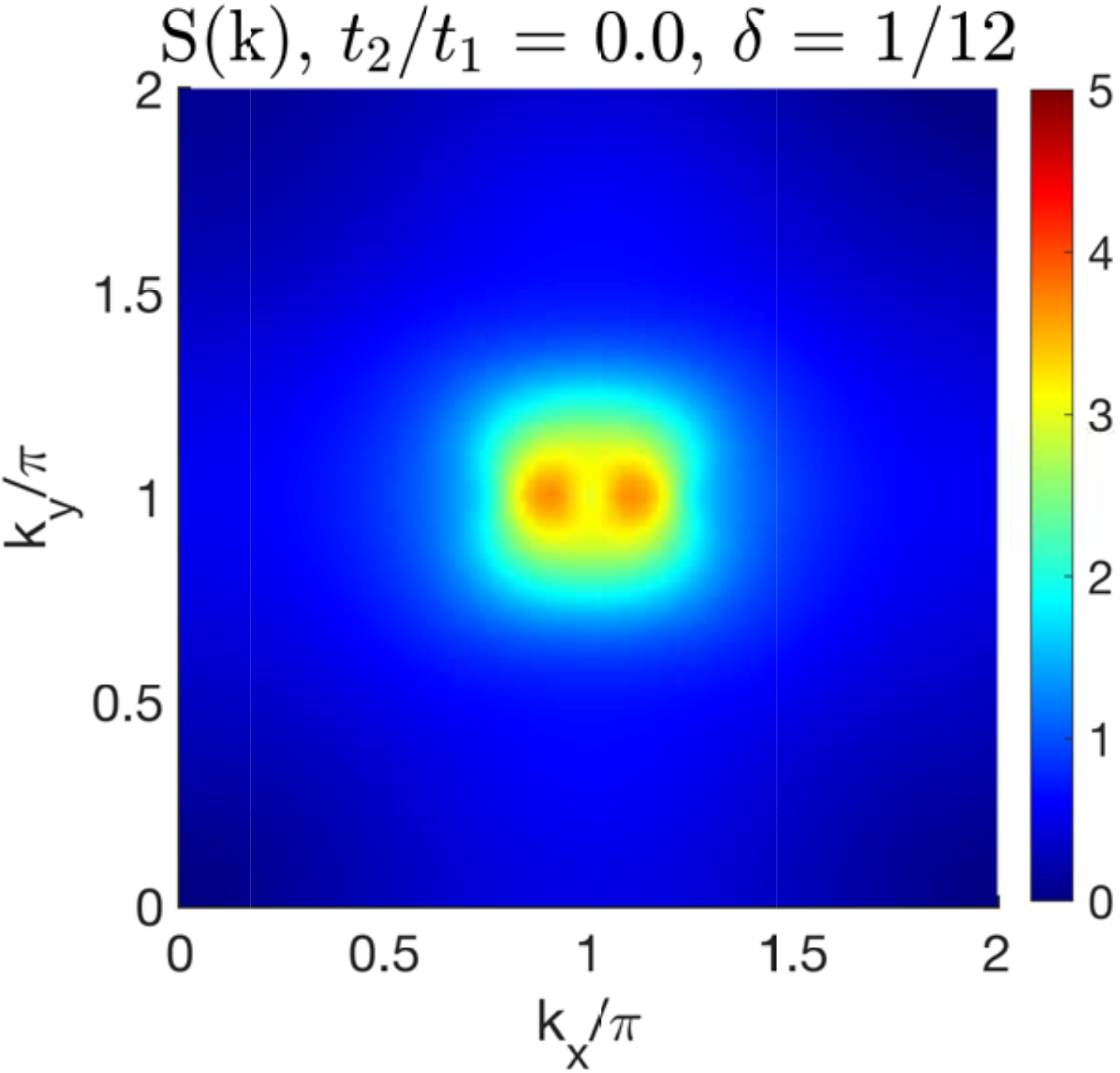}
\includegraphics[width=0.245\linewidth]{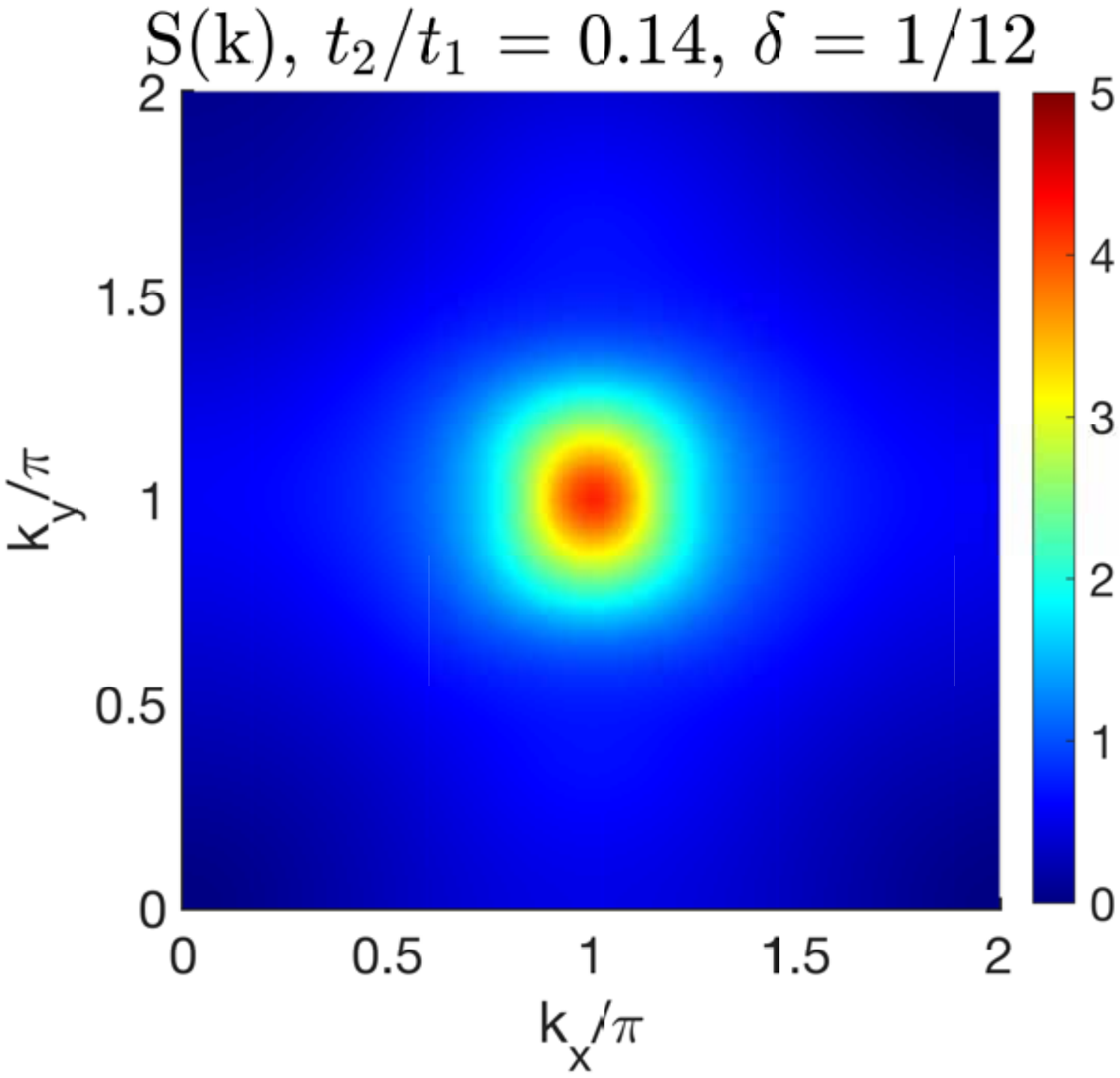}
\includegraphics[width=0.245\linewidth]{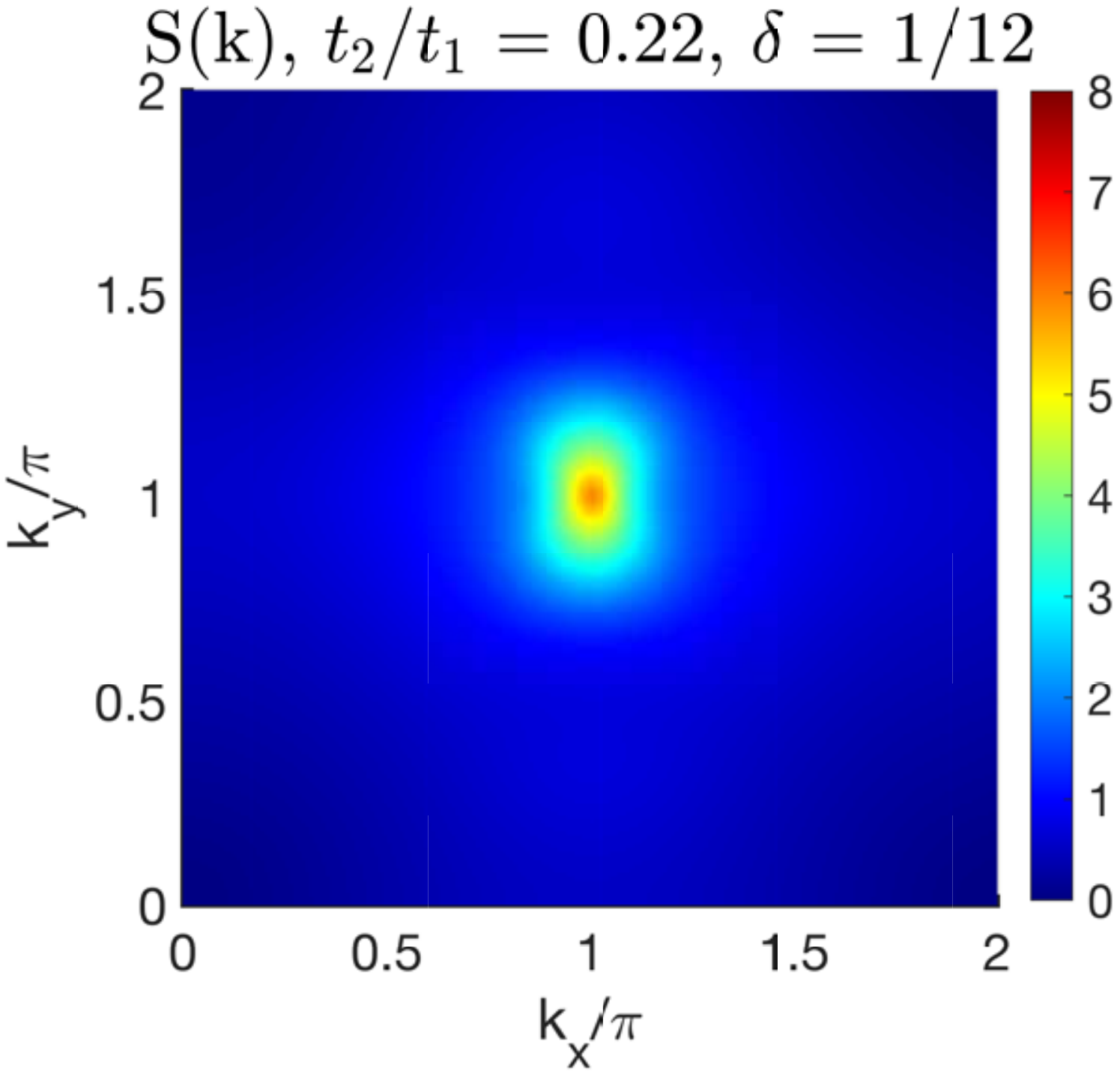}
\includegraphics[width=0.245\linewidth]{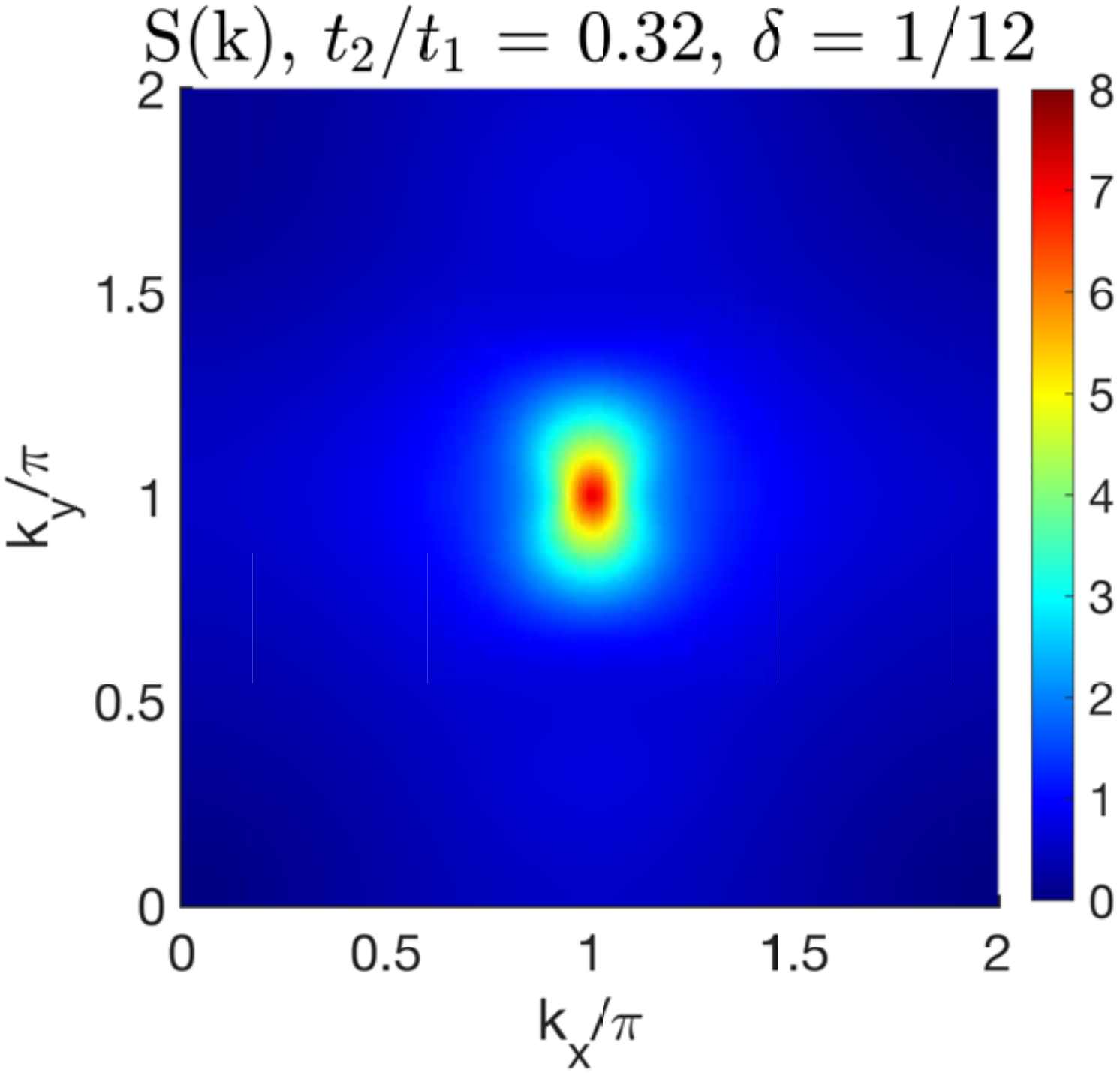}
\caption{Spin structure factors $S(\bf k)$ for different $t_2$ and doping ratios $\delta$.
$S(\bf k)$ are obtained by taking the Fourier transformation for the spin correlations of the middle $6\times 24$ sites on the $6 \times 48$ cylinder. In the CDW phase, $S(\bf k)$ shows two round peaks near ${\bf k} = (\pi, \pi)$. With growing $t_2 / t_1$, the two peaks gradually move towards ${\bf k} = (\pi, \pi)$ and change slowly in the d-wave SC phase. In the SC + CDW coexistent phase, $S(\bf k)$ shows an enhanced peak at ${\bf k} = (\pi, \pi)$, as depicted by $t_2 / t_1 = 0.32, \delta = 1/12$. These results are obtained by keeping the bond dimension $D = 10000$.
}
\label{supfig:sq_spin}
\end{figure}

\begin{figure}[htp]
\includegraphics[width=0.8\linewidth]{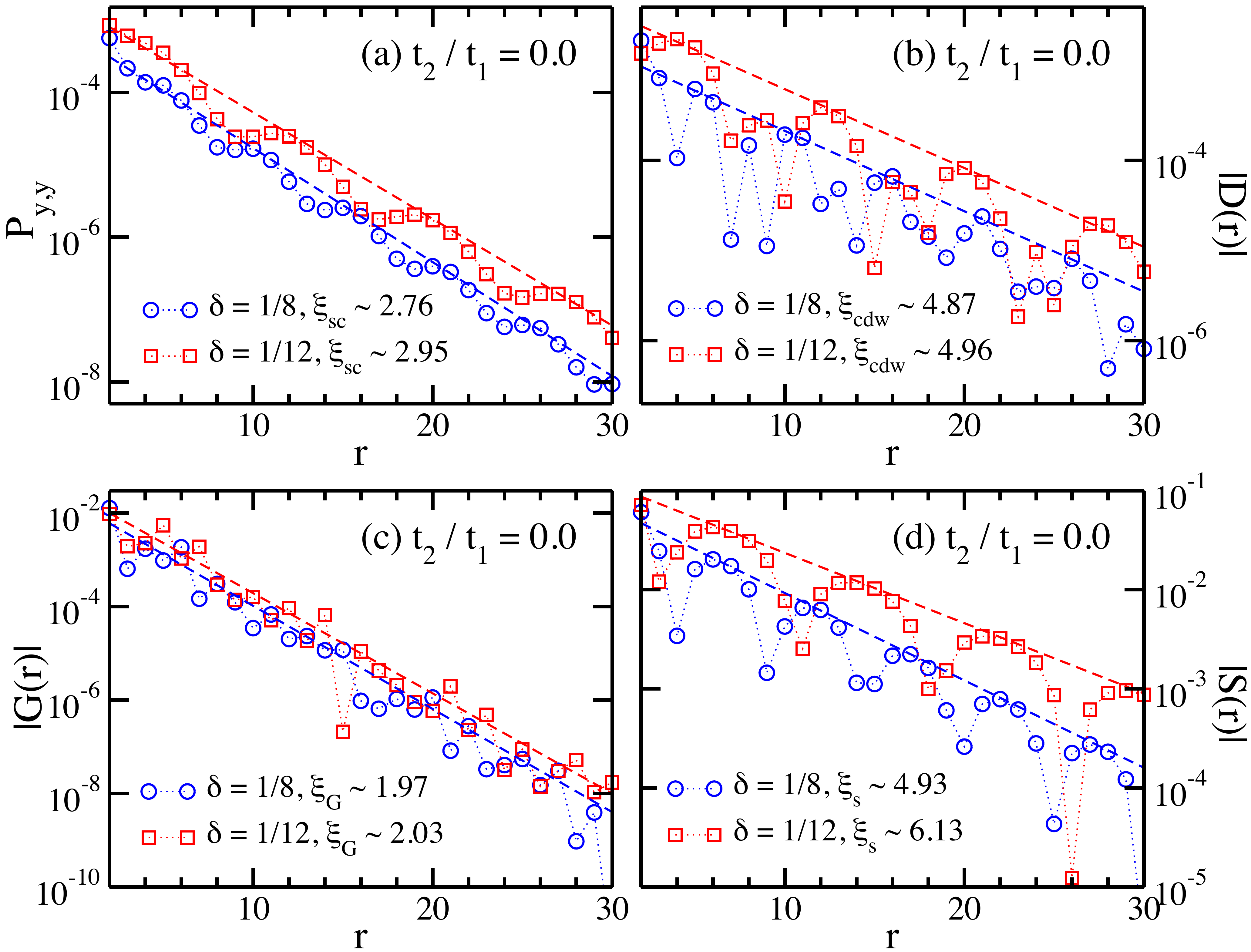}
\caption{Correlation functions in the charge density wave phase.
The different correlation functions are shown for $t_2 / t_1 = 0.0$ at the doping ratios $\delta = 1/8$ and $1/12$, including (a) the SC pairing correlation $P_{y,y}$, (b) the density correlation $D(r)$, (c) the single-particle correlation $G(r)$, and (d) the spin correlation $S(r)$. These results are obtained by keeping the bond dimension $D = 10000$.
}
\label{subfig:correlation_cdw}
\end{figure}

\begin{figure}[htp]
\includegraphics[width=0.8\linewidth]{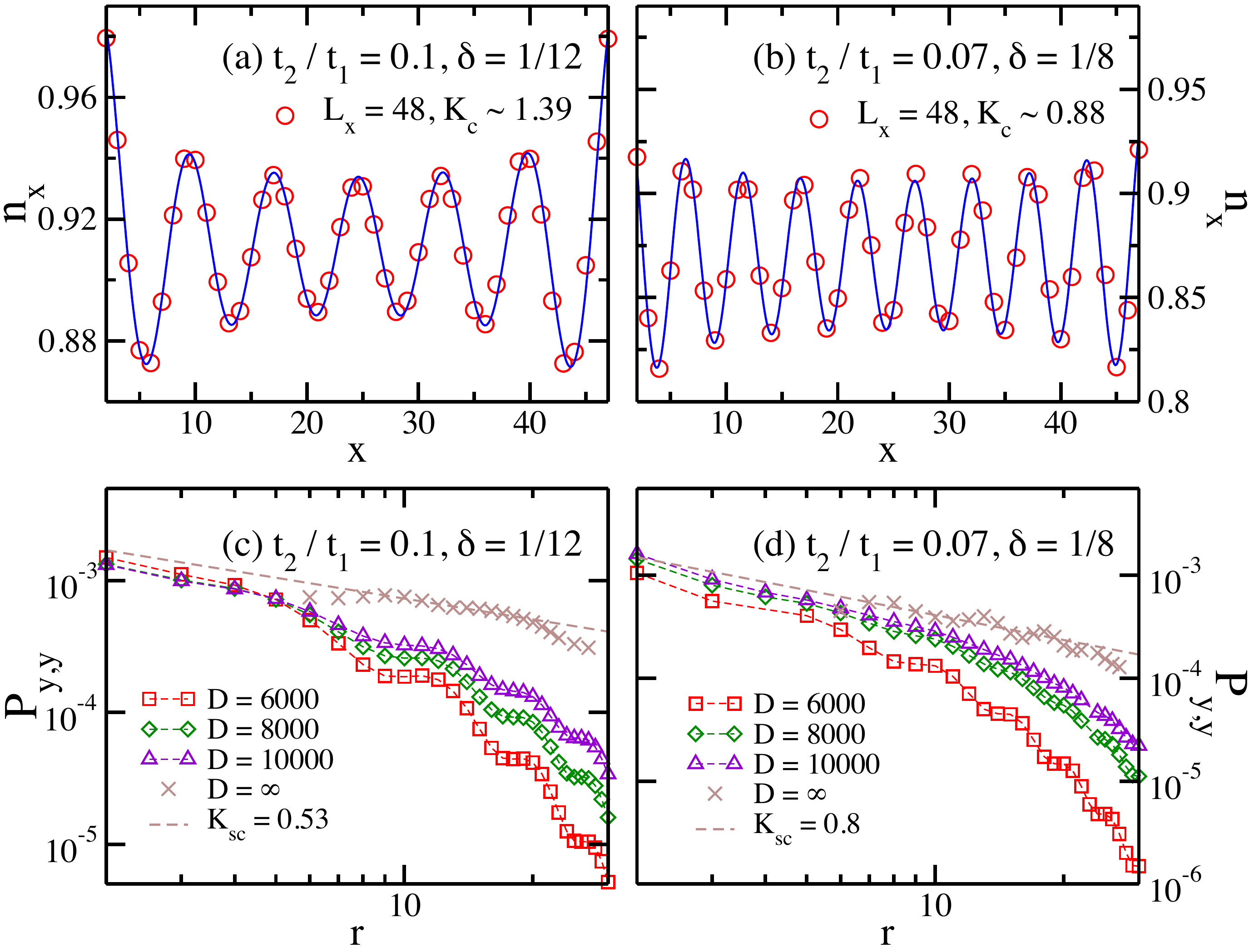}
\caption{Charge density profile and SC pairing correlation near the boundary between the CDW and the uniform d-wave SC phase. (a) and (b) are the charge density profiles for $t_2 / t_1 = 0.1, \delta = 1/12$ and $t_2 / t_1 = 0.07, \delta = 1/8$ on the $L_y = 6, L_x = 48$ cylinder. The DMRG data are shown as the red circles, which are also fitted by the formula $n(x) = n_0(x) + A_0 \cos(Qx + \phi) [x^{-K_c / 2} + (L_x + 1 - x)^{-K_c / 2}]$. (c) and (d) are the double-logarithmic plots of the SC pairing correlations $P_{y,y}(r)$ by keeping the bond dimensions $D = 6000, 8000, 10000$. The $D = \infty$ data are obtained by using the polynomial extrapolation of the DMRG data, which gives the Luttinger exponent $K_{sc} \simeq 0.53$ and $K_{sc} \simeq 0.8$, respectively.
}
\label{subfig:boundary}
\end{figure}

\begin{figure}[htp]
\includegraphics[width=0.8\linewidth]{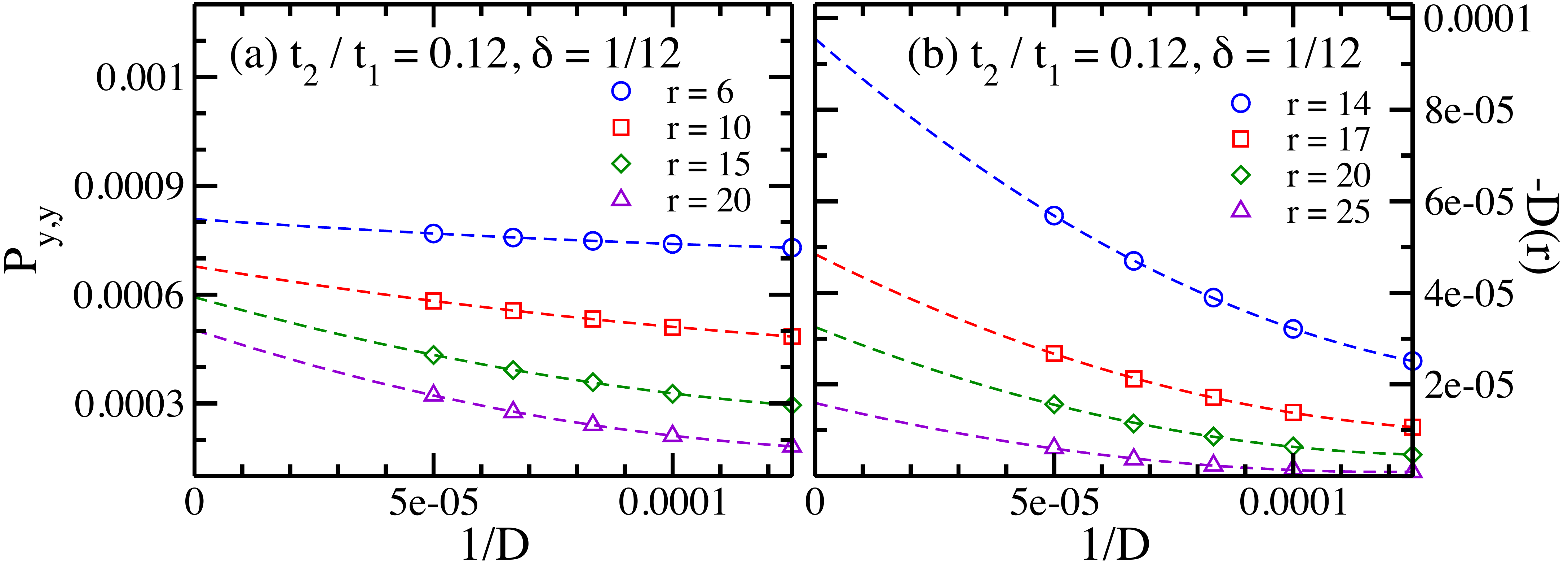}
\caption{Extrapolations of correlation functions versus the bond dimension.
(a) and (b) show the extrapolations of the SC pairing correlation function $P_{y,y}(r)$ and the density correlation function $-D(r)$ for $t_2 / t_1 = 0.12, \delta = 1/12$ in the d-wave SC phase. $D$ is the $SU(2)$ bond dimension, which corresponds to $D = 8000, 10000, 12000, 15000, 20000$ here. The different symbols denote the correlations at the different given distance $r$. For each given distance $r$, the correlations obtained by different bond dimensions are extrapolated by the polynomial function $C(1/D) = C(0) + a / D + b / D^2$. Please note that the plotting scales in (a) and (b) are different.
}
\label{subfig:m_scaling}
\end{figure}

\begin{figure}[htp]
\includegraphics[width=0.8\linewidth]{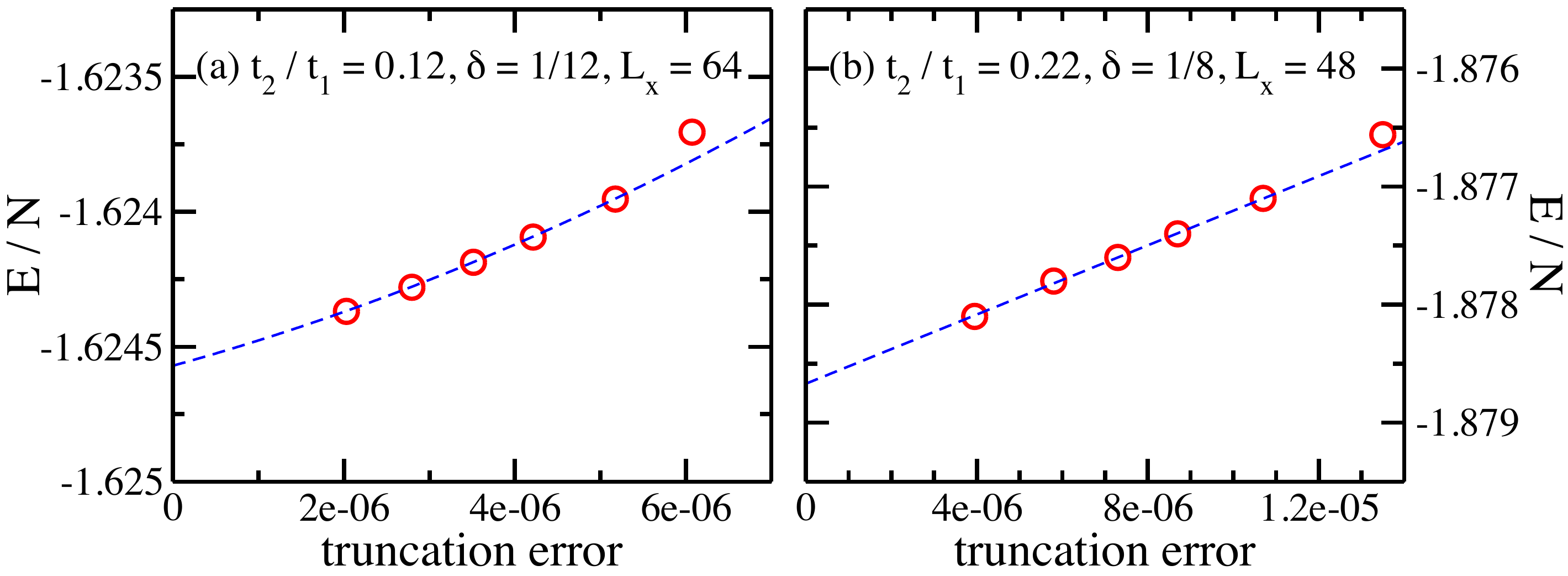}
\caption{Extrapolations of total energy per site versus the DMRG truncation error.
(a) and (b) show the extrapolations of the total energy per site $E / N$ versus the DMRG truncation error, for $t_2 / t_1 = 0.12, \delta = 1/12$ on the $L_x = 64$ cylinder and $t_2 / t_1 = 0.22, \delta = 1/8$ on the $L_x = 48$ cylinder, respectively. The dashed lines indicate the energy extrapolations using the polynomial fitting up to the second order of truncation error. 
}
\label{subfig:energy}
\end{figure}

\end{document}